\documentclass[fleqn,usenatbib,useAMS]{mnras}

\usepackage{graphicx}	
\usepackage{amsmath}	
\usepackage{amssymb}	
\usepackage{multicol}        
\usepackage{pdflscape}	
\usepackage{verbatim}
\usepackage{siunitx}
\usepackage{placeins}
\usepackage{gensymb}
\usepackage[dvipsnames]{xcolor}

\usepackage[T1]{fontenc}
\usepackage{ae,aecompl}

\usepackage{newtxtext,newtxmath}

\title[Characterisation of Cool TESS Candidate Planet Hosts]{Characterisation of 92 Southern TESS Candidate Planet Hosts and a New Photometric [Fe/H] Relation for Cool Dwarfs}

\author[Adam D. Rains et al.]{Adam D. Rains,$^{1}$\thanks{E-mail: adam.rains@anu.edu.au (ADR)}
Maru\v{s}a \v{Z}erjal,$^{1}$
Michael J. Ireland,$^{1}$
Thomas Nordlander,$^{1,2}$ \newauthor
Michael S. Bessell,$^{1}$
Luca Casagrande,$^{1,2}$
Christopher A. Onken,$^{1,3}$
Meridith Joyce,$^{1,2}$ \newauthor
Jens Kammerer,$^{1,4}$ 
and Harrison Abbot$^{1}$
\\
$^{1}$Research School of Astronomy and Astrophysics, Australian National University, Canberra, ACT 2611, Australia\\
$^{2}$ARC Centre of Excellence for All Sky Astrophysics in 3 Dimensions (ASTRO 3D)\\
$^{3}$Centre for Gravitational Astrophysics, Research Schools of Physics, and Astronomy and Astrophysics, Australian National University\\
$^{4}$European Southern Observatory, Karl-Schwarzschild-Str 2, 85748, Garching, Germany
}

\date{Last updated 2015 May 22; in original form 2013 September 5}

\pubyear{2015}

\begin{document}
\label{firstpage}
\pagerange{\pageref{firstpage}--\pageref{lastpage}}
\maketitle

\begin{abstract}
We present the results of a medium resolution optical spectroscopic survey of 92 cool ($3,000 \lesssim T_{\rm eff} \lesssim 4,500\,$K) southern TESS candidate planet hosts, and describe our spectral fitting methodology used to recover stellar parameters. We quantify model deficiencies at predicting optical fluxes, and while our technique works well for $T_{\rm eff}$, further improvements are needed for [Fe/H]. 
To this end, we developed an updated photometric [Fe/H] calibration for isolated main sequence stars built upon a calibration sample of 69 cool dwarfs in binary systems, precise to $\pm0.19\,$dex, from super-solar to metal poor, over $1.51 < {\rm Gaia}~(B_P-R_P) < 3.3$.
Our fitted $T_{\rm eff}$ and $R_\star$ have median precisions of 0.8\% and 1.7\%,  respectively and are consistent with our sample of standard stars. We use these to model the transit light curves and determine exoplanet radii for 100 candidate planets to 3.5\% precision and see evidence that the planet-radius gap is also present for cool dwarfs.
Our results are consistent with the sample of confirmed TESS planets, with this survey representing one of the largest uniform analyses of cool TESS candidate planet hosts to date.

\end{abstract}



\section{Introduction}
Low mass stars are the most common kind of star in the Galaxy, comprising more than two thirds of all stars \citep{chabrier_galactic_2003}, and dominating the Solar Neighbourhood population \citep[e.g.][]{henry_solar_1994, henry_solar_2006, winters_solar_2015, henry_solar_2018}. This abundance alone makes them prime targets for planet searches, with microlensing surveys, which have very little bias on host star masses, revealing that there is at least one bound planet per Milky Way star \citep{cassan_one_2012}. Results from the Kepler Mission \citep{borucki_kepler_2010} also bear this out, showing that a large number of planets remain undiscovered around cool dwarfs \citep{morton_radius_2014}, and that such cool stars are actually more likely to host small planets ($2 < R_P < 4\,R_\oplus$, where $R_P$ and $R_\oplus$ are the planet and earth radius respectively) than their hotter counterparts \citep{howard_planet_2012, dressing_occurrence_2015}. 

However, the inherent faintness of these stars complicates the study of both them and their planets. While we now know of over 4,000 confirmed planets orbiting stars other than our own (overwhelmingly discovered by transiting exoplanet surveys), almost an equal number await confirmation\footnote{\url{https://exoplanetarchive.ipac.caltech.edu/}}. Exoplanet transit surveys like Kepler and TESS \citep{ricker_transiting_2015} are able to place tight constraints on planetary radii given a known stellar radius, but follow-up precision radial velocity observations are required to provide planetary mass constraints. This is the second reason why planet searches around low mass stars are critical: their smaller radii and lower masses make the transit signals and radial velocities of higher amplitudes for any planets they host as compared to the same planets around more massive host stars. This is especially important when looking for planets with terrestrial radii or masses respectively. 

Many planet host stars have never been targeted by a spectroscopic survey, leaving their properties to be estimated through photometry alone. For instance, the TESS input catalogue \citep{stassun_tess_2018, stassun_revised_2019} based its stellar parameters primarily on photometry, having spectroscopic properties for only about 4 million stars of the nearly 700 million with photometrically estimated equivalents. While stars warmer than 4,000$\,$K are well suited to bulk estimation of properties from photometry \citep[see e.g.][]{carrillo_know_2020}, special care must be taken for cool dwarfs whose faintness and complex atmospheres make such relations more complex to develop and implement (e.g. see \citealt{muirhead_catalog_2018} for the K and M dwarf specific approach taken from the TESS input catalogue).

NASA's TESS Mission, by virtue of being all sky, has given us a wealth of bright candidates which are now being actively followed up by ground based spectroscopic surveys.
While multi-epoch radial velocity observations are required to determine planetary masses, these surveys are typically biased towards the brightest stars and smallest planets. As such, there remains a need for single-epoch spectroscopic follow-up of fainter targets to provide reliable host star properties (primarily $T_{\rm eff}$, $\log g$, [Fe/H], and the stellar radius $R_\star$) and allow radial constraints to be placed on transiting planet candidates. Indeed, the LAMOST Survey \citep{zhao_lamost_2012} undertook targeted low resolution spectroscopic follow-up of stars in the Kepler field \citep{de_cat_lamost_2015} with the goal of deriving spectroscopic stellar properties. Considering the goal of planet radii determination specifically, \citet{dressing_characterizing_2019} used medium-resolution near-infrared (NIR) spectra, and \citet{wittenmyer_k2-hermes_2020} high-resolution optical spectra to follow-up K2 \citep{howell_k2_2014} transiting planet candidate hosts and place radius constraints on both planets and their hosts.

Even without mass estimates, much can be learned about exoplanet demographics from their radii alone. As demonstrated by \citet{fulton_california-kepler_2017}, \citet{fulton_california-kepler_2018}, \citet{van_eylen_asteroseismic_2018}, \citet{kruse_detection_2019}, \citet{hardegree-ullman_scaling_2020}, \citet{cloutier_evolution_2020}, and \citet{hansen_confirming_2021}, having a large sample of precise planet radii allows insight into the exoplanet radius distribution, which appears to be bimodal with an observable gap in the super-Earth regime ($\sim1.8\,R_\oplus$). This is thought to be the result of physical phenomena like photoevaporation \citep[where flux from the parent star strips away weakly held atmospheres, e.g.][]{owen_kepler_2013,lee_make_2014,lopez_understanding_2014,lee_breeding_2016,owen_evaporation_2017,lopez_how_2018}, or core-powered mass loss \citep[where a cooling rocky core erodes light planetary atmospheres via its cooling luminosity, e.g.][]{ikoma_situ_2012, ginzburg_core-powered_2018,gupta_sculpting_2019, gupta_signatures_2020}, and its location likely has a dependence on stellar host mass \citep[e.g.][]{cloutier_evolution_2020}. As such, improving the sample of planets with radius measurements allows us to place observational constraints on planet formation channels and the mechanisms that sculpt planets throughout their lives. 

The scientific importance of searching for planets around low-mass stars to study their demographics is thus clear. However, the exact approach for understanding the stars themselves is less obvious, as cool dwarfs are not as well understood as their prevalence would suggest. Their inherent faintness and atmospheric complexity has lead to long standing issues observing representative sets of standard stars, generating synthetic spectra accounting for molecular absorption as well as consistently modelling their evolution \citep[see e.g.][]{allard_model_1997, chabrier_galactic_2003}.

Analysis of spectra from warmer stars is made simpler by the existence of regions of spectral continuum where atomic or molecular line absorption is minimal, allowing one to disentangle within reasonable uncertainties the effect of [Fe/H] and $T_{\rm eff}$ on an emerging spectrum. 
This is not the case for cool dwarfs for which there is no continuum at shorter wavelengths, with the deepest absorption caused by most notably TiO in the optical and water in the NIR, but also various other oxides or hydrides. The strength of these features is a function of \textit{both} temperature and [Fe/H], making it difficult to ascribe a unique $T_{\rm eff}$-[Fe/H] pair to a given star. 

Despite this complexity, it is possible to take advantage of the relative [Fe/H]-insensitivity of NIR $K$ band magnitudes alongside [Fe/H]-sensitive optical photometry to probe cool dwarf [Fe/H]. This was predicted by theory (see e.g. \citealt{allard_model_1997}, \citealt{baraffe_evolutionary_1998}, and \citealt{chabrier_theory_2000} for summaries), confirmed observationally \citep[][]{delfosse_accurate_2000}, and later formalised into various empirical calibrations \citep{bonfils_metallicity_2005, johnson_metal_2009, schlaufman_physically-motivated_2010, neves_metallicity_2012, hejazi_optical-near_2015, dittmann_calibration_2016}.

The last decade has seen a number of studies using low-medium resolution (mostly NIR) spectra, often focused on the development of [Fe/H] relations based on spectral indices (e.g. optical-NIR: \citealt{mann_full_2013, mann_spectro-thermometry_2013, mann_how_2015, kuznetsov_characterization_2019}; NIR: \citealt{newton_near-infrared_2014}; H band: \citealt{terrien_h-band_2012}; K band: \citealt{rojas-ayala_metal-rich_2010, rojas-ayala_metallicity_2012}). Other studies have opted to use high-resolution spectra which gives access to unblended atomic lines that are not accessible to lower resolution observations (e.g. optical: \citealt{bean_accurate_2006, bean_metallicities_2006, rajpurohit_high-resolution_2014, passegger_fundamental_2016}; Y band: \citealt{veyette_physically_2017}; optical-NIR: \citealt{woolf_metallicity_2005,woolf_calibrating_2006,passegger_carmenes_2018}; J band: \citealt{onehag_m-dwarf_2012}; H band: \citealt{souto_chemical_2017}).

Finally, on the point of M-dwarf evolutionary models (and low-mass, cool main sequence stars more generally), there has long been contention between model radii and observed radii \citep[e.g.][]{kraus_mass-radius_2015}. This is often attributed to magnetic fields (and/or the mixing length parameter, which simplistically parameterizes the effects of magnetic fields among other energy transport mechanisms in 1D stellar structure and evolution programs) and is related to the difficulty in accurately modelling convection  \citep[e.g.][]{feiden_reevaluating_2012, joyce_not_2018}. Fortunately, due to the aforementioned insensitivity of NIR $K$ band photometry to [Fe/H], empirical mass and radius relations have been developed and calibrated on interferometric diameters and dynamical masses \citep[e.g.][]{henry_mass-luminosity_1993, delfosse_accurate_2000, benedict_solar_2016, mann_how_2015, mann_how_2019}.

Here we conduct a moderate resolution spectroscopic survey of 92 southern cool ($T_{\rm eff} \lesssim 4,500\,$K) TESS candidate planet hosts with the WiFeS instrument \citep{dopita_wide_2007} on the ANU 2.3 m Telescope at Siding Spring Observatory (NSW, Australia). We combine our spectroscopic observations with literature optical photometry and trigonometric parallaxes from Gaia DR2 \citep{gaia_collaboration_gaia_2016, brown_gaia_2018}, infrared photometry from 2MASS \citep[][]{skrutskie_two_2006}, optical photometry from SkyMapper DR3 (\citealt{keller_skymapper_2007}, \citealt{onken_skymapper_2019}, DR3 DOI: 10.25914/5f14eded2d116), empirical relations from \citep{mann_how_2015, mann_how_2019}, and synthetic MARCS model atmospheres \citep{gustafsson_grid_2008} in order to produce stellar $T_{\rm eff}$, $\log g$, [Fe/H], bolometric flux ($f_{\rm bol}$), $R_\star$, and stellar mass ($M_\star$). By modelling the transit light curves of these host stars, we are additionally able to produce precision planetary radii for 100 candidate planets, which represents one of the largest uniform analyses of cool TESS hosts to date. Our observations and data reduction are described in Section \ref{sec:observations}, our photometric [Fe/H] relation in Section \ref{sec:feh_rel}, our host star characterisation methodology and resulting parameters in Section \ref{sec:spectroscopic_analysis}, our transit light curve fitting and results in Section \ref{sec:planet_fitting}, discussion of results in Section \ref{sec:discussion}, and concluding remarks in Section \ref{sec:conclusion}.

\section{Observations and data reduction}\label{sec:observations}
\subsection{Target Selection}

\begin{figure}
    \centering
    \includegraphics[width=\columnwidth]{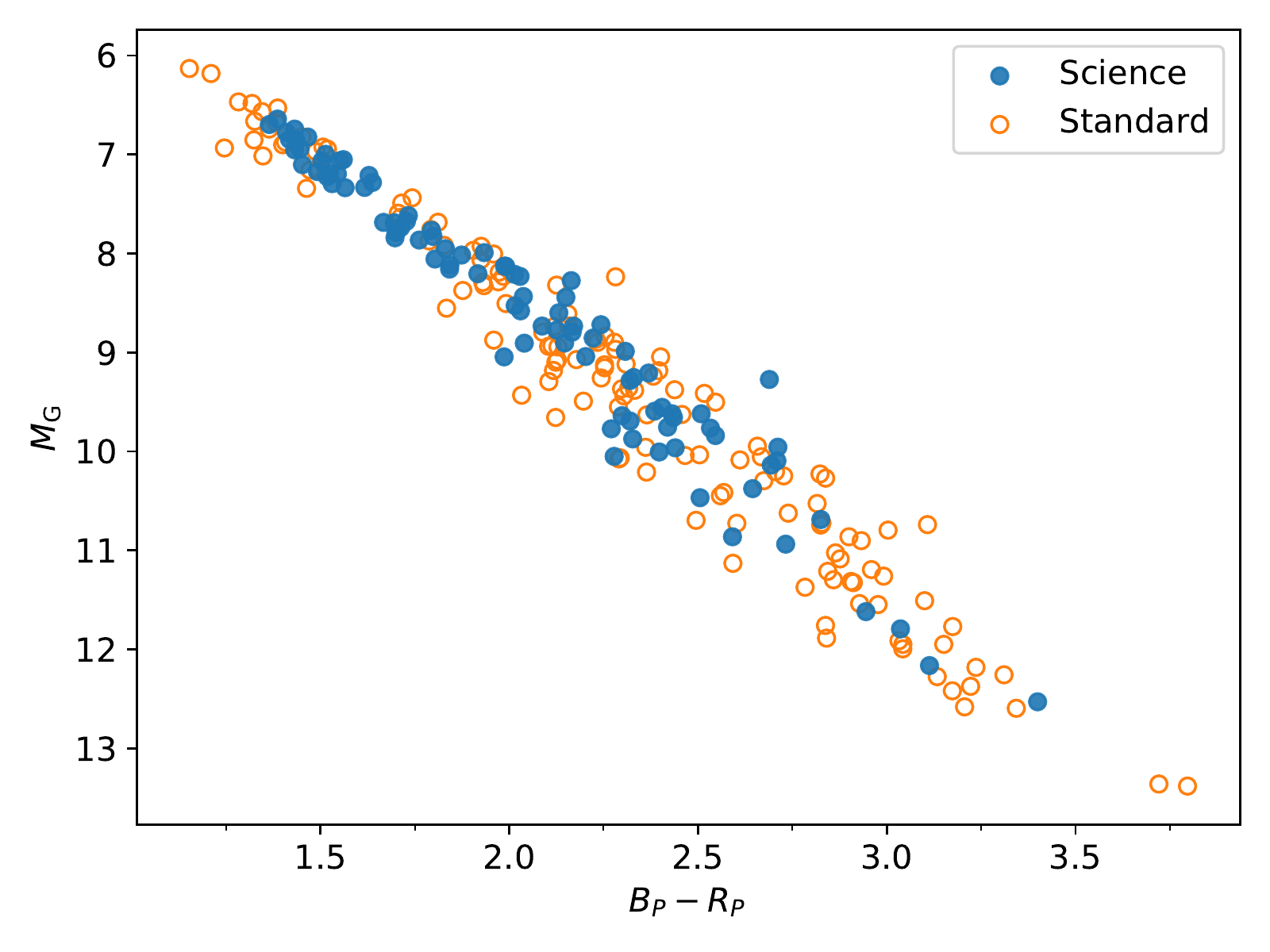}
    \caption{Gaia DR2 $M_G$ versus $(B_p-R_p)$ colour magnitude diagram for science targets (filled blue circles) and cool dwarf standards (orange open circles).}
    \label{fig:hr_diagram}
\end{figure}

Our initial target selection of southern cool dwarf TOIs was done in August 2019, including stars with $T_{\rm eff} \leq 4500\,$K in the TESS input catalogue and unblended 2MASS photometry. In order to have reliable parallaxes, we impose the additional requirement that our stars have a Gaia DR2 Renormalised Unit Weight Error (RUWE)\footnote{Expected to be approximately 1.0 in case where the single star model provides a good fit for the astrometric data. See: \url{https://gea.esac.esa.int/archive/documentation/GDR2/Gaia_archive/chap_datamodel/sec_dm_main_tables/ssec_dm_ruwe.html}} of $<1.4$, as recommended by the Gaia team\footnote{Though we do accept TIC 158588995 with a marginal RUWE$\sim$1.47 as it sits on the main sequence and does not appear overluminous.}. Adding extra targets sourced in August 2020, and removing those identified as false-positives through community follow-up observations (as listed on NASA's Exoplanet Follow-up Observing Program for TESS, ExoFOP-TESS, website\footnote{\url{https://exofop.ipac.caltech.edu/tess/}}), we are left with a sample of 92 southern candidate planet hosts spread across the sky with $8.7 < {\rm apparent~Gaia}~G < 15.8$. These targets are listed in Table \ref{tab:science_targets}, and plotted on a colour-magnitude diagram in Figure \ref{fig:hr_diagram}, noting that a few appear distinctly above the main sequence. These stars are thus overluminous because they are young and still contracting to the main sequence, or because they are unresolved binaries.

All our targets have Gaia DR2 $G$, $B_P$, $R_P$, and 2MASS $J$, $H$, $K_S$ photometry, and most have at least one of SkyMapper DR3 $r$, $i$, $z$ (noting that the survey is still ongoing, so not all bands are available for all targets). We calculate distances from Gaia DR2 parallaxes, incorporating the systematic parallax offset of $-82 \pm 33\,\mu$as found by \citet{stassun_evidence_2018}. 

To correct for reddening we use the 3D dust map of \citet{leike_resolving_2020}, implemented within the python package \texttt{dustmaps} \citep{green_dustmaps_2018}. Targeting bright, cool dwarfs as we do here automatically means our stars will be relatively close, and we take those within the Local Bubble, two thirds of our sample, to be unreddened \cite[$\lesssim 70\,$pc, e.g.][]{leroy_polarimetric_1993, lallement_3d_2003} so long as the Gaia $G$ band extinction reported by the dust map is consistent with zero ($A_G < 0.01$). Nominal extinction coefficients were sourced from \citet{casagrande_synthetic_2014} for the 2MASS $JHK_S$ bands and \citet{casagrande_skymapper_2019} for SkyMapper $uvgriz$, with Gaia $G$, $B_P$, and $R_P$ coefficients computed from the relation given in \citet{casagrande_effective_2020} for $B_P-R_P=2.03$, the median value for our sample.

\subsection{Standard Selection}
Given the complexities involved in determining the properties of cool dwarfs, we also observed a set of 136 well characterised late K/M-dwarf standards from the literature. Broadly these standards have parameters from at least one of the following sources:
\begin{enumerate}
    \item {[Fe/H]} from an FGK companion,
    \item {[Fe/H]} from low resolution NIR spectra,
    \item $T_{\rm eff}$ from interferometry.
\end{enumerate}
With the exception of available interferometric $T_{\rm eff}$ standards, we additionally wanted to source standards from large uniform catalogues due to the known problem of systematics between different spectroscopic techniques \citep[e.g.][]{lebzelter_comparative_2012, hinkel_comparison_2016}. With this in mind, the bulk of our M/late-K dwarf standards come from the works of \citet{rojas-ayala_metallicity_2012} and \citet{mann_how_2015}, with interferometric targets from \citet{von_braun_gj_2012}, \citet{boyajian_stellar_2012-1}, \citet{von_braun_stellar_2014}, \citet{rabus_discontinuity_2019}, and \citet{rains_precision_2020}; and FGK companion [Fe/H] compiled by \citet{newton_near-infrared_2014} from \citet{valenti_spectroscopic_2005}, \citet{sousa_spectroscopic_2006} and \citet{sozzetti_keck_2009}. Our mid-K dwarf calibrators do not come from a single uniform catalogue; they are instead pulled from the works of \citet{woolf_metallicity_2005}, \citet{sousa_spectroscopic_2008}, \citet{prugniel_atmospheric_2011}, \citet{sousa_spectroscopic_2011}, \citet{tsantaki_deriving_2013}, \citet{luck_abundances_2017}, \citet{luck_abundances_2018} and \citet{montes_calibrating_2018}.

These stars were observed with the same instrument settings as our science targets (but at higher SNR), with the intent to provide checks against our analysis techniques for this notoriously complex set of stars.

\subsection{Spectroscopic Observations}
Observations were conducted using the WiFeS instrument \citep[Wide-Field Spectrograph,][]{dopita_wide_2007} on the ANU 2.3$\,$m Telescope at Siding Spring Observatory, Australia between August 2019 and September 2020. WiFeS, a dual camera integral field spectrograph, is an effective stellar survey instrument due to its high throughput and broad wavelength coverage. Using the B3000 and R7000 gratings, and RT480 beam splitter, we obtain low resolution blue spectra ($3500 \leq \lambda \leq 5700\,$\SI{}{\angstrom}, $\lambda/\Delta\lambda\sim3000$) and moderate resolution red spectra ($5400 \leq \lambda \leq 7000\,$\SI{}{\angstrom}, $\lambda/\Delta\lambda\sim7000$) with median signal to noise ratios (SNR) per spectral pixel of 16 and 58 respectively. Exposure times ranged from 20$\,$sec to 30$\,$min, and were chosen on the basis of 0.5 magnitude bins in Gaia $G$. 

Target observations were bracketed hourly with NeAr Arc lamp exposures, telluric standards were observed every few hours, and flux standards were observed several times throughout each night.  Data reduction was done using the standard PyWiFeS pipeline \citep{childress_pywifes:_2014} with the exception of custom flux calibration due to PyWiFeS' poor performance with R7000 spectra. Science target observations are listed in Table \ref{tab:observing_log_tess}, and standard star observations in Table \ref{tab:observing_log_std}.

\begin{landscape}
\begin{table}
\centering
\caption{Science targets}
\label{tab:science_targets}
\begin{tabular}{cccccccccccc}
\hline
TOI$^a$ & TIC$^b$ & 2MASS$^c$ & Gaia DR2$^d$ & RA$^d$ & DEC$^d$ & $G^d$ & ${B_p-R_p}^d$ & Plx$^d$ & ruwe$^d$ & E($B-V$) & N$_{\rm pc}^e$ \\
 &  &  &  & (hh mm ss.ss) & (dd mm ss.ss) & (mag) & (mag) & (mas) &  &  &  \\
\hline
741 & 359271092 & 09213761-6016551 & 5299440441521812992 & 09 21 35.86 & -61 43 07.68 & 8.68 & 2.03 & 95.63 $\pm$ 0.03 & 1.1 & 0.00 & 1 \\
731 & 34068865 & 09442986-4546351 & 5412250540681250560 & 09 44 29.16 & -46 13 15.60 & 9.15 & 2.32 & 106.21 $\pm$ 0.03 & 1.1 & 0.00 & 1 \\
260 & 37749396 & 00190556-0957530 & 2428162410789155328 & 00 19 05.52 & -10 02 01.68 & 9.31 & 1.70 & 49.51 $\pm$ 0.06 & 0.9 & 0.00 & 1 \\
836 & 440887364 & 15001942-2427147 & 6230733559097425152 & 15 00 19.18 & -25 32 44.88 & 9.39 & 1.55 & 36.33 $\pm$ 0.04 & 1.0 & 0.01 & 2 \\
562 & 413248763 & 09360161-2139371 & 5664814198431308288 & 09 36 01.80 & -22 20 05.64 & 9.88 & 2.40 & 105.88 $\pm$ 0.06 & 1.1 & 0.00 & 3 \\
455 & 98796344 & 03015142-1635356 & 5153091836072107136 & 03 01 51.00 & -17 24 19.80 & 10.05 & 2.59 & 145.55 $\pm$ 0.08 & 1.1 & 0.00 & 2 \\
139 & 62483237 & 22253655-3454346 & 6598814657249555328 & 22 25 36.58 & -35 05 25.08 & 10.08 & 1.45 & 23.55 $\pm$ 0.04 & 1.0 & 0.00 & 1 \\
253 & 322063810 & 00571629-5135048 & 4928367189956040960 & 00 57 16.44 & -52 24 52.92 & 10.18 & 1.71 & 32.39 $\pm$ 0.03 & 1.0 & 0.00 & 1 \\
134 & 234994474 & 23200751-6003545 & 6491962296196145664 & 23 20 06.86 & -61 56 03.48 & 10.23 & 2.03 & 39.73 $\pm$ 0.04 & 1.0 & 0.00 & 1 \\
544 & 50618703 & 05290957-0020331 & 3220926542276901888 & 05 29 09.62 & -1 39 25.56 & 10.40 & 1.57 & 24.29 $\pm$ 0.04 & 1.1 & 0.01 & 1 \\
486 & 260708537 & 06334998-5831426 & 5482827676662168832 & 06 33 49.18 & -59 28 30.00 & 10.53 & 2.43 & 65.70 $\pm$ 0.03 & 1.2 & 0.00 & 1 \\
177 & 262530407 & 01214538-4642518 & 4933912198893332224 & 01 21 45.22 & -47 17 07.08 & 10.55 & 2.17 & 44.46 $\pm$ 0.05 & 1.0 & 0.00 & 1 \\
129 & 201248411 & 00004490-5449498 & 4923860051276772608 & 00 00 44.54 & -55 10 09.12 & 10.59 & 1.39 & 16.16 $\pm$ 0.02 & 1.1 & 0.00 & 1 \\
175 & 307210830 & 08180763-6818468 & 5271055243163629056 & 08 18 07.90 & -69 41 07.80 & 10.60 & 2.51 & 94.14 $\pm$ 0.03 & 1.1 & 0.00 & 3 \\
824 & 193641523 & 14483982-5735175 & 5880886001621564928 & 14 48 39.72 & -58 24 39.96 & 10.72 & 1.36 & 15.61 $\pm$ 0.03 & 1.1 & 0.02 & 1 \\
133 & 219338557 & 23373497-5857166 & 6489346046933733632 & 23 37 35.38 & -59 02 41.64 & 10.72 & 1.53 & 20.53 $\pm$ 0.03 & 1.0 & 0.00 & 1 \\
1130 & 254113311 & 19053021-4126151 & 6715688452614516736 & 19 05 30.24 & -42 33 44.64 & 10.88 & 1.55 & 17.14 $\pm$ 0.05 & 1.1 & 0.01 & 2 \\
198 & 12421862 & 00090428-2707196 & 2333676738049780352 & 00 09 05.16 & -28 52 41.88 & 10.92 & 1.99 & 42.12 $\pm$ 0.05 & 1.0 & 0.00 & 1 \\
833 & 362249359 & 09423526-6228346 & 5250780970316845696 & 09 42 34.92 & -63 31 26.76 & 11.05 & 1.83 & 23.94 $\pm$ 0.02 & 0.8 & 0.01 & 1 \\
178 & 251848941 & 00291228-3027133 & 2318295979126499200 & 00 29 12.48 & -31 32 45.24 & 11.15 & 1.49 & 15.92 $\pm$ 0.05 & 1.2 & 0.00 & 3 \\
279 & 122613513 & 02444524-3212391 & 5063070558501465856 & 02 44 45.24 & -33 47 20.40 & 11.20 & 1.44 & 13.42 $\pm$ 0.04 & 1.1 & 0.00 & 1 \\
704 & 260004324 & 06042035-5518468 & 5500061456275483776 & 06 04 21.60 & -56 41 18.60 & 11.23 & 2.22 & 33.48 $\pm$ 0.03 & 1.0 & 0.00 & 1 \\
1078 & 370133522 & 20274210-5627262 & 6468968316900356736 & 20 27 42.86 & -57 32 15.72 & 11.24 & 2.32 & 49.06 $\pm$ 0.05 & 1.0 & 0.00 & 1 \\
969 & 280437559 & 07403284+0205561 & 3087206553745290624 & 07 40 32.81 & +2 05 54.96 & 11.25 & 1.47 & 12.92 $\pm$ 0.05 & 1.4 & 0.01 & 1 \\
620 & 296739893 & 09284158-1209551 & 5738284016370287616 & 09 28 41.62 & -13 49 58.08 & 11.31 & 2.24 & 30.25 $\pm$ 0.05 & 1.0 & 0.00 & 1 \\
910 & 369327947 & 19205439-8233170 & 6347643496607835520 & 19 20 57.10 & -83 26 24.72 & 11.42 & 2.73 & 80.09 $\pm$ 0.04 & 1.1 & 0.00 & 1 \\
713 & 167600516 & 06480517-6537252 & 5285060409961261696 & 06 48 05.14 & -66 22 32.52 & 11.42 & 1.52 & 14.39 $\pm$ 0.02 & 0.9 & 0.01 & 2 \\
932 & 260417932 & 06234590-5434414 & 5499671713762981248 & 06 23 45.82 & -55 25 19.20 & 11.42 & 1.41 & 11.68 $\pm$ 0.02 & 0.8 & 0.01 & 2 \\
240 & 101948569 & 00590112-4409389 & 4982951791883929472 & 00 59 01.18 & -45 50 20.76 & 11.43 & 1.50 & 13.33 $\pm$ 0.03 & 1.1 & 0.00 & 1 \\
696 & 77156829 & 04324261-3947112 & 4864160624337973248 & 04 32 42.96 & -40 12 32.76 & 11.54 & 2.28 & 50.28 $\pm$ 0.02 & 1.1 & 0.00 & 2 \\
244 & 118327550 & 00421695-3643053 & 5001098681543159040 & 00 42 16.75 & -37 16 55.20 & 11.55 & 2.55 & 45.36 $\pm$ 0.07 & 1.3 & 0.00 & 1 \\
270 & 259377017 & 04333970-5157222 & 4781196115469953024 & 04 33 39.86 & -52 02 33.36 & 11.63 & 2.33 & 44.46 $\pm$ 0.03 & 1.0 & 0.00 & 3 \\
912 & 406941612 & 15172165-8028225 & 5772442647192375808 & 15 17 18.86 & -81 31 36.12 & 11.64 & 2.40 & 38.27 $\pm$ 0.02 & 1.1 & 0.01 & 1 \\
475 & 100608026 & 05465951-3231592 & 2901089987127041920 & 05 46 59.59 & -33 28 03.00 & 11.70 & 1.76 & 16.99 $\pm$ 0.02 & 1.0 & 0.01 & 1 \\
442 & 70899085 & 04164560-1205023 & 3189306030970782208 & 04 16 45.65 & -13 54 54.36 & 11.73 & 1.99 & 18.98 $\pm$ 0.04 & 1.2 & 0.00 & 1 \\
761 & 165317334 & 11570326-3806169 & 3460168250866990848 & 11 57 03.12 & -39 53 42.72 & 11.73 & 1.73 & 14.94 $\pm$ 0.04 & 1.2 & 0.01 & 1 \\
870 & 219229644 & 04131645-5056400 & 4782093729275660160 & 04 13 16.63 & -51 03 20.52 & 11.78 & 1.99 & 18.56 $\pm$ 0.02 & 1.1 & 0.00 & 1 \\
904 & 261257684 & 05572938-8307486 & 4620844400530949376 & 05 57 29.11 & -84 52 13.08 & 11.84 & 2.02 & 21.67 $\pm$ 0.02 & 1.0 & 0.01 & 1 \\
732 & 36724087 & 10183516-1142599 & 3767281845873242112 & 10 18 34.78 & -12 16 55.92 & 11.85 & 2.69 & 45.46 $\pm$ 0.08 & 1.1 & 0.00 & 2 \\
656 & 36734222 & 10193800-0948225 & 3767805209112436736 & 10 19 37.97 & -10 11 36.96 & 11.89 & 1.63 & 11.50 $\pm$ 0.04 & 1.1 & 0.01 & 1 \\
1075 & 351601843 & 20395334-6526579 & 6426188308031756288 & 20 39 53.09 & -66 33 01.08 & 12.05 & 1.84 & 16.24 $\pm$ 0.02 & 1.2 & 0.00 & 1 \\
700 & 150428135 & 06282325-6534456 & 5284517766615492736 & 06 28 22.97 & -66 25 17.04 & 12.06 & 2.39 & 32.10 $\pm$ 0.02 & 1.1 & 0.00 & 3 \\
727 & 149788158 & 08425684-0229529 & 3072157538091829120 & 08 42 56.86 & -3 30 05.04 & 12.07 & 2.15 & 23.24 $\pm$ 0.03 & 1.1 & 0.00 & 1 \\
249 & 179985715 & 00561930-3856552 & 4987729474846997248 & 00 56 19.20 & -39 03 02.88 & 12.08 & 1.70 & 14.13 $\pm$ 0.03 & 1.1 & 0.00 & 1 \\
1201 & 29960110 & 02485926-1432152 & 5157183324996790272 & 02 48 59.45 & -15 27 45.72 & 12.10 & 2.37 & 26.37 $\pm$ 0.04 & 1.1 & 0.00 & 1 \\
\hline
\end{tabular}
\begin{minipage}{\linewidth}
\vspace{0.1cm}
\textbf{Notes:} $^a$ TESS Object of Interest ID, $^b$ TESS Input Catalogue ID \citep{stassun_tess_2018, stassun_revised_2019},$^c$2MASS \citep{skrutskie_two_2006}, $^c$Gaia \citep{brown_gaia_2018} -  note that Gaia parallaxes listed here have been corrected for the zeropoint offset, $^d$Number of candidate planets, NASA Exoplanet Follow-up Observing Program for TESS \\
\end{minipage}
\end{table}
\end{landscape}

\begin{landscape}
\begin{table}
\centering
\contcaption{Science targets}
\begin{tabular}{cccccccccccc}
\hline
TOI$^a$ & TIC$^b$ & 2MASS$^c$ & Gaia DR2$^d$ & RA$^d$ & DEC$^d$ & $G^d$ & ${B_p-R_p}^d$ & Plx$^d$ & ruwe$^d$ & E($B-V$) & N$_{\rm pc}^e$ \\
 &  &  &  & (hh mm ss.ss) & (dd mm ss.ss) & (mag) & (mag) & (mas) &  &  &  \\
\hline
875 & 14165625 & 05120890-3742313 & 4820828591913853696 & 05 12 08.93 & -38 17 29.40 & 12.12 & 1.51 & 9.39 $\pm$ 0.02 & 1.0 & 0.01 & 1 \\
929 & 175532955 & 03033741-3955515 & 5044287532642519680 & 03 03 37.73 & -40 04 09.12 & 12.13 & 1.42 & 8.71 $\pm$ 0.02 & 1.1 & 0.01 & 1 \\
493 & 19025965 & 07583071+1253005 & 3151371883379694720 & 07 58 30.65 & +12 52 59.88 & 12.19 & 1.56 & 9.29 $\pm$ 0.04 & 1.1 & 0.01 & 1 \\
1216 & 141527965 & 05505139-7541200 & 4648441970589471104 & 05 50 51.55 & -76 18 41.04 & 12.31 & 1.62 & 10.02 $\pm$ 0.03 & 1.2 & 0.01 & 1 \\
233 & 415969908 & 22545039-1854426 & 2402715141877299584 & 22 54 50.06 & -19 05 15.36 & 12.41 & 2.27 & 29.58 $\pm$ 0.04 & 1.0 & 0.00 & 2 \\
711 & 38510224 & 04100386-6156326 & 4676789514954240768 & 04 10 03.86 & -62 03 26.28 & 12.45 & 1.45 & 8.43 $\pm$ 0.02 & 1.0 & 0.01 & 1 \\
876 & 32497972 & 05362611-2414377 & 2963392606627366912 & 05 36 26.23 & -25 45 20.16 & 12.47 & 1.87 & 12.76 $\pm$ 0.03 & 1.1 & 0.01 & 1 \\
785 & 374829238 & 05532099-6538022 & 4755884700667639296 & 05 53 20.95 & -66 21 59.40 & 12.51 & 2.04 & 15.23 $\pm$ 0.02 & 1.0 & 0.01 & 1 \\
406 & 153065527 & 03170297-4214323 & 4851053999056603904 & 03 17 03.02 & -43 45 21.24 & 12.55 & 2.71 & 32.17 $\pm$ 0.04 & 1.2 & 0.00 & 2 \\
714 & 219195044 & 06093401-5349245 & 5500474185452572032 & 06 09 34.18 & -54 10 37.20 & 12.56 & 2.04 & 18.54 $\pm$ 0.02 & 1.2 & 0.01 & 2 \\
900 & 210873792 & 16233735-3122228 & 6037266684232926208 & 16 23 37.22 & -32 37 35.04 & 12.59 & 1.52 & 8.23 $\pm$ 0.05 & 0.9 & 0.03 & 1 \\
557 & 55488511 & 03560411-1016192 & 3193508849745633280 & 03 56 04.27 & -11 43 40.80 & 12.60 & 1.92 & 13.14 $\pm$ 0.04 & 1.1 & 0.01 & 1 \\
864 & 231728511 & 05254662-5121253 & 4772266186971169792 & 05 25 46.42 & -52 38 34.80 & 12.66 & 2.42 & 26.22 $\pm$ 0.03 & 1.1 & 0.01 & 1 \\
256 & 92226327 & 00445930-1516166 & 2371032916186181760 & 00 44 59.66 & -16 43 33.24 & 12.67 & 3.04 & 66.70 $\pm$ 0.07 & 1.1 & 0.00 & 2 \\
702 & 237914496 & 03444203-6511567 & 4672700190692201088 & 03 44 41.98 & -66 48 06.12 & 12.68 & 1.80 & 11.82 $\pm$ 0.03 & 1.1 & 0.01 & 1 \\
1082 & 261108236 & 05330624-8048563 & 4621526273835900288 & 05 33 06.19 & -81 11 04.20 & 12.68 & 1.70 & 9.95 $\pm$ 0.03 & 1.1 & 0.01 & 1 \\
672 & 151825527 & 11115769-3919400 & 5396580575830873728 & 11 11 57.82 & -40 40 18.84 & 12.72 & 2.13 & 14.92 $\pm$ 0.03 & 1.1 & 0.01 & 1 \\
806 & 33831980 & 04134003-7605515 & 4627952094666051072 & 04 13 39.86 & -77 54 07.20 & 12.77 & 1.67 & 9.55 $\pm$ 0.02 & 1.3 & 0.01 & 1 \\
797 & 271596225 & 07141480-7436089 & 5262540590756812032 & 07 14 15.14 & -75 23 48.84 & 12.78 & 2.20 & 17.77 $\pm$ 0.02 & 1.1 & 0.01 & 2 \\
663 & 54962195 & 10401596-0830385 & 3762515188088861184 & 10 40 15.82 & -9 29 20.04 & 12.81 & 2.13 & 15.54 $\pm$ 0.04 & 1.0 & 0.01 & 2 \\
540 & 200322593 & 05051443-4756154 & 4785886941312921344 & 05 05 14.33 & -48 03 45.00 & 12.89 & 3.11 & 71.39 $\pm$ 0.04 & 1.0 & 0.00 & 1 \\
285 & 220459976 & 04584731-5623385 & 4764216563561182336 & 04 58 47.33 & -57 36 22.32 & 13.07 & 1.79 & 8.61 $\pm$ 0.02 & 0.9 & 0.01 & 1 \\
674 & 158588995 & 10582099-3651292 & 5400949450924312576 & 10 58 20.78 & -37 08 30.84 & 13.08 & 2.53 & 21.67 $\pm$ 0.05 & 1.5 & 0.01 & 1 \\
789 & 300710077 & 07410444-7118157 & 5264306681309492864 & 07 41 04.85 & -72 41 46.32 & 13.15 & 2.44 & 23.01 $\pm$ 0.03 & 1.3 & 0.01 & 1 \\
873 & 237920046 & 03465622-6320142 & 4673392195823039744 & 03 46 56.78 & -64 39 47.16 & 13.16 & 2.09 & 12.97 $\pm$ 0.02 & 1.3 & 0.01 & 1 \\
698 & 141527579 & 05505661-7637132 & 4647922867959139072 & 05 50 57.38 & -77 22 49.44 & 13.26 & 2.33 & 15.75 $\pm$ 0.03 & 1.2 & 0.01 & 1 \\
136 & 410153553 & 22415815-6910089 & 6385548541499112448 & 22 41 59.09 & -70 49 40.44 & 13.39 & 3.40 & 67.15 $\pm$ 0.05 & 1.1 & 0.00 & 1 \\
269 & 220479565 & 05032306-5410378 & 4770828304936109056 & 05 03 23.11 & -55 49 20.28 & 13.41 & 2.30 & 17.51 $\pm$ 0.02 & 1.1 & 0.01 & 1 \\
654 & 35009898 & 10585379-0532468 & 3788670679927991296 & 10 58 53.90 & -6 27 09.00 & 13.42 & 2.51 & 17.29 $\pm$ 0.05 & 1.1 & 0.01 & 1 \\
782 & 429358906 & 12154108-1854365 & 3518374197418907648 & 12 15 40.90 & -19 05 22.92 & 13.55 & 2.71 & 19.01 $\pm$ 0.07 & 1.2 & 0.01 & 1 \\
521 & 27649847 & 08132251+1213181 & 649852779797683968 & 08 13 22.63 & +12 13 19.56 & 13.58 & 2.44 & 16.38 $\pm$ 0.06 & 1.3 & 0.00 & 1 \\
203 & 259962054 & 02520450-6741155 & 4647534190597951232 & 02 52 04.34 & -68 18 46.80 & 13.59 & 2.94 & 40.31 $\pm$ 0.05 & 1.4 & 0.00 & 1 \\
532 & 144700903 & 05401918+1133463 & 3340265717587057536 & 05 40 19.22 & +11 33 45.36 & 13.63 & 1.93 & 7.38 $\pm$ 0.03 & 1.1 & 0.02 & 1 \\
756 & 73649615 & 12482523-4528140 & 6129327525817451648 & 12 48 24.89 & -46 31 46.20 & 13.66 & 2.31 & 11.58 $\pm$ 0.04 & 1.2 & 0.02 & 1 \\
302 & 229111835 & 01095538-5214219 & 4927215760764862976 & 01 09 55.51 & -53 45 37.80 & 13.71 & 1.64 & 5.10 $\pm$ 0.01 & 1.0 & 0.01 & 1 \\
435 & 44647437 & 03573850-2511238 & 5082797618168232320 & 03 57 38.54 & -26 48 36.36 & 13.75 & 1.73 & 6.02 $\pm$ 0.02 & 1.1 & 0.01 & 1 \\
1067 & 201642601 & 19144126-5934458 & 6638412919991750912 & 19 14 41.28 & -60 25 14.16 & 13.82 & 1.43 & 3.76 $\pm$ 0.02 & 1.0 & 0.03 & 1 \\
210 & 141608198 & 05555049-7359046 & 4650160717726370816 & 05 55 50.83 & -74 00 52.92 & 13.84 & 2.83 & 23.33 $\pm$ 0.05 & 1.3 & 0.01 & 1 \\
1073 & 158297421 & 19095625-4939538 & 6658373007402886400 & 19 09 56.26 & -50 20 06.36 & 14.31 & 1.43 & 3.30 $\pm$ 0.04 & 1.0 & 0.04 & 1 \\
468 & 33521996 & 05523523-1901539 & 2966680597368750720 & 05 52 35.23 & -20 58 06.24 & 14.34 & 2.01 & 5.88 $\pm$ 0.04 & 1.2 & 0.01 & 1 \\
122 & 231702397 & 22114728-5856422 & 6411096106487783296 & 22 11 47.57 & -59 03 14.04 & 14.34 & 2.64 & 16.07 $\pm$ 0.06 & 1.2 & 0.00 & 1 \\
507 & 348538431 & 08063103-1545526 & 5724250571514167424 & 08 06 31.10 & -16 14 07.08 & 14.48 & 2.69 & 8.99 $\pm$ 0.06 & 1.3 & 0.01 & 1 \\
551 & 192826603 & 05305145-3637508 & 4821739369794767744 & 05 30 51.41 & -37 22 08.40 & 14.83 & 1.84 & 4.56 $\pm$ 0.02 & 1.0 & 0.02 & 1 \\
552 & 44737596 & 04034783-2524320 & 5082914338199586560 & 04 03 47.86 & -26 35 27.96 & 14.87 & 2.15 & 5.10 $\pm$ 0.03 & 1.1 & 0.01 & 1 \\
234 & 12423815 & 00101648-2616566 & 2335244779070099200 & 00 10 16.54 & -27 43 03.36 & 15.69 & 2.17 & 3.98 $\pm$ 0.07 & 1.1 & 0.01 & 1 \\
555 & 170849515 & 04412154-3219128 & 4877426575724467456 & 04 41 21.55 & -33 40 46.56 & 15.71 & 1.80 & 2.56 $\pm$ 0.04 & 1.0 & 0.04 & 1 \\
142 & 425934411 & 00182539-6250523 & 4901321849613348736 & 00 18 25.42 & -63 09 07.56 & 15.77 & 2.16 & 3.09 $\pm$ 0.04 & 1.0 & 0.01 & 1 \\
\hline
\end{tabular}
\end{table}
\end{landscape}

\FloatBarrier

\subsection{Radial velocity determination}\label{sec:rv_fitting}
Radial velocities of the WiFeS R7000 spectra were determined from a least squares minimisation of a set of synthetic template spectra varying in temperature (see Section \ref{sec:model_limitations} for details of model grid). We use a coarsely sampled version of this grid, computed at R$\sim7000$ over $5400 \leq \lambda \leq 7000$ for $3000 \leq T_{\rm eff} \leq 5500\,$K, $\log g = 4.5$, and [Fe/H]$= 0.0$, with $T_{\rm eff}$ steps of $100\,$K for radial velocity determination. For further information on our RV fitting formalism, see \citet{zerjal_spectroscopically_2021}\footnote{Our RV fitting code, along with all other code for this project, can be found at \url{https://github.com/adrains/plumage}}.

\begin{figure}
    \centering
    \includegraphics[width=\columnwidth]{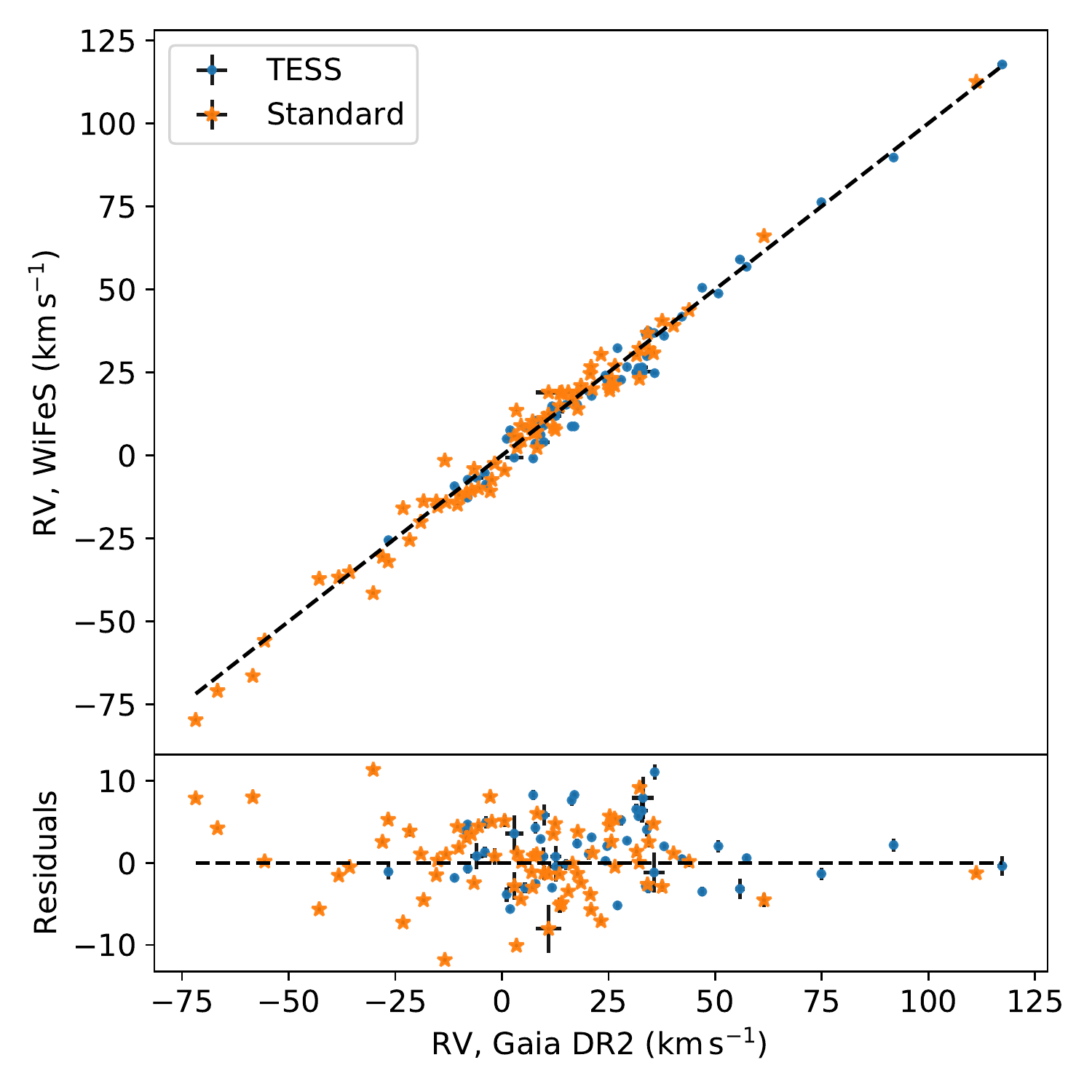}
    \caption{Comparison between those stars with radial velocities in Gaia DR2 and our work here, from which we determine a scatter of $\sim$$4.5\,$km$\,$s$^{-1}$.}
    \label{fig:rv_comp}
\end{figure}

Statistical uncertainties on this approach are median $\sim$410$\,$m$\,$s$^{-1}$, though comparison to Gaia DR2 in Figure \ref{fig:rv_comp} reveals a larger scatter with standard deviation $\sim$$4.5\,$km$\,$s$^{-1}$, computed from a median absolute deviation, which we add in quadrature with our statistical uncertainties. Higher uncertainties are consistent with the work of \citet{kuruwita_multiplicity_2018} who found that WiFeS varies on shorter timescales than our hourly arcs can account for. While they additionally improved precision by calibrating using oxygen B-band absorption, RV uncertainties of $\sim$$4.5\,$km$\,$s$^{-1}$ are sufficient for this work. Our final values are reported in Table \ref{tab:observing_log_tess} for science targets, and Table \ref{tab:observing_log_std} for standards.

\section{Photometric Metallicity Calibration}\label{sec:feh_rel}
As established earlier, cool dwarf metallicities are notoriously difficult to determine, particularly when working with optical spectra. \citet{bonfils_metallicity_2005} initially proposed empirical calibrations to determine [Fe/H] from a star's position in $M_{K}-(V-K)$ space, a technique which was later iterated on by \citet{johnson_metal_2009}, \citet{schlaufman_physically-motivated_2010}, and \citet{neves_metallicity_2012}. Such relations are based on the fact that once on the main sequence, low mass stars do not evolve (and hence change in brightness and temperature) appreciably on moderate timescales as compared to their higher mass and faster evolving counterparts. Thus, assuming no extra scatter from unresolved binaries and standard helium enrichment \citep[e.g.][]{pagel_yz_1998}, a star's position above or below the mean main sequence is directly correlated with its chemical composition \citep{baraffe_evolutionary_1998}.

These relations are benchmarked on what is considered the gold standard for M-dwarf metallicites: [Fe/H] from a hotter FGK companion taken to have formed at the same time and thus have the same chemical composition. This chemical homogeneity is now well established for FGK-FGK pairs \citep[e.g.][]{desidera_abundance_2004, hawkins_identical_2020}. The process of determining which stars on the sky are likely associated has now been greatly simplified with the release of Gaia DR2, which has provided precision parallax measurements and proper motions for nearly all nearby M-dwarfs, with our sample of secondaries having median $0.17\,$\% parallax precision.

We take as input the sample of FGK-KM-dwarf pairs compiled by \citet{mann_prospecting_2013} and \citet{newton_near-infrared_2014}. These combine primary star [Fe/H] measurements from high resolution spectra sourced from a variety of previous surveys \citep{mishenina_correlation_2004,luck_stars_2005, valenti_spectroscopic_2005,bean_accurate_2006,ramirez_oxygen_2007,robinson_n2k_2007,fuhrmann_nearby_2008, casagrande_new_2011,da_silva_homogeneous_2011,mann_prospecting_2013}, with \citet{mann_prospecting_2013} correcting for inter-survey systematics to place them on a common [Fe/H] scale. To this set we add the metal-poor, cool subdwarf VB12 to extend our metallicity coverage, taking the [Fe/H] reported by \citet{ramirez_oxygen_2007} for its primary HD 219617 AB (and correcting for the systematic reported by \citealt{mann_prospecting_2013}). This provided 128 total pairs, which was reduced to 69 after crossmatching with both Gaia DR2 and 2MASS, and removing those stars with missing or poor photometry (2MASS Qflg$\ne$`AAA', where `AAA' is the highest photometric quality rating and corresponds to $JHK_S$ respectively); those  flagged on SIMBAD\footnote{\url{http://simbad.u-strasbg.fr/simbad/}} as spectroscopic binaries; those with poor Gaia astrometry (Gaia dup flag=1, RUWE $>1.4$); those pairs with M dwarf primaries; or whose parallaxes, astrometry, and RVs indicate they aren't associated with the putative primary. These 69 stars are listed in Table \ref{tab:cpm_feh}, and span $-1.28 < {\rm [Fe/H]} < +0.56$.

\begin{figure*}
    \centering
    \includegraphics[width=0.75\textwidth,page=1]{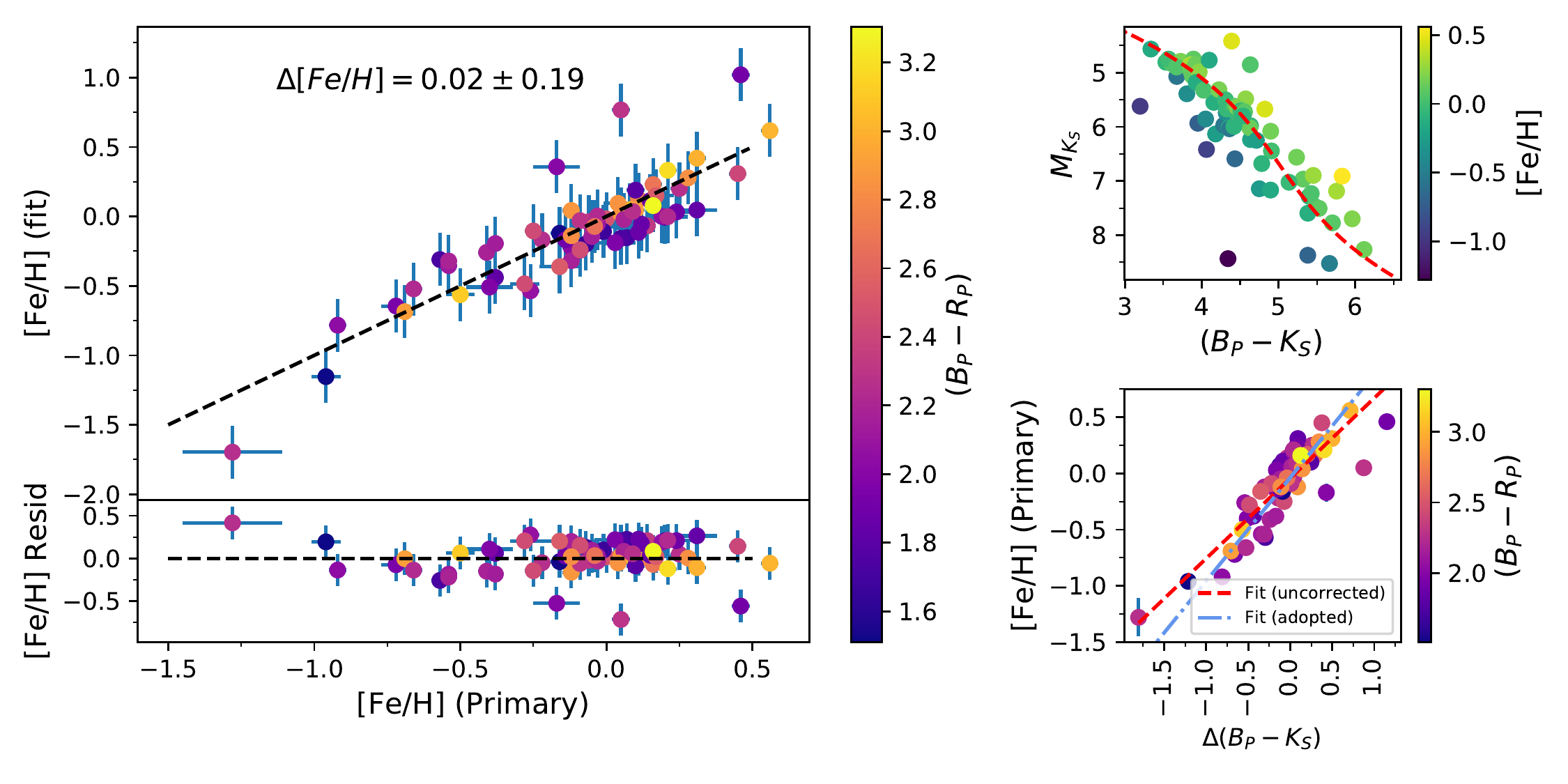}
    \caption{\textbf{Left:} Cool dwarf secondary [Fe/H] calculated from our photometric calibration vs [Fe/H] from the associated FGK primary star, colour coded by Gaia $(B_P-R_P)$. The standard deviation of the residuals, and our adopted uncertainty for the relation, is $\pm0.19\,$dex. See Table \ref{tab:cpm_feh} for further information on this FGK-KM binary calibration sample. \textbf{Top Right:} $M_{K_S}-(B_P-K_S)$ colour magnitude diagram for the calibration sample of cool dwarf secondaries colour coded by host star [Fe/H]. The dashed red line is a third order polynomial representing the main sequence, fitted to the \citet{mann_how_2015} sample of cool dwarfs. \textbf{Bottom Right:} Fitted [Fe/H] as a function $\Delta(B_P-K_S)$ offset from the mean main sequence polynomial. The dashed red line is the initial uncorrected linear least squares fit, and the dash-dotted blue line is the adopted fit after correcting for the remaining trend in the residuals}

    \label{fig:phot_feh_rel}
\end{figure*}

From this sample we follow the approaches of \citet{johnson_metal_2009} and \citet{schlaufman_physically-motivated_2010} and use a polynomial to trace the mean main sequence in $M_{K_S}-$colour space, though using $(B_P-K_S)$ instead of $(V-K_S)$. For our main sequence fit, we use the complete \citet{mann_how_2015} sample of cool dwarfs with Gaia parallaxes, which spans a wider range in $(B_P-K_S)$ and is less sparse than the assembled sample of M-dwarf secondaries. We find the following third order polynomial sufficient to describe the main sequence:
\begin{equation}
    (B_P-K_S) = a_3 M_{K_S}^3 + a_2 M_{K_S}^2 + a_1 M_{K_S} + a_0
\end{equation}

where $a_3=0.05385$, $a_2=-1.08356$, $a_1=7.76175$, and $a_0=-14.54705$. We then calculate the offset in $(B_P-K_S)$ from this polynomial (as a colour offers greater discriminatory power than $M_{K_S}$, \citealt{schlaufman_physically-motivated_2010}), and use least squares to find the best fitting linear relation for [Fe/H]:
\begin{equation}
    {\rm[Fe/H]} = b_1\Delta(B_P-K_S) + b_0
\end{equation}
where $b_1$ and $b_0$ are the linear polynomial coefficients. After correcting for a remaining trend in the residuals, our adopted coefficients are $b_1=0.71339$, and $b_0=-0.04301$. This relation is valid for stars with $1.51 < (B_P-R_P) < 3.3$ (based on the hottest and coolest secondaries respectively), and has an uncertainty of $\pm0.19\,$dex (from the standard deviation in the residuals). We stress that the relation should only be used for stars that pass the same quality cuts we use to build the relation: unsaturated photometry, not flagged as a duplicate source in Gaia, RUWE $<1.4$, and not a known/suspected spectroscopic binary or pre-main sequence star. Our [Fe/H] recovery and fits can be seen in Figure \ref{fig:phot_feh_rel}. 

\section{Spectroscopic Analysis}\label{sec:spectroscopic_analysis}
The TESS candidate planet host observing program described here developed from an ANU 2.3$\,$m/WiFeS survey of potential young stars \citep{zerjal_spectroscopically_2021} to identify signs of youth (via Balmer Series and Ca II H\&K emission, and Li 6708\SI{}{\angstrom} absorption) and determine RVs to enable kinematic analysis with \texttt{Chronostar} \citep{crundall_chronostar_2019} when combined with Gaia astrometry. While their spectral type coverage ($1.27 < B_P < 2.6$) was relatively similar to our own, instrument setup however prioritised higher spectral resolution for improved velocity precision and coverage of the key wavelength regions of interest. These regions are firmly in the optical, where M-dwarf spectral features are strongly blended and heavily dominated by molecular absorption from hydrides (e.g. MgH, CaH, SiH) and oxides (e.g. TiO, VO, ZrO). This is in contrast to most of the previous low-medium resolution studies of M-dwarfs which work in the NIR where the absorption is less severe and many more [Fe/H] sensitive features are available. 

Here we describe our attempts to derive reliable atmospheric parameters from our spectra using a model based approach. Our investigation ultimately revealed substantial systematics and degeneracies when fitting to model optical spectra, resulting in our inability to recover $\log g$ or [Fe/H]. While the spectra are included in our temperature fitting routine, they are primarily used for RV determination, identification of peculiarities (such as signs of youth), and for testing model fluxes. The details of our findings are covered below, and we await follow-up work to explore a standard-based or data-driven approach (e.g. similar to the work of \citealt{birky_temperatures_2020}, but in the optical) to take full advantage of the information in our now large library of optical cool dwarf spectra.

\subsection{Selection of Model Atmosphere Grid}\label{sec:model_limitations}
While synthetic spectra show better agreement for FGK stars, the onset of strong molecular features such as TiO and H$_2$O in the atmospheres of late K and M dwarf atmospheres make the task of modelling their spectra far more complex. There are known historical issues, for instance, when computing optical colours from synthetic spectra (e.g. difficulties in computing accurate $V$ band magnitudes, \citealt{leggett_infrared_1996}), and the line lists required are considerably more complicated. 
Thus, before using models in our automatic fitting routine, we first investigate their performance at different wavelengths to flag regions requiring special consideration.
For the purposes of this comparison, we check the MARCS grid of stellar atmospheres against the BT-Settl grid \citep{allard_model_2011}, both of which are described in detail below.

Our template grid of 1D LTE MARCS spectra was previously described by \citet{nordlander_lowest_2019} and computed using the TURBOSPECTRUM code (v15.1; \citealt{alvarez_near-infrared_1998, plez_turbospectrum_2012}) and MARCS model atmospheres \citep{gustafsson_grid_2008}. 
The spectra are computed with a sampling resolution of $1\,$km$\,$s$^{-1}$, corresponding to a resolving power of $R\sim300\,000$, with a microturbulent velocity of $1\,$km$\,$s$^{-1}$. We adopt the solar chemical composition and isotopic ratios from \citet{asplund_chemical_2009}, except for an alpha enhancement that varies linearly from $\text[\alpha / \text{Fe}] = 0$ when $\rm [Fe/H] \ge 0$ to $\text[\alpha/\text{Fe}] = +0.4$ when $\rm [Fe/H] \le -1$. 
We use a selection of atomic lines from VALD3 \citep{ryabchikova_major_2015} together with roughly 15 million molecular lines representing 18 different molecules, the most important of which for this work are CaH (Plez, priv. comm.), MgH \citep{kurucz_kurucz_1995,skory_new_2003}, and TiO \citep[with updates via VALD3]{plez_new_1998}.

MARCS model fluxes were developed for usage over a range of spectral types including both cool giants and, critically for our work here, cool dwarfs. Recent work fitting cool dwarf stellar atmospheres however have mostly used high-resolution NIR spectra (J band: \citealt{onehag_m-dwarf_2012, lindgren_metallicity_2016, lindgren_metallicity_2017}; H band: \citealt{souto_chemical_2017, souto_stellar_2018}) rather than the medium resolution optical spectra we use here.

For BT-Settl, we use the most recently published grid \citep{allard_atmospheres_2012, allard_models_2012, baraffe_new_2015}\footnote{\url{https://phoenix.ens-lyon.fr/Grids/BT-Settl/CIFIST2011_2015/}} which uses abundances from \citet{caffau_solar_2011} and covers $1200 < T_{\rm eff} < 7000\,$K, $2.5 < \log g < 5.5$, [M/H]$ = 0.0$. Note that while older grids have a wider range of [M/H], they are also less complete in terms of physics and line lists, so we opt for the newest grid for our comparison here, and limit ourselves to testing on stars with approximately Solar [Fe/H]. 

BT-Settl atmospheres have been developed with a focus on cool dwarf atmospheres and have a strong history of use for studying cool dwarfs at a variety of wavelengths and resolutions \citep[e.g.][]{rojas-ayala_metallicity_2012, muirhead_characterizing_2012, mann_they_2012, rajpurohit_effective_2013, lepine_spectroscopic_2013, gaidos_trumpeting_2014, mann_spectro-thermometry_2013, mann_how_2015, veyette_physical_2016, veyette_physically_2017, souto_stellar_2018}. Most noteworthy for our comparison are tests by \citet{reyle_effective_2011} and \citet{mann_spectro-thermometry_2013}, which examined model performance at optical wavelength regions $>5500\,$\SI{}{\angstrom} common to our WiFeS R7000 spectra.

\begin{figure*}
    \centering
    \includegraphics[width=\textwidth,page=1]{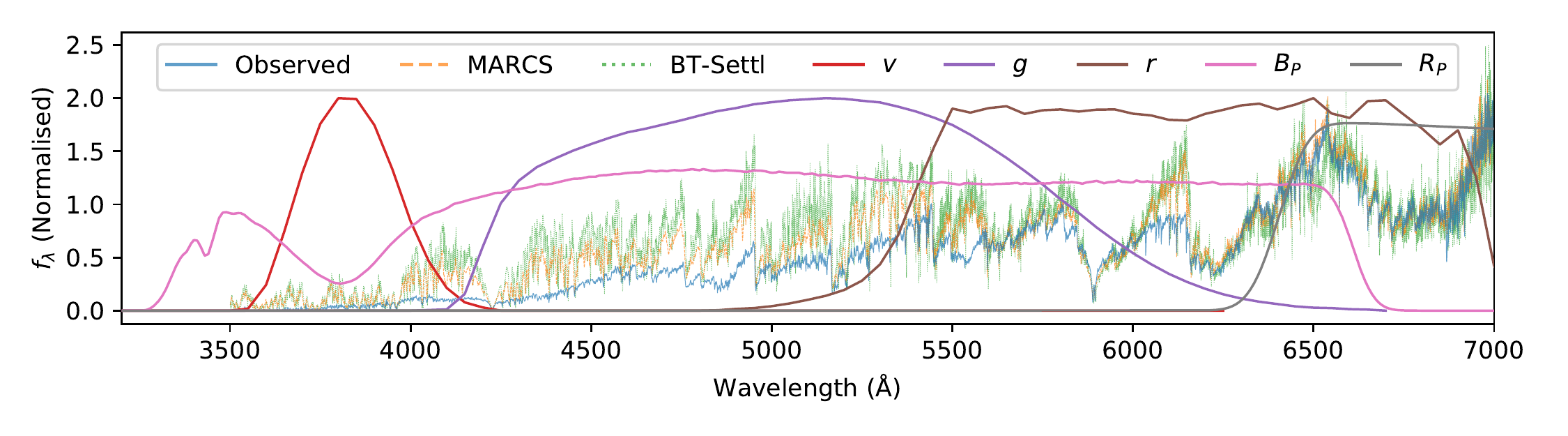}
    \caption{Observed WiFeS B3000 and R7000 spectra for GJ 447, along with a MARCS synthetic spectrum interpolated to the parameters from \citealt{mann_how_2015} ($T_{\rm eff}=3192\,$K, $\log g=5.04$, [Fe/H] $=-0.02$), and a PHOENIX/BT-Settl spectrum at the closest grid point available ($T_{\rm eff}=3200\,$K, $\log g=5.0$, [Fe/H] $=0.0$). SkyMapper $v$, $g$, $r$, and Gaia $B_P$ and $R_P$ filters are overplotted for reference. Note the severe model disagreement below 5400$\,$\SI{}{\angstrom}.}
    \label{fig:model_limitations}
\end{figure*}

For each of our standard stars we combined and normalised our flux calibrated B3000 and R7000 spectra to give a single spectrum with $3500 < \lambda < 7000\,$\SI{}{\angstrom}. To this we compared synthetic MARCS fluxes interpolated to literature values of $T_{\rm eff}$, $\log g$, and [Fe/H], as well as the BT-Settl equivalent for those with close to Solar [Fe/H]. Given our large library of standards we were able to observe model performance as a function of both stellar parameters and wavelength. A representative comparison (with overplotted filter bandpasses) is shown in Figure \ref{fig:model_limitations}, and our main conclusions are summarised as follows:
\begin{itemize}
    \item Both MARCS and BT-Settl models severely overpredict (worsening with decreasing $T_{\rm eff}$) flux blueward of $\sim$5400$\,$\SI{}{\angstrom}. The MARCS systematic offset is also a strong function of [Fe/H], an effect also observed in \citet{joyce_investigating_2015}, and while this is likely also true for BT-Settl, we cannot comment definitively while limited to the Solar [Fe/H] grid. 
    \item BT-Settl additionally underpredicts flux at $\sim$6500$\,$\SI{}{\angstrom} (as expected from \citealt{reyle_effective_2011} and \citealt{mann_spectro-thermometry_2013}).
    \item Synthetic photometry generated in SkyMapper $v$, $g$, $r$, and Gaia $B_P$ is thus systematically brighter than the observed equivalents for reasonable assumptions of $T_{\rm eff}$, $\log g$, and [Fe/H] for the star under consideration.
\end{itemize}

\begin{figure}
 \includegraphics[width=\columnwidth]{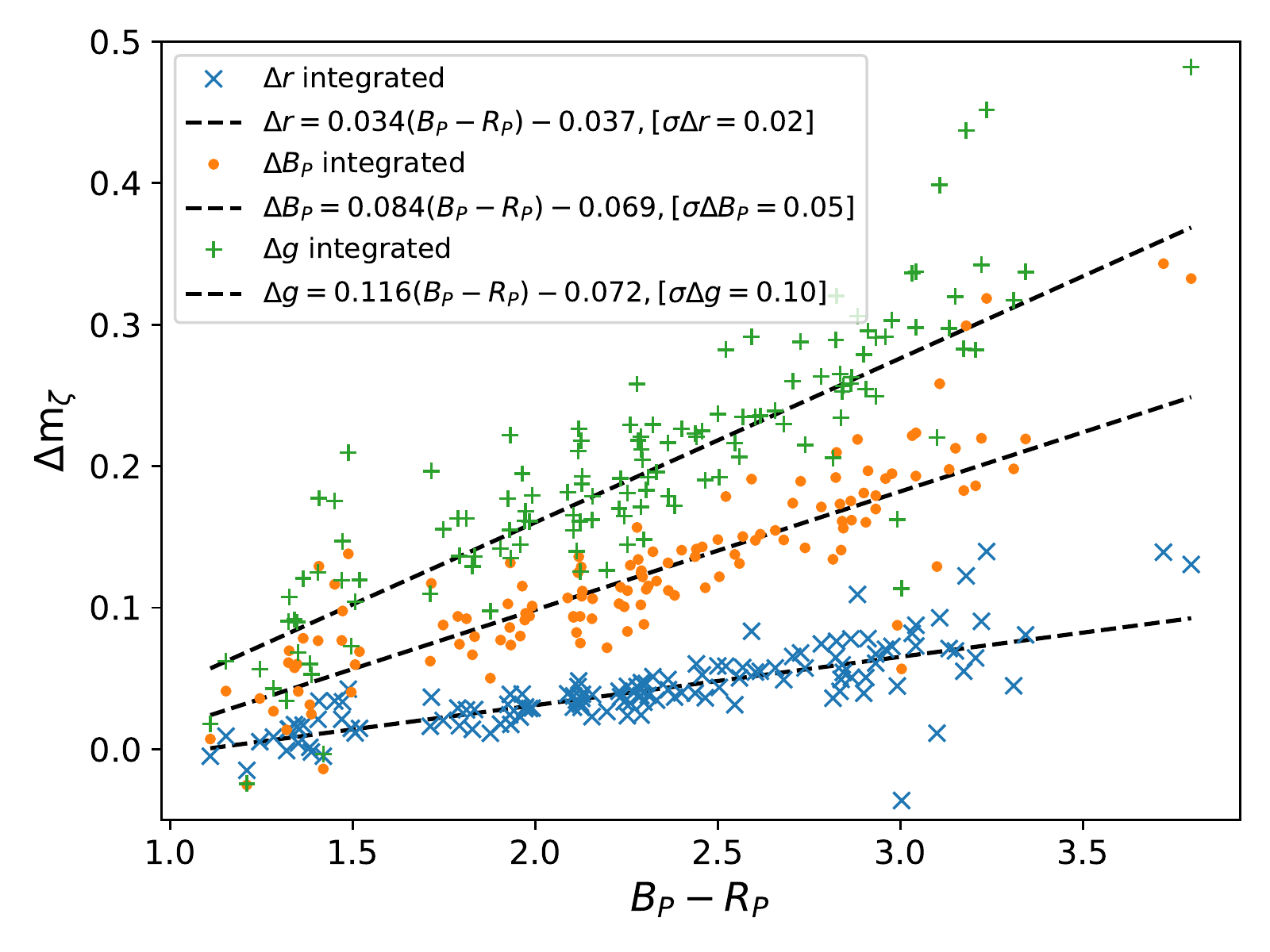}
 \caption{Gaia $B_P$, and SkyMapper $gr$ systematic offsets between integrated flux calibrated WiFeS spectra and MARCS model integrated spectra at literature parameters for our standard stars, plotted as a function of observed Gaia $B_P-R_P$. Stars redder in $B_P-R_P$ have systematically more flux at bluer wavelengths, with the best fit linear magnitude offset plotted for each filter, and the standard deviation in magnitude noted.}
 \label{fig:mag_offsets}
\end{figure}

We are able to quantify these systematics by integrating photometry from our flux calibrated observed spectra and comparing to the MARCS synthetic equivalents generated at the literature parameters for each star. Our wavelength coverage allows us to check the magnitude offsets $\Delta v$, $\Delta g$, $\Delta r$ and $\Delta B_P$, corresponding to $v$, $g$, $r$, and $B_P$ respectively. We note that for the purpose of this comparison we do not account for inaccuracies in our flux calibration, telluric absorption, nor for WiFeS not covering the bluest $\sim200\,$\SI{}{\angstrom} of $B_P$. However, checks with synthetic spectra show that this region accounts for less than $0.25\,$\% of $B_P$ flux at 3000$\,$K where our correction is greatest, and remains less than $0.5\,$\% of flux at 4,500$\,$K where our correction is more modest. These offsets are shown for $g$, $r$, and $B_P$ in Figure \ref{fig:mag_offsets}, and fit separately for each filter by the following linear relation in observed Gaia DR2 $(B_P-R_P)$:
\begin{equation}\label{eqn:bp_offset}
    \Delta m_\zeta = a_1 (B_P-R_P) + a_0
\end{equation}
where $\Delta m_\zeta$ is the magnitude offset in filter $\zeta$; $a_1$ equals 0.116, 0.084, and 0.034 for $g$, $r$, and $B_P$ fits respectively; and $a_0$ equals -0.072, -0.069, and -0.037 for $g$, $r$, and $B_P$ fits respectively. Computing the standard deviation for the residuals shows 0.10, 0.05, and 0.02 uncertainties in magnitude (equivalent to roughly 10\%, 5\%, and 2\% uncertainties in flux) for $g$, $r$, and $B_P$ respectively. From this we conclude that while the corrections to $r$, and $B_P$ are modest, $g$ is likely too affected to prove useful.

Following this both qualitative and quantitative investigation comparing model fluxes to our library of standard star spectra, we make the following decisions for our synthetic fitting methodology:
\begin{itemize}
    \item Given similar observed systematics for both MARCS and BT-Settl model fluxes, we adopt the MARCS grid to enable fitting for [Fe/H] as well as $T_{\rm eff}$ and $\log g$. 
    \item Only use our R7000 spectra ($5400 \leq \lambda \leq 7000\,$\SI{}{\angstrom}) for fitting, additionally masking out the two regions worst affected by missing opacities (5498-5585$\,$\SI{}{\angstrom} and 6029-6159$\,$\SI{}{\angstrom}).
    \item Apply an \textit{observed} $(B_P-R_P)$ dependent systematic offset to our generated synthetic $B_P$ and $r$ photometry per Equation \ref{eqn:bp_offset}.
    \item Given the widespread historical use and success of studying M-dwarfs at NIR wavelengths, we use $R_P$, $i$, $z$, $J$, $H$, and $K_S$ photometry assuming no substantial model systematics.
    \item However, to account for remaining model uncertainties, we add conservative $\pm0.011$ magnitude (1\% in flux) uncertainties in quadrature with the observed uncertainties for $R_P$, $i$, $z$; and the fitted $\pm0.02$ for $r$, and $\pm0.05$ for $B_P$.

\end{itemize}

\subsection{Synthetic Fitting}\label{sec:synthetic_fitting}
Our approach to spectral fitting was developed specifically to work with the complicated spectra of our cool star sample and incorporates nine distinct sources of information. While it was hoped that this methodology would be sufficient to disentangle the strong degeneracy between $T_{\rm eff}$ and [Fe/H] and accurately recover \textit{distant-independent} [Fe/H] for our standard sample, this ultimately proved not to be the case. While we are able to tightly constrain $T_{\rm eff}$, we must resort to using the photometric [Fe/H] relation developed in Section \ref{sec:feh_rel} to fix [Fe/H] during the fit. The information included in our fit is as follows:
\begin{enumerate}
    \item Medium resolution R7000 optical spectra from WiFeS,
    \item Observed Gaia $B_P$, $R_P$; 2MASS $J$, $H$, and $K_S$; and SkyMapper DR3 $r$, $i$, $z$ photometry,
    \item Empirical cool dwarf radius relations from \citet{mann_how_2015} - valid for K7-M7 stars, and used to estimate $\log g$,
    \item Empirical cool dwarf mass relations from \citet{mann_how_2019} - valid for $0.075 M_\odot < M_\star < 0.70 M_\odot$, and used to estimate $\log g$,
    \item Synthetic MARCS model spectra (for spectral fitting, interpolated to the resolution and wavelength grid of WiFeS)
    \item MARCS model fluxes (for photometric fitting),
    \item Stellar parallaxes from Gaia DR2,
    \item The interstellar dust map from \citet{leike_charting_2019},
    \item A set of reference stellar standards with known parameters for testing and validation purposes (see Section \ref{sec:standards} for details).
\end{enumerate}

We found that least squares fitting between real and synthetic spectra alone consistently underestimated expected $\log g$ values of our sample by up to $0.3\,$dex - physical for a set of young stars, but not realistic for our overwhelmingly main sequence sample. To counter this, we calculate $\log g$ using the absolute $K_S$ band radius and mass relations of \citet{mann_how_2015} and \citet{mann_how_2019}\footnote{Calculated using the Python code available at: \url{https://github.com/awmann/M_-M_K-}} respectively, and fix it during fitting. We then use a two step iterative procedure, with the first fit fixing $\log g$ to the value from empirical relations, and a second and final fit using our interim measured radius and a mass from \citet{mann_how_2019}. All of our TESS targets fall within the stated $4 < M_{K_S} < 11$ limits for the mass relation. Although the relation is only valid for main sequence stars, we employ it with caution for two suspected young stars TOI 507 (TIC 348538431) and TOI 142 (425934411), both discussed in more detail in Section \ref{sec:emission}, on the assumption that the resulting value of $\log g$ will still be more accurate than an unconstrained synthetic fit. Additionally, we suspect TOI 507 of being a near-equal mass binary, and as such treat it as $0.75$ magnitudes fainter (or half as bright) for the purpose of using the relation, equivalent to determining the mass for only a single component.

While this now solves the $\log g$ issue, we are still left with two issues arising from the spectra themselves. The first is that certain wavelength regions of our MARCS model spectra are a poor match  compared to our reference sample with known $T_{\rm eff}$, $\log g$, and [Fe/H] - particularly at cooler temperatures. As discussed in Section \ref{sec:model_limitations}, we account for this by using only spectra from the red arm of WiFeS with $\lambda > 5400\,$A, and masking out remaining regions with poor agreement.

The second remaining issue is that of the degeneracy between $T_{\rm eff}$ and [Fe/H] when fitting spectra. This effect is caused by both the temperature and metallicity influencing the strength of atmospheric molecular absorbers or opacity sources (predominantly TiO in the optical, but also various hydrides). What this means in practice is that there often isn't a single minimum or optimal set of atmospheric parameters when fitting synthetic spectra, but instead there exists a range of good fits (or even multiple minima) at different combinations of $T_{\rm eff}$ and [Fe/H] - possibly separated by several 100$\,$K in $T_{\rm eff}$ or several 0.1$\,$dex in [Fe/H].

In an attempt to overcome this, we include photometry from redder wavelengths that are less dominated by absorption than optical wavelengths, meaning that $T_{\rm eff}$ and [Fe/H] are less degenerate. While we do not have NIR spectra for our science or reference sample, we do have Gaia, SkyMapper, and 2MASS photometry in the form of $B_P$, $R_P$, $r$, $i$, $z$, $J$, $H$, and $K_S$ which together give us almost continuous wavelength coverage out to nearly 2.4$\,\mu$m and covers the bulk of stellar emission for our cool stars.

We thus modified our fitting methodology to also compute the uncertainty weighted residuals between observed and synthetic stellar photometry. In order to compare synthetic photometry to its observed equivalent we formulate the fit as follows:

\begin{equation}
    m_{\zeta, m} = {\rm BC}_{\zeta}(T_{\rm eff},\log g, {\rm [Fe/H]}) + m_{\rm bol} 
\end{equation}
where $m_{\zeta, m}$ is the model magnitude in filter $\zeta$; ${\rm BC}_{\zeta}$ is the bolometric correction (i.e. the total flux outside of a filter $\zeta$) as a function of $T_{\rm eff}$, $\log g$, and [Fe/H] in filter $\zeta$; and $m_{\rm bol}$ is the apparent bolometric magnitude (i.e. the apparent magnitude of the star over all wavelengths). 
In this implementation $m_{\rm bol}$ serves as a physically meaningful free parameter used to scale synthetic magnitudes to their observed equivalents and ultimately allow computation of the apparent bolometric flux $f_{\rm bol}$.
This is done using the well tested \texttt{bolometric-corrections}\footnote{\url{https://github.com/casaluca/bolometric-corrections}} software \citep{casagrande_synthetic_2014, casagrande_synthetic_2018,casagrande_use_2018} to interpolate a grid of bolometric corrections from MARCS fluxes in different filters for the stellar parameters at each fitting call. By fitting for $m_{\rm bol}$ and using bolometric corrections, we are thus directly able to compare an observed magnitude, $m_{\zeta,o}$, from Gaia, SkyMapper, or 2MASS directly with its MARCS synthetic equivalent. With $\log g$ fixed, we now have a three term fit in terms of $T_{\rm eff}$, [Fe/H], and $m_{\rm bol}$, the latter of which allows for direct computation of the bolometric flux (and thus the stellar radius).

This fitting procedure is equivalent to minimising the following relation (performed using the least\_squares function from \texttt{scipy}'s optimize module):
\begin{equation}
    R(\theta) = \displaystyle\sum_{i=1}^{M}\Bigg(\frac{1}{C\sqrt{\chi_f^2}}\frac{{f_{o,i} - f_{m,i}}}{\sigma_{f_{o, i}}}\Bigg)^2 + \displaystyle\sum_{\zeta=1}^{N}\Bigg(\frac{1}{\sqrt{\chi_m^2}}\frac{{m_{\zeta,o} - (m_{\zeta,m}+\Delta m_\zeta)}}{\sigma_{m_{\zeta}}}\Bigg)^2
\end{equation}
with model uncertainties taken into account via:
\begin{equation}
    \sigma_{m_{\zeta}} = \sqrt{\sigma_{m_{\zeta,o}}^2 + \sigma_{m_{\zeta,m}}^2}
\end{equation}

where $R(\theta)$ are the combined spectral and photometric squared residuals as a function of $\theta$, a vector of $T_{\rm eff}$, $\log g$, [Fe/H], $m_{\rm bol}$); $M$ is the total number of spectral pixels, $i$ is the spectral pixel index, $f_{\rm o, i}$ and $f_{\rm m,i}$ are the observed and model spectral fluxes respectively at pixel $i$, normalised by their respective medians in the range $6200 \leq \lambda \leq 7000\,$\SI{}{\angstrom}; $\sigma_{f_{\rm o,i}}$ is the observed flux uncertainty at pixel $i$; $N$ is the total number of photometric filters; $\zeta$ is the filter index, $m_{\zeta,o}$ and $m_{\zeta,m}$ are the observed and model magnitudes respectively in filter $\zeta$; $\Delta m_\zeta$ is the systematic model magnitude offset in filter $\zeta$ (per Equation \ref{eqn:bp_offset} for $r$ and $B_P$, and 0 for all other filters); $\sigma_{m_{\zeta, o}}$ and $\sigma_{m_{\zeta, m}}$ are the uncertainties on the observed and model magnitudes respectively, added in quadrature to give the total magnitude uncertainty $\sigma_{m_\zeta}$; $\chi_f^2$ and $\chi_m^2$ are the \textit{global} minimum $\chi^2$ values computed from the spectral and photometry residuals respectively  (i.e. global fit using only R7000 spectra, without photometry, and a separate global photometric fit without spectra) used to normalise the two sets of residuals in the case of poor fits and place them on a similar scale; and $C$, set to 20, is a constant used to account for the spectra having many more pixels than the number of photometric points. This value of $C$ was chosen by visually inspecting the residuals of our spectral fits and means that we assume, on average, every 20 spectral pixels are correlated and do not contain unique information.

\begin{figure*}
 \includegraphics[width=\textwidth]{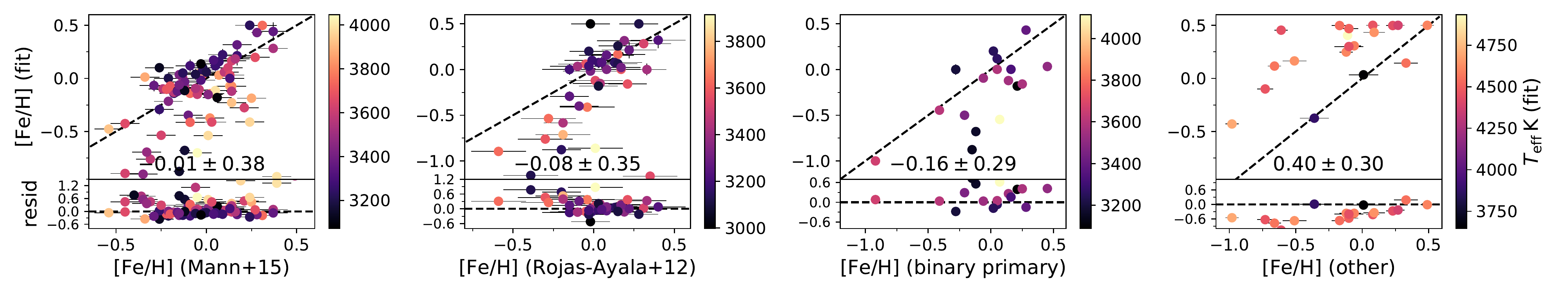}
 \caption{[Fe/H] recovery for our 3 parameter fit in $T_{\rm eff}$, [Fe/H], $m_{\rm bol}$ for our four sets of [Fe/H] standards: \citealt{mann_how_2015}, \citealt{rojas-ayala_metallicity_2012}, primary star [Fe/H] for cool dwarfs in binaries, and mid-K dwarfs. The median and standard deviation of each set of residuals is annotated. Note the inability of the 3 parameter fit to reliably recover [Fe/H], with the scatter on our recovered [Fe/H] for the binary sample (the most reliable set of [Fe/H] standards) being larger than the scatter on our photometric [Fe/H] relation.}
 \label{fig:standard_feh_comp}
\end{figure*}

We test the accuracy of our fitted [Fe/H] using a set of cool star stellar standards in Figure \ref{fig:standard_feh_comp}. It is immediately clear that, despite the tight constraint on $T_{\rm eff}$ that our broad wavelength coverage from photometry allows, we are unable to recover [Fe/H] for our standard sample to better precision than our photometric [Fe/H] relation from Section \ref{sec:feh_rel}. Our fits systematically overpredict [Fe/H] for the coolest stars in our sample, which might be similar to what was observed in Figure 3 of \citealt{rojas-ayala_metallicity_2012} (using BT-Settl models), where they find even metal-rich models fail to reproduce the depth of certain features. This has also previously been observed for cool, metal-poor clusters when using evolutionary models \citep[e.g.][]{joyce_investigating_2015}, and observed for isochrones \citep[e.g.][]{joyce_not_2018}. From this we conclude that a simple least squares fit to our medium resolution optical spectra, unweighted to [Fe/H] sensitive regions, and using models with both known and unknown systematics is not sufficient to accurately determine [Fe/H] for cool dwarfs. 

\begin{figure*}
 \includegraphics[width=\textwidth]{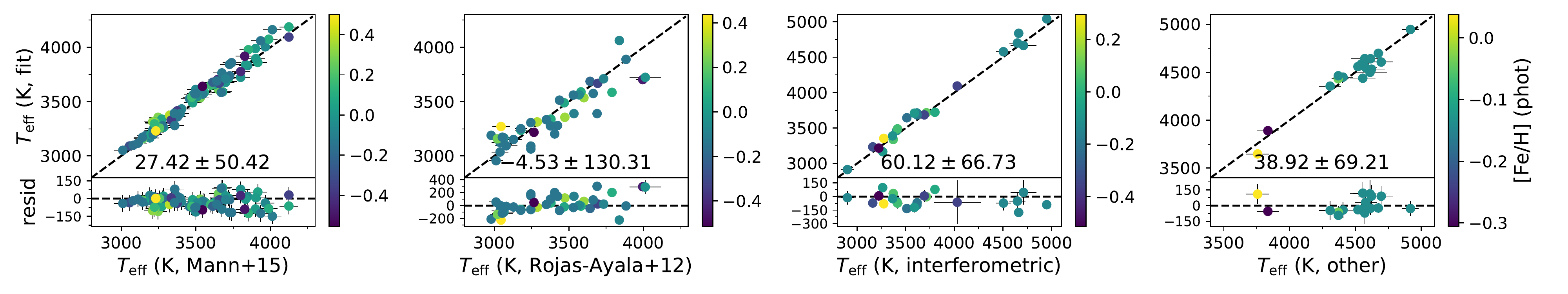}
 \caption{$T_{\rm eff}$ recovery for our 2 parameter fit in $T_{\rm eff}$, and $m_{\rm bol}$ for our four sets of $T_{\rm eff}$ standards: \citealt{mann_how_2015}, \citealt{rojas-ayala_metallicity_2012}, interferometry, and mid-K dwarfs. [Fe/H] is from our photometric [Fe/H] relation where appropriate, or fixed to the mean Solar Neighbourhood [Fe/H] if not. The median and standard deviation of each set of residuals is annotated (note that these values have not yet been corrected for the systematic, as discussed in Section \ref{sec:discussion:param_recovery}).}
 \label{fig:standard_teff_comp}
\end{figure*}

Given this, it is clear a three parameter fit to $T_{\rm eff}$, [Fe/H], and $m_{\rm bol}$ is unreasonable. Our final reported parameters are thus a two parameter fit to $T_{\rm eff}$, and $m_{\rm bol}$, fixing [Fe/H] to the value from our relation in Section \ref{sec:feh_rel} for those stars falling within the $(B_P-R_P)$ range, and the mean value for the Solar Neighbourhood of [Fe/H] $=-0.14$ \citep{schlaufman_physically-motivated_2010} for stars outside this range, or suspected of binarity or being young. To further account for both model and zeropoint uncertainties, we add a 1\% flux uncertainty in quadrature with our fitted statistical uncertainties on $m_{\rm bol}$. Our standard star $T_{\rm eff}$ recovery for the two parameter fit is shown in Figure \ref{fig:standard_teff_comp}.

\begin{figure}
 \includegraphics[width=\columnwidth]{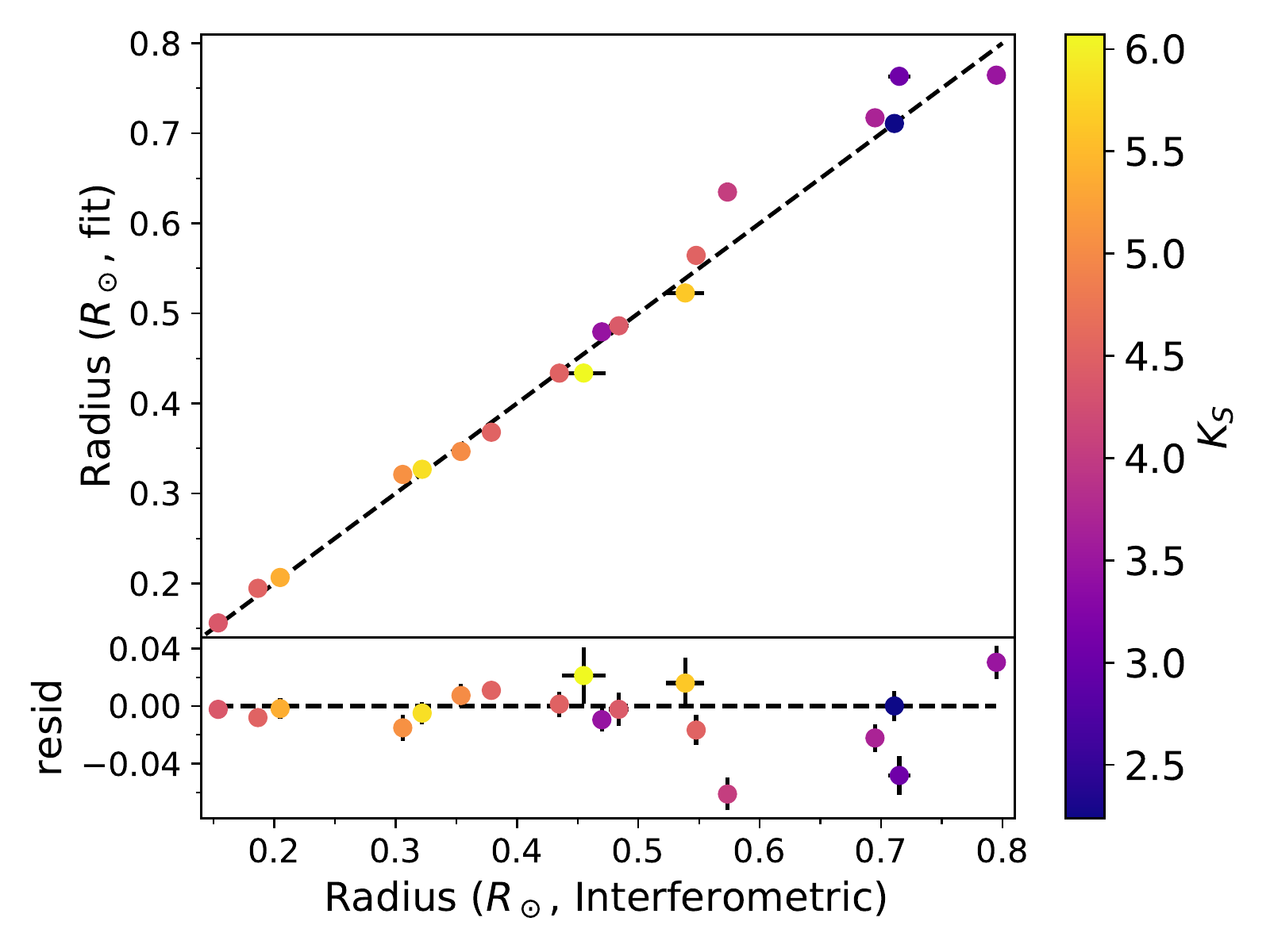}
 \caption{Radius comparison for those targets with interfometric radii to better than 5\% precision. The median distance precision for these targets is $0.04\,$\%. We find generally good agreement between literature measurements and our own, though noting that the brightness of this sample (see apparent 2MASS $K_S$ magnitude on the colour bar) results in photometry that is either saturated or has lower precision and thus may be the cause of some of the scatter observed.}
 \label{fig:radius_comp_interferometric}
\end{figure}

We compute the apparent bolometric flux $f_{\rm bol}$ from our fitted value of $m_{\rm bol}$ using Equation 3 from \citet{casagrande_synthetic_2018}, from which we then compute the stellar radius $R_\star$. Figure \ref{fig:radius_comp_interferometric} shows a comparison between our radii and those from our interferometric standard sample, and final values for TESS science targets and stellar standards are reported in Tables \ref{tab:final_results_tess} and \ref{tab:final_results_std} respectively.

\begin{table*}
\centering
\caption{Final results for TESS candidate exoplanet hosts}
\label{tab:final_results_tess}
\begin{tabular}{ccccccccccc}
\hline
TOI & TIC & $T_{\rm eff}$ & $\log g$ & [Fe/H] & $M$ & $R_\star$ & $m_{\rm bol}$ & $f_{\rm bol}$ & EW(H$\alpha$) & $\log R^{'}_{\rm HK}$ \\
 &  & (K) &  &  & ($M_\odot$) & ($R_\odot$) &  & {{\footnotesize(10$^{-12}\,$ergs s$^{-1}$ cm $^{-2}$)}} & \SI{}{\angstrom} &  \\
\hline
136 & 410153553 & 2988 $\pm$ 30 & 5.06 $\pm$ 0.02 &- &0.155 $\pm$ 0.004 &0.192 $\pm$ 0.004 &12.05 $\pm$ 0.01 &384.4 $\pm$ 3.9 &-0.05 & -5.37 \\
540 & 200322593 & 3104 $\pm$ 30 & 5.07 $\pm$ 0.02 &-0.10 &0.164 $\pm$ 0.004 &0.197 $\pm$ 0.004 &11.71 $\pm$ 0.01 &528.3 $\pm$ 5.3 &2.49 & - \\
256 & 92226327 & 3150 $\pm$ 30 & 5.01 $\pm$ 0.02 &-0.13 &0.182 $\pm$ 0.005 &0.220 $\pm$ 0.004 &11.55 $\pm$ 0.01 &611.8 $\pm$ 6.1 &-0.22 & -5.53 \\
203 & 259962054 & 3169 $\pm$ 30 & 5.01 $\pm$ 0.02 &-0.07 &0.200 $\pm$ 0.005 &0.232 $\pm$ 0.004 &12.49 $\pm$ 0.01 &255.4 $\pm$ 2.6 &0.56 & -4.98 \\
507 & 348538431 & 3279 $\pm$ 30 & 4.76 $\pm$ 0.02 &- &0.383 $\pm$ 0.010 &0.424 $\pm$ 0.008 &14.28 $\pm$ 0.01 & 49.2 $\pm$ 0.5 &2.24 & -4.49 \\
910 & 369327947 & 3282 $\pm$ 30 & 4.97 $\pm$ 0.02 &-0.04 &0.262 $\pm$ 0.008 &0.278 $\pm$ 0.005 &10.46 $\pm$ 0.01 &1656.2 $\pm$ 16.6 &-0.27 & -5.47 \\
210 & 141608198 & 3284 $\pm$ 30 & 4.90 $\pm$ 0.02 &0.21 &0.312 $\pm$ 0.008 &0.326 $\pm$ 0.006 &12.78 $\pm$ 0.01 &195.8 $\pm$ 2.0 &-0.23 & -5.84 \\
122 & 231702397 & 3326 $\pm$ 30 & 4.86 $\pm$ 0.02 &-0.07 &0.316 $\pm$ 0.008 &0.345 $\pm$ 0.006 &13.42 $\pm$ 0.01 &109.3 $\pm$ 1.1 &-0.21 & - \\
455 & 98796344 & 3330 $\pm$ 30 & 4.97 $\pm$ 0.02 &-0.27 &0.248 $\pm$ 0.006 &0.271 $\pm$ 0.005 &9.16 $\pm$ 0.01 &5507.5 $\pm$ 54.8 &-0.29 & -5.39 \\
732 & 36724087 & 3354 $\pm$ 30 & 4.83 $\pm$ 0.02 &0.13 &0.364 $\pm$ 0.009 &0.382 $\pm$ 0.007 &10.91 $\pm$ 0.01 &1103.5 $\pm$ 11.0 &-0.23 & -5.55 \\
674 & 158588995 & 3355 $\pm$ 30 & 4.77 $\pm$ 0.02 &- &0.419 $\pm$ 0.010 &0.443 $\pm$ 0.008 &12.19 $\pm$ 0.01 &338.7 $\pm$ 3.4 &-0.45 & -5.58 \\
406 & 153065527 & 3369 $\pm$ 30 & 4.83 $\pm$ 0.02 &0.21 &0.380 $\pm$ 0.009 &0.392 $\pm$ 0.007 &11.58 $\pm$ 0.01 &594.6 $\pm$ 5.9 &-0.15 & -5.40 \\
175 & 307210830 & 3381 $\pm$ 30 & 4.94 $\pm$ 0.02 &-0.24 &0.293 $\pm$ 0.007 &0.304 $\pm$ 0.005 &9.78 $\pm$ 0.01 &3099.8 $\pm$ 31.1 &-0.31 & -5.47 \\
782 & 429358906 & 3390 $\pm$ 30 & 4.81 $\pm$ 0.02 &0.25 &0.401 $\pm$ 0.010 &0.413 $\pm$ 0.008 &12.57 $\pm$ 0.01 &237.3 $\pm$ 2.4 &-0.37 & -6.38 \\
244 & 118327550 & 3422 $\pm$ 30 & 4.82 $\pm$ 0.02 &0.10 &0.402 $\pm$ 0.010 &0.407 $\pm$ 0.007 &10.69 $\pm$ 0.01 &1349.2 $\pm$ 13.5 &-0.30 & -5.36 \\
789 & 300710077 & 3434 $\pm$ 30 & 4.86 $\pm$ 0.02 &-0.16 &0.360 $\pm$ 0.009 &0.371 $\pm$ 0.007 &12.34 $\pm$ 0.01 &294.0 $\pm$ 2.9 &-0.28 & - \\
654 & 35009898 & 3436 $\pm$ 30 & 4.77 $\pm$ 0.02 &0.07 &0.425 $\pm$ 0.010 &0.445 $\pm$ 0.008 &12.56 $\pm$ 0.01 &239.6 $\pm$ 2.4 &-0.33 & -8.10 \\
864 & 231728511 & 3452 $\pm$ 30 & 4.82 $\pm$ 0.02 &-0.10 &0.392 $\pm$ 0.010 &0.403 $\pm$ 0.007 &11.86 $\pm$ 0.01 &460.0 $\pm$ 4.6 &-0.37 & -5.21 \\
562 & 413248763 & 3456 $\pm$ 30 & 4.88 $\pm$ 0.02 &-0.22 &0.347 $\pm$ 0.008 &0.353 $\pm$ 0.006 &9.11 $\pm$ 0.01 &5758.7 $\pm$ 57.3 &-0.25 & -5.93 \\
486 & 260708537 & 3467 $\pm$ 30 & 4.81 $\pm$ 0.02 &0.03 &0.424 $\pm$ 0.010 &0.424 $\pm$ 0.007 &9.74 $\pm$ 0.01 &3239.2 $\pm$ 32.4 &-0.40 & -5.39 \\
700 & 150428135 & 3467 $\pm$ 30 & 4.80 $\pm$ 0.02 &- &0.416 $\pm$ 0.010 &0.426 $\pm$ 0.007 &11.28 $\pm$ 0.01 &783.3 $\pm$ 7.8 &-0.35 & -5.36 \\
521 & 27649847 & 3468 $\pm$ 30 & 4.80 $\pm$ 0.02 &-0.01 &0.416 $\pm$ 0.010 &0.424 $\pm$ 0.008 &12.74 $\pm$ 0.01 &203.5 $\pm$ 2.0 &-0.30 & -5.49 \\
1078 & 370133522 & 3486 $\pm$ 30 & 4.83 $\pm$ 0.02 &-0.24 &0.383 $\pm$ 0.009 &0.392 $\pm$ 0.007 &10.51 $\pm$ 0.01 &1583.2 $\pm$ 15.8 &-0.34 & -6.54 \\
912 & 406941612 & 3488 $\pm$ 30 & 4.80 $\pm$ 0.02 &-0.03 &0.424 $\pm$ 0.010 &0.427 $\pm$ 0.007 &10.87 $\pm$ 0.01 &1143.7 $\pm$ 11.4 &-0.31 & -5.19 \\
270 & 259377017 & 3493 $\pm$ 30 & 4.88 $\pm$ 0.02 &-0.25 &0.364 $\pm$ 0.009 &0.361 $\pm$ 0.006 &10.90 $\pm$ 0.01 &1111.3 $\pm$ 11.1 &-0.29 & -5.40 \\
696 & 77156829 & 3496 $\pm$ 30 & 4.92 $\pm$ 0.02 &-0.45 &0.321 $\pm$ 0.008 &0.327 $\pm$ 0.006 &10.84 $\pm$ 0.01 &1170.5 $\pm$ 11.7 &-0.22 & -6.08 \\
269 & 220479565 & 3518 $\pm$ 30 & 4.83 $\pm$ 0.02 &-0.25 &0.391 $\pm$ 0.010 &0.396 $\pm$ 0.007 &12.69 $\pm$ 0.01 &214.2 $\pm$ 2.1 &-0.30 & - \\
233 & 415969908 & 3527 $\pm$ 30 & 4.88 $\pm$ 0.02 &-0.31 &0.371 $\pm$ 0.009 &0.365 $\pm$ 0.006 &11.72 $\pm$ 0.01 &522.8 $\pm$ 5.2 &-0.31 & -5.21 \\
698 & 141527579 & 3540 $\pm$ 30 & 4.76 $\pm$ 0.02 &0.04 &0.473 $\pm$ 0.011 &0.474 $\pm$ 0.008 &12.49 $\pm$ 0.01 &255.5 $\pm$ 2.5 &-0.39 & -5.12 \\
731 & 34068865 & 3543 $\pm$ 30 & 4.77 $\pm$ 0.02 &-0.03 &0.458 $\pm$ 0.011 &0.463 $\pm$ 0.008 &8.41 $\pm$ 0.01 &11008.1 $\pm$ 109.5 &-0.39 & -5.24 \\
1201 & 29960110 & 3546 $\pm$ 30 & 4.75 $\pm$ 0.02 &0.11 &0.484 $\pm$ 0.012 &0.488 $\pm$ 0.008 &11.31 $\pm$ 0.01 &757.8 $\pm$ 7.6 &-0.13 & -4.74 \\
756 & 73649615 & 3581 $\pm$ 30 & 4.71 $\pm$ 0.02 &0.11 &0.509 $\pm$ 0.012 &0.522 $\pm$ 0.009 &12.90 $\pm$ 0.01 &175.4 $\pm$ 1.7 &-0.42 & -4.90 \\
704 & 260004324 & 3596 $\pm$ 30 & 4.69 $\pm$ 0.02 &-0.05 &0.506 $\pm$ 0.012 &0.533 $\pm$ 0.009 &10.54 $\pm$ 0.01 &1540.8 $\pm$ 15.4 &-0.41 & -5.01 \\
177 & 262530407 & 3613 $\pm$ 30 & 4.70 $\pm$ 0.02 &- &0.519 $\pm$ 0.013 &0.532 $\pm$ 0.009 &9.91 $\pm$ 0.01 &2762.3 $\pm$ 27.5 &0.00 & -4.45 \\
797 & 271596225 & 3618 $\pm$ 30 & 4.76 $\pm$ 0.02 &- &0.475 $\pm$ 0.011 &0.476 $\pm$ 0.008 &12.13 $\pm$ 0.01 &356.5 $\pm$ 3.5 &-0.45 & -4.92 \\
727 & 149788158 & 3634 $\pm$ 30 & 4.74 $\pm$ 0.02 &-0.11 &0.499 $\pm$ 0.012 &0.501 $\pm$ 0.008 &11.42 $\pm$ 0.01 &686.3 $\pm$ 6.8 &-0.40 & -5.02 \\
142 & 425934411 & 3647 $\pm$ 30 & 4.55 $\pm$ 0.02 &- &0.594 $\pm$ 0.016 &0.675 $\pm$ 0.016 &15.09 $\pm$ 0.01 & 23.4 $\pm$ 0.2 &0.98 & -4.25 \\
663 & 54962195 & 3658 $\pm$ 30 & 4.72 $\pm$ 0.02 &-0.10 &0.514 $\pm$ 0.012 &0.521 $\pm$ 0.009 &12.18 $\pm$ 0.01 &341.4 $\pm$ 3.4 &-0.49 & - \\
234 & 12423815 & 3668 $\pm$ 30 & 4.70 $\pm$ 0.02 &0.13 &0.545 $\pm$ 0.015 &0.546 $\pm$ 0.013 &14.99 $\pm$ 0.01 & 25.7 $\pm$ 0.3 &-0.51 & -4.97 \\
620 & 296739893 & 3669 $\pm$ 30 & 4.70 $\pm$ 0.02 &0.19 &0.547 $\pm$ 0.014 &0.547 $\pm$ 0.009 &10.62 $\pm$ 0.01 &1437.2 $\pm$ 14.3 &-0.45 & -5.07 \\
672 & 151825527 & 3678 $\pm$ 30 & 4.67 $\pm$ 0.02 &0.01 &0.544 $\pm$ 0.013 &0.563 $\pm$ 0.009 &12.08 $\pm$ 0.01 &375.7 $\pm$ 3.7 &-0.34 & -4.62 \\
873 & 237920046 & 3682 $\pm$ 30 & 4.72 $\pm$ 0.02 &-0.09 &0.521 $\pm$ 0.013 &0.519 $\pm$ 0.009 &12.55 $\pm$ 0.01 &243.2 $\pm$ 2.4 &-0.39 & -4.81 \\
714 & 219195044 & 3698 $\pm$ 30 & 4.77 $\pm$ 0.02 &-0.33 &0.474 $\pm$ 0.011 &0.468 $\pm$ 0.008 &11.98 $\pm$ 0.01 &410.0 $\pm$ 4.1 &-0.47 & -5.22 \\
134 & 234994474 & 3722 $\pm$ 30 & 4.60 $\pm$ 0.02 &- &0.590 $\pm$ 0.015 &0.633 $\pm$ 0.010 &9.65 $\pm$ 0.01 &3512.5 $\pm$ 35.0 &-0.47 & -4.69 \\
468 & 33521996 & 3738 $\pm$ 30 & 4.60 $\pm$ 0.02 &- &0.588 $\pm$ 0.015 &0.634 $\pm$ 0.012 &13.75 $\pm$ 0.01 & 80.4 $\pm$ 0.8 &-0.55 & -4.93 \\
785 & 374829238 & 3740 $\pm$ 30 & 4.67 $\pm$ 0.02 &-0.02 &0.559 $\pm$ 0.014 &0.572 $\pm$ 0.009 &11.92 $\pm$ 0.01 &432.7 $\pm$ 4.3 &-0.39 & -4.52 \\
552 & 44737596 & 3742 $\pm$ 30 & 4.65 $\pm$ 0.02 &0.21 &0.579 $\pm$ 0.014 &0.594 $\pm$ 0.011 &14.19 $\pm$ 0.01 & 53.6 $\pm$ 0.5 &-0.42 & -4.19 \\
904 & 261257684 & 3752 $\pm$ 30 & 4.71 $\pm$ 0.02 &-0.17 &0.533 $\pm$ 0.013 &0.537 $\pm$ 0.009 &11.28 $\pm$ 0.01 &778.7 $\pm$ 7.8 &-0.39 & -4.60 \\
741 & 359271092 & 3763 $\pm$ 30 & 4.73 $\pm$ 0.02 &-0.17 &0.527 $\pm$ 0.013 &0.520 $\pm$ 0.008 &8.12 $\pm$ 0.01 &14329.1 $\pm$ 142.6 &-0.46 & -5.03 \\
198 & 12421862 & 3770 $\pm$ 30 & 4.86 $\pm$ 0.02 &- &0.447 $\pm$ 0.011 &0.413 $\pm$ 0.007 &10.39 $\pm$ 0.01 &1769.6 $\pm$ 17.7 &-0.42 & -5.14 \\
442 & 70899085 & 3831 $\pm$ 30 & 4.63 $\pm$ 0.02 &0.09 &0.598 $\pm$ 0.015 &0.620 $\pm$ 0.010 &11.17 $\pm$ 0.01 &867.7 $\pm$ 8.6 &-0.40 & -4.53 \\
557 & 55488511 & 3845 $\pm$ 30 & 4.67 $\pm$ 0.02 &-0.15 &0.570 $\pm$ 0.014 &0.581 $\pm$ 0.009 &12.09 $\pm$ 0.01 &371.3 $\pm$ 3.7 &-0.51 & -4.69 \\
870 & 219229644 & 3847 $\pm$ 30 & 4.63 $\pm$ 0.02 &0.11 &0.601 $\pm$ 0.015 &0.618 $\pm$ 0.010 &11.21 $\pm$ 0.01 &836.3 $\pm$ 8.3 &-0.38 & -4.54 \\
551 & 192826603 & 3871 $\pm$ 30 & 4.67 $\pm$ 0.02 &-0.28 &0.563 $\pm$ 0.014 &0.577 $\pm$ 0.010 &14.35 $\pm$ 0.01 & 46.4 $\pm$ 0.5 &-0.54 & -4.60 \\
876 & 32497972 & 3888 $\pm$ 30 & 4.63 $\pm$ 0.02 &-0.08 &0.595 $\pm$ 0.015 &0.615 $\pm$ 0.010 &11.98 $\pm$ 0.01 &410.9 $\pm$ 4.1 &-0.55 & -4.85 \\
532 & 144700903 & 3903 $\pm$ 30 & 4.62 $\pm$ 0.02 &0.07 &0.610 $\pm$ 0.015 &0.630 $\pm$ 0.011 &13.09 $\pm$ 0.01 &147.8 $\pm$ 1.5 &-0.68 & -5.39 \\
1075 & 351601843 & 3916 $\pm$ 30 & 4.69 $\pm$ 0.02 &- &0.575 $\pm$ 0.014 &0.571 $\pm$ 0.009 &11.59 $\pm$ 0.01 &588.2 $\pm$ 5.9 &-0.50 & -4.66 \\
833 & 362249359 & 3945 $\pm$ 30 & 4.65 $\pm$ 0.02 &-0.13 &0.595 $\pm$ 0.015 &0.601 $\pm$ 0.009 &10.60 $\pm$ 0.01 &1458.0 $\pm$ 14.5 &-0.49 & -4.49 \\
702 & 237914496 & 3966 $\pm$ 30 & 4.69 $\pm$ 0.02 &-0.24 &0.577 $\pm$ 0.014 &0.566 $\pm$ 0.009 &12.24 $\pm$ 0.01 &323.3 $\pm$ 3.2 &-0.56 & -4.65 \\
555 & 170849515 & 3993 $\pm$ 30 & 4.63 $\pm$ 0.02 &-0.06 &0.615 $\pm$ 0.017 &0.626 $\pm$ 0.015 &15.25 $\pm$ 0.01 & 20.1 $\pm$ 0.2 &-0.40 & -4.76 \\
285 & 220459976 & 3995 $\pm$ 30 & 4.62 $\pm$ 0.02 &0.00 &0.625 $\pm$ 0.015 &0.643 $\pm$ 0.010 &12.61 $\pm$ 0.01 &229.2 $\pm$ 2.3 &-0.60 & -4.71 \\
475 & 100608026 & 4003 $\pm$ 30 & 4.66 $\pm$ 0.02 &-0.18 &0.601 $\pm$ 0.015 &0.599 $\pm$ 0.009 &11.29 $\pm$ 0.01 &775.1 $\pm$ 7.7 &-0.54 & -4.62 \\
253 & 322063810 & 4039 $\pm$ 30 & 4.65 $\pm$ 0.02 &- &0.617 $\pm$ 0.015 &0.618 $\pm$ 0.009 &9.79 $\pm$ 0.01 &3087.3 $\pm$ 30.8 &-0.60 & -4.87 \\
435 & 44647437 & 4079 $\pm$ 30 & 4.64 $\pm$ 0.02 &-0.13 &0.621 $\pm$ 0.015 &0.627 $\pm$ 0.010 &13.34 $\pm$ 0.01 &116.9 $\pm$ 1.2 &-0.56 & -4.57 \\
\hline
\end{tabular}
\end{table*}

\begin{table*}
\centering
\contcaption{Final results for TESS candidate exoplanet hosts}
\begin{tabular}{ccccccccccc}
\hline
TOI & TIC & $T_{\rm eff}$ & $\log g$ & [Fe/H] & $M$ & $R_\star$ & $m_{\rm bol}$ & $f_{\rm bol}$ & EW(H$\alpha$) & $\log R^{'}_{\rm HK}$ \\
 &  & (K) &  &  & ($M_\odot$) & ($R_\odot$) &  & {{\footnotesize(10$^{-12}\,$ergs s$^{-1}$ cm $^{-2}$)}} & \SI{}{\angstrom} &  \\
\hline
1082 & 261108236 & 4096 $\pm$ 30 & 4.65 $\pm$ 0.02 &-0.24 &0.610 $\pm$ 0.015 &0.609 $\pm$ 0.009 &12.31 $\pm$ 0.01 &302.8 $\pm$ 3.0 &-0.61 & -4.76 \\
260 & 37749396 & 4097 $\pm$ 30 & 4.70 $\pm$ 0.02 &-0.28 &0.598 $\pm$ 0.015 &0.575 $\pm$ 0.009 &8.96 $\pm$ 0.01 &6614.9 $\pm$ 65.9 &-0.60 & -4.75 \\
761 & 165317334 & 4121 $\pm$ 30 & 4.64 $\pm$ 0.02 &-0.15 &0.623 $\pm$ 0.015 &0.629 $\pm$ 0.010 &11.34 $\pm$ 0.01 &740.9 $\pm$ 7.4 &-0.60 & -4.49 \\
249 & 179985715 & 4128 $\pm$ 30 & 4.71 $\pm$ 0.02 &- &0.589 $\pm$ 0.015 &0.561 $\pm$ 0.008 &11.70 $\pm$ 0.01 &531.3 $\pm$ 5.3 &-0.56 & -4.61 \\
806 & 33831980 & 4137 $\pm$ 30 & 4.68 $\pm$ 0.02 &-0.29 &0.607 $\pm$ 0.015 &0.592 $\pm$ 0.009 &12.42 $\pm$ 0.01 &274.5 $\pm$ 2.7 &-0.62 & -4.65 \\
1216 & 141527965 & 4217 $\pm$ 30 & 4.61 $\pm$ 0.02 &-0.13 &0.649 $\pm$ 0.016 &0.664 $\pm$ 0.010 &11.98 $\pm$ 0.01 &410.5 $\pm$ 4.1 &-0.67 & -4.68 \\
302 & 229111835 & 4227 $\pm$ 30 & 4.59 $\pm$ 0.02 &-0.03 &0.661 $\pm$ 0.016 &0.684 $\pm$ 0.011 &13.36 $\pm$ 0.01 &115.2 $\pm$ 1.2 &-0.73 & -4.60 \\
656 & 36734222 & 4241 $\pm$ 30 & 4.58 $\pm$ 0.02 &-0.07 &0.662 $\pm$ 0.016 &0.690 $\pm$ 0.010 &11.57 $\pm$ 0.01 &595.9 $\pm$ 6.0 &-0.61 & -4.46 \\
1130 & 254113311 & 4275 $\pm$ 30 & 4.55 $\pm$ 0.02 &-0.12 &0.670 $\pm$ 0.017 &0.716 $\pm$ 0.011 &10.60 $\pm$ 0.01 &1463.9 $\pm$ 14.6 &-0.73 & -4.78 \\
133 & 219338557 & 4276 $\pm$ 30 & 4.64 $\pm$ 0.02 &-0.22 &0.646 $\pm$ 0.016 &0.639 $\pm$ 0.009 &10.45 $\pm$ 0.01 &1673.5 $\pm$ 16.7 &-0.71 & -5.06 \\
544 & 50618703 & 4292 $\pm$ 30 & 4.66 $\pm$ 0.02 &-0.24 &0.641 $\pm$ 0.016 &0.623 $\pm$ 0.009 &10.13 $\pm$ 0.01 &2253.9 $\pm$ 22.4 &-0.64 & -4.56 \\
836 & 440887364 & 4308 $\pm$ 30 & 4.62 $\pm$ 0.02 &-0.15 &0.658 $\pm$ 0.016 &0.659 $\pm$ 0.009 &9.12 $\pm$ 0.01 &5710.9 $\pm$ 56.9 &-0.68 & -4.61 \\
240 & 101948569 & 4333 $\pm$ 30 & 4.59 $\pm$ 0.02 &- &0.676 $\pm$ 0.017 &0.693 $\pm$ 0.010 &11.16 $\pm$ 0.01 &876.1 $\pm$ 8.8 &-0.68 & -4.60 \\
713 & 167600516 & 4340 $\pm$ 30 & 4.64 $\pm$ 0.02 &-0.25 &0.648 $\pm$ 0.016 &0.637 $\pm$ 0.009 &11.17 $\pm$ 0.01 &868.0 $\pm$ 8.7 &-0.74 & -5.09 \\
900 & 210873792 & 4347 $\pm$ 30 & 4.63 $\pm$ 0.02 &-0.23 &0.653 $\pm$ 0.016 &0.649 $\pm$ 0.010 &12.32 $\pm$ 0.01 &300.0 $\pm$ 3.0 &-0.71 & -5.46 \\
493 & 19025965 & 4360 $\pm$ 30 & 4.59 $\pm$ 0.02 &-0.05 &0.677 $\pm$ 0.017 &0.690 $\pm$ 0.010 &11.92 $\pm$ 0.01 &434.7 $\pm$ 4.3 &-0.73 & -4.87 \\
178 & 251848941 & 4366 $\pm$ 30 & 4.64 $\pm$ 0.02 &- &0.649 $\pm$ 0.016 &0.639 $\pm$ 0.009 &10.91 $\pm$ 0.01 &1095.3 $\pm$ 10.9 &-0.75 & -4.85 \\
875 & 14165625 & 4404 $\pm$ 30 & 4.59 $\pm$ 0.02 &-0.13 &0.675 $\pm$ 0.017 &0.689 $\pm$ 0.010 &11.85 $\pm$ 0.01 &463.0 $\pm$ 4.6 &-0.69 & -4.57 \\
711 & 38510224 & 4433 $\pm$ 30 & 4.64 $\pm$ 0.02 &- &0.653 $\pm$ 0.016 &0.638 $\pm$ 0.009 &12.22 $\pm$ 0.01 &327.6 $\pm$ 3.3 &-0.73 & -4.77 \\
139 & 62483237 & 4434 $\pm$ 30 & 4.61 $\pm$ 0.02 &- &0.673 $\pm$ 0.016 &0.674 $\pm$ 0.009 &9.88 $\pm$ 0.01 &2827.2 $\pm$ 28.2 &-0.69 & -4.47 \\
1073 & 158297421 & 4481 $\pm$ 30 & 4.62 $\pm$ 0.02 &- &0.666 $\pm$ 0.017 &0.664 $\pm$ 0.014 &14.09 $\pm$ 0.01 & 58.6 $\pm$ 0.6 &-0.69 & -4.60 \\
929 & 175532955 & 4482 $\pm$ 30 & 4.59 $\pm$ 0.02 &- &0.684 $\pm$ 0.017 &0.693 $\pm$ 0.010 &11.92 $\pm$ 0.01 &432.7 $\pm$ 4.3 &-0.79 & -10.15 \\
969 & 280437559 & 4503 $\pm$ 30 & 4.59 $\pm$ 0.02 &- &0.682 $\pm$ 0.017 &0.693 $\pm$ 0.010 &11.06 $\pm$ 0.01 &960.4 $\pm$ 9.6 &-0.72 & -4.60 \\
932 & 260417932 & 4508 $\pm$ 30 & 4.58 $\pm$ 0.02 &- &0.689 $\pm$ 0.017 &0.705 $\pm$ 0.010 &11.23 $\pm$ 0.01 &817.5 $\pm$ 8.1 &-0.80 & -5.22 \\
1067 & 201642601 & 4510 $\pm$ 30 & 4.57 $\pm$ 0.02 &- &0.698 $\pm$ 0.018 &0.721 $\pm$ 0.012 &13.61 $\pm$ 0.01 & 91.7 $\pm$ 0.9 &-0.76 & - \\
279 & 122613513 & 4512 $\pm$ 30 & 4.61 $\pm$ 0.02 &- &0.677 $\pm$ 0.017 &0.673 $\pm$ 0.009 &11.03 $\pm$ 0.01 &986.9 $\pm$ 9.8 &-0.76 & -4.61 \\
129 & 201248411 & 4569 $\pm$ 30 & 4.57 $\pm$ 0.02 &- &0.697 $\pm$ 0.018 &0.721 $\pm$ 0.010 &10.42 $\pm$ 0.01 &1721.3 $\pm$ 17.2 &-0.77 & -4.50 \\
824 & 193641523 & 4589 $\pm$ 30 & 4.60 $\pm$ 0.02 &- &0.688 $\pm$ 0.018 &0.691 $\pm$ 0.009 &10.57 $\pm$ 0.01 &1505.7 $\pm$ 15.0 &-0.81 & -4.76 \\
\hline
\end{tabular}
\end{table*}

\section{Candidate Planet Parameters}\label{sec:planet_fitting}

\subsection{Transit Light Curve Analysis}
We now present results for all TOIs not ruled out as false positives (e.g. due to background stars, or eclipsing binaries) by the TESS Team and exoplanet community, as listed on the NASA ExoFOP-TESS website.

Transit light curves for targets across all TESS sectors were downloaded from NASA's Mikulski Archive for Space Telescopes (MAST) service. For all high-cadence data, we used the Pre-search Data Conditioning Simple Aperture Photometry (PDCSAP) fluxes, which have already had some measure of processing to remove systematics. 

All light curves were downloaded and manipulated using the python package \texttt{LightKurve} \citep{lightkurve_collaboration_lightkurve_2018}.

Many stars in our sample show some amount of stellar variability, with periods ranging from days to many weeks. We remove this using \texttt{LightKurve}'s flatten function, which applies a Savitzky-Golay filter \citep{savitzky_smoothing_1964} to the data to remove low frequency trends. When applying the filter we mask out all planetary transits by known TOIs. 
Once flattened, the light curves are then phase folded using either the period provided by NASA ExoFOP-TESS (for most stars), or our own fitted period (for stars revisited in the TESS extended mission whose long time baseline reveals the ExoFOP-TESS period to be incorrect). We use the provided measurement of transit duration to select only photometry from the transit itself, plus 10\% of a duration either side for use in model fitting.

Model fitting is implemented using the python package \texttt{BATMAN} \citep{kreidberg_batman_2015}, which is capable of generating model transit light curves for a given set of orbital elements (scaled by the stellar radius $R_\star$) and limb darkening coefficients. We use a four term limb darkening law, interpolating the PHOENIX grid provided by \citet{claret_limb_2017} using values of $T_{\rm eff}$ and $\log g$ from Table \ref{tab:final_results_tess}. The resulting coefficients are in Table \ref{tab:ld_coeff}.

Transit photometry alone is not sufficient to uniquely constrain the planet orbit and radius when fitting for the scaled semi-major axis $a_{R_\star}=\frac{a}{R_\star}$, the planetary radius ratio $R_{P,R_\star}=\frac{R_P}{R_\star}$, the inclination $i$, the eccentricity $e$, and the longitude of periastron $\omega$ \citep{kipping_transiting_2008}. While we can use our measurements of $M_\star$, $R_\star$, and $T$ to constrain the semi-major axis of a circular orbit (Equation \ref{eqn:sma}), we do not have the precision required to fit for eccentric orbits. As such, we fix $e=0$ and $\omega=0$ during our fit, and include our calculated value $a_{R_\star,c}$ - the value $a_{R_\star}$ assuming a circular orbit, as a prior during fitting. In cases where $e\sim0$, we expect the fitted semi-major axis $a_{R_\star,f}$ to approach $a_{R_\star,c}$. For cases with a discrepancy between the two, we flag the planet as an indication of a possibly eccentric orbit in Table \ref{tab:planet_params}.

This measured semi-major axis, calculated using our \citet{mann_how_2015} absolute $K_S$ band $M_\star$, and $T$ from NASA ExoFOP, can be constrained as follows:

\begin{equation}\label{eqn:sma}
    a = \sqrt[3]{\frac{GM_\star T^2}{4\pi^2}}
\end{equation}
where $a$ is the semi-major axis, $G$ is the gravitational constant, $M_\star$ is the stellar mass (with $M_\star >> M_p$, the planetary mass), and $T$ is the planet orbital period - all of which we assume are independent quantities.

Now with a prior on the semi-major axis, we again use the least\_squares function from \texttt{scipy}'s optimize module to perform least squares fitting to minimise the following expression: 
\begin{equation}
    R_t = \Bigg(\frac{{a_{R_\star, m} - a_{R_\star, f}}}{\sigma_{a_{R_\star, m}}} + \displaystyle\sum_{j}^{N}\frac{{t_{\rm obs,~j} - t_{\rm model,~j}}}{\sigma_{t_{\rm obs,~j}}}\Bigg)^2
\end{equation}
where $R_t$ are the light curve and prior residuals (as a function of $R_{p, R_\star}$, $a_{R_\star, f}$, and $i$), $a_{R_\star,m}$ the measured scaled semi-major axis, $a_{R_\star,f}$ the fitted scaled semi-major axis, $\sigma_{a_{R_\star, m}}$ the uncertainty on the measured scaled semi-major axis, $j$ is the time step, $N$ the total number of epochs, $t_{\rm obs,~j}$ is the observed flux at time step $j$, $t_{\rm model,~j}$ the model flux at time step $j$, and $\sigma_{t_{\rm obs,~j}}$ is the measured flux uncertainty at time step $j$.

Results from this fitting procedure are presented in Table \ref{tab:planet_params}, a comparison with confirmed planets in Figure \ref{fig:lit_planet_comp}, and a histogram of the resulting planet candidate radii in Figure \ref{fig:planet_radii_hist}. Note that we do not fit the light curves for some candidates: TOIs 256.01 and 415969908.02 have only two and one transits respectively; TOI 507.01 is a suspected equal mass binary; TOIs 302.01 and 969.01 do not have PDCSAP two minute cadence data; and TOIs 203.01, 253.01, 285.01, 696.02, 785.01, 864.01, 1216.01, 260417932.02, and 98796344.02 have transits observed only at low SNR.

\begin{table*}
\centering
\caption{Final results for TESS candidate exoplanets. }
\label{tab:planet_params}
\begin{tabular}{ccccccccc}
\hline
TOI & TIC & Sector/s & Period & $R_p/R_{\star}$ & $a/R_{\star}$ & $e$ flag & $i$ & $R_p$ \\
 &  &  & (days) &  &  &  & ($^{\circ}$) & ($R_{\oplus}$) \\
\hline
122.01 & 231702397 & 1,27-28 & 5.07803 $\dagger$ &0.0797 $\pm$ 0.0022 &24.63 $\pm$ 0.49 &0 & 88.337 $\pm$ 0.001 &3.00 $\pm$ 0.10\\
129.01 & 201248411 & 1-2,28-29 & 0.98097 $\dagger$ &0.3223 $\pm$ 0.0884 &5.15 $\pm$ 0.04 &0 & 76.381 $\pm$ 0.018 &25.35 $\pm$ 6.96\\
133.01 & 219338557 & 1,28 & 8.19918 $\dagger$ &0.0269 $\pm$ 0.0010 &23.15 $\pm$ 0.41 &0 & 88.470 $\pm$ 0.002 &1.88 $\pm$ 0.07\\
134.01 & 234994474 & 1,28 & 1.40153 $\dagger$ &0.0223 $\pm$ 0.0006 &6.98 $\pm$ 0.16 &0 & 84.566 $\pm$ 0.005 &1.54 $\pm$ 0.05\\
136.01 & 410153553 & 1,27-28 & 0.46293 $\dagger$ &0.0587 $\pm$ 0.0006 &6.80 $\pm$ 0.17 &1 & 90.000 $\pm$ 5.000 &1.23 $\pm$ 0.03\\
139.01 & 62483237 & 1,28 & 11.07083 $\dagger$ &0.0346 $\pm$ 0.0008 &27.20 $\pm$ 0.56 &0 & 88.549 $\pm$ 0.001 &2.54 $\pm$ 0.07\\
142.01 & 425934411 & 1-2,28-29 & 0.85335 $\dagger$ &0.1809 $\pm$ 0.0184 &4.74 $\pm$ 0.11 &0 & 79.385 $\pm$ 0.012 &13.31 $\pm$ 1.39\\
175.01 & 307210830 & 2,5,8-12,28-29,32 & 3.69066 $\dagger$ &0.0397 $\pm$ 0.0003 &21.32 $\pm$ 0.47 &0 & 88.809 $\pm$ 0.002 &1.32 $\pm$ 0.03\\
175.02 & 307210830 & 2,5,8-12,28-29,32 & 7.45075 $\dagger$ &0.0446 $\pm$ 0.0006 &34.59 $\pm$ 0.75 &0 & 88.483 $\pm$ 0.001 &1.48 $\pm$ 0.03\\
175.03 & 307210830 & 2,5,8-12,28-29,32 & 2.25310 $\dagger$ &0.0238 $\pm$ 0.0003 &15.76 $\pm$ 0.34 &0 & 88.133 $\pm$ 0.003 &0.79 $\pm$ 0.02\\
177.01 & 262530407 & 2-3,29 & 2.85310 $\dagger$ &0.0385 $\pm$ 0.0005 &12.70 $\pm$ 0.26 &0 & 86.765 $\pm$ 0.002 &2.24 $\pm$ 0.05\\
178.01 & 251848941 & 2,29 & 6.55770 $\dagger$ &0.0365 $\pm$ 0.0009 &19.97 $\pm$ 0.36 &0 & 88.506 $\pm$ 0.002 &2.54 $\pm$ 0.07\\
178.02 & 251848941 & 2,29 & 20.70950 $\dagger$ &0.0439 $\pm$ 0.0023 &42.94 $\pm$ 0.70 &0 & 88.821 $\pm$ 0.001 &3.06 $\pm$ 0.16\\
178.03 & 251848941 & 2,29 & 9.96188 $\dagger$ &0.0252 $\pm$ 0.0013 &26.40 $\pm$ 0.45 &0 & 88.855 $\pm$ 0.002 &1.76 $\pm$ 0.09\\
178 b & 251848941 & 2,29 & 1.91456 $\dagger$ &0.0197 $\pm$ 0.0008 &8.78 $\pm$ 0.15 &0 & 89.745 $\pm$ 0.094 &1.37 $\pm$ 0.06\\
178 c & 251848941 & 2,29 & 3.23845 $\dagger$ &0.0231 $\pm$ 0.0008 &12.47 $\pm$ 0.22 &0 & 88.423 $\pm$ 0.007 &1.61 $\pm$ 0.06\\
178 f & 251848941 & 2,29 & 15.23191 $\dagger$ &0.0314 $\pm$ 0.0011 &35.01 $\pm$ 0.59 &0 & 88.904 $\pm$ 0.001 &2.19 $\pm$ 0.08\\
198.01 & 12421862 & 2,29 & 20.43021 $\dagger$ &0.0291 $\pm$ 0.0012 &58.19 $\pm$ 1.14 &0 & 89.374 $\pm$ 0.001 &1.31 $\pm$ 0.06\\
210.01 & 141608198 & 1-5,7-13,27-32 & 9.01056 $\dagger$ &0.0629 $\pm$ 0.0007 &37.78 $\pm$ 0.66 &0 & 89.531 $\pm$ 0.002 &2.24 $\pm$ 0.05\\
233.01 & 415969908 & 2,29 & 11.66993 $\dagger$ &0.0457 $\pm$ 0.0013 &42.65 $\pm$ 0.82 &0 & 89.606 $\pm$ 0.002 &1.82 $\pm$ 0.06\\
234.01 & 12423815 & 2,29 & 2.83927 $\dagger$ &0.1932 $\pm$ 0.0045 &12.63 $\pm$ 0.28 &0 & 86.641 $\pm$ 0.003 &11.50 $\pm$ 0.39\\
240.01 & 101948569 & 2,29 & 19.47241 $\dagger$ &0.0388 $\pm$ 0.0013 &38.59 $\pm$ 0.59 &0 & 89.278 $\pm$ 0.001 &2.93 $\pm$ 0.10\\
244.01 & 118327550 & 2,29 & 7.39719 $\dagger$ &0.0321 $\pm$ 0.0012 &29.05 $\pm$ 0.56 &0 & 88.382 $\pm$ 0.001 &1.42 $\pm$ 0.06\\
249.01 & 179985715 & 2,29 & 6.61542 $\dagger$ &0.0329 $\pm$ 0.0018 &22.18 $\pm$ 0.38 &0 & 88.757 $\pm$ 0.003 &2.01 $\pm$ 0.11\\
256.02 & 92226327 & 3,30 & 3.77796 $\dagger$ &0.0480 $\pm$ 0.0010 &25.61 $\pm$ 0.59 &0 & 90.000 $\pm$ 5.000 &1.15 $\pm$ 0.03\\
260.01 & 37749396 & 3 & 13.470018 &0.0265 $\pm$ 0.0010 &34.87 $\pm$ 0.60 &0 & 88.758 $\pm$ 0.001 &1.66 $\pm$ 0.07\\
269.01 & 220479565 & 3-6,10,13,30-32 & 3.69770 $\dagger$ &0.0691 $\pm$ 0.0012 &18.65 $\pm$ 0.31 &0 & 87.384 $\pm$ 0.001 &2.99 $\pm$ 0.07\\
270.01 & 259377017 & 3-5,30,32 & 5.66054 $\dagger$ &0.0581 $\pm$ 0.0004 &26.06 $\pm$ 0.55 &0 & 89.210 $\pm$ 0.002 &2.29 $\pm$ 0.04\\
270.02 & 259377017 & 3-5,30,32 & 11.37960 $\dagger$ &0.0535 $\pm$ 0.0004 &41.87 $\pm$ 0.78 &0 & 89.707 $\pm$ 0.002 &2.11 $\pm$ 0.04\\
270.03 & 259377017 & 3-5,30,32 & 3.36014 $\dagger$ &0.0301 $\pm$ 0.0005 &18.65 $\pm$ 0.37 &0 & 89.218 $\pm$ 0.005 &1.18 $\pm$ 0.03\\
279.01 & 122613513 & 3-4 & 11.494122 &0.0361 $\pm$ 0.0013 &27.95 $\pm$ 0.45 &0 & 88.583 $\pm$ 0.001 &2.65 $\pm$ 0.10\\
406.01 & 153065527 & 3-4,30-31 & 13.17573 $\dagger$ &0.0430 $\pm$ 0.0013 &43.35 $\pm$ 0.87 &0 & 89.303 $\pm$ 0.001 &1.84 $\pm$ 0.06\\
435.01 & 44647437 & 4-5,31 & 3.35293 $\dagger$ &0.0583 $\pm$ 0.0016 &12.82 $\pm$ 0.23 &0 & 88.622 $\pm$ 0.007 &3.99 $\pm$ 0.12\\
442.01 & 70899085 & 5,32 & 4.05203 $\dagger$ &0.0741 $\pm$ 0.0007 &14.48 $\pm$ 0.25 &0 & 86.865 $\pm$ 0.002 &5.02 $\pm$ 0.09\\
455.01 & 98796344 & 4,31 & 5.35880 $\dagger$ &0.0467 $\pm$ 0.0009 &29.82 $\pm$ 1.22 &0 & 89.391 $\pm$ 0.005 &1.38 $\pm$ 0.04\\
468.01 & 33521996 & 6,32 & 3.32527 $\dagger$ &0.1735 $\pm$ 0.0014 &12.63 $\pm$ 0.25 &0 & 87.382 $\pm$ 0.003 &12.01 $\pm$ 0.24\\
475.01 & 100608026 & 5-6,32 & 8.26159 $\dagger$ &0.0307 $\pm$ 0.0014 &24.22 $\pm$ 0.42 &0 & 89.006 $\pm$ 0.003 &2.00 $\pm$ 0.10\\
486.01 & 260708537 & 1-6,8-13,27-32 & 1.74468 $\dagger$ &0.0128 $\pm$ 0.0003 &10.82 $\pm$ 0.21 &0 & 88.585 $\pm$ 0.008 &0.59 $\pm$ 0.02\\
493.01 & 19025965 & 7 & 5.947773 &0.0518 $\pm$ 0.0020 &17.60 $\pm$ 0.25 &0 & 87.987 $\pm$ 0.003 &3.90 $\pm$ 0.16\\
521.01 & 27649847 & 7 & 1.542131 &0.0463 $\pm$ 0.0028 &9.89 $\pm$ 0.17 &0 & 86.478 $\pm$ 0.006 &2.15 $\pm$ 0.14\\
532.01 & 144700903 & 6 & 2.326811 &0.0876 $\pm$ 0.0020 &9.96 $\pm$ 0.14 &0 & 87.050 $\pm$ 0.004 &6.02 $\pm$ 0.17\\
540.01 & 200322593 & 4-6,31-32 & 1.23914 $\dagger$ &0.0366 $\pm$ 0.0010 &13.48 $\pm$ 0.32 &0 & 87.063 $\pm$ 0.003 &0.78 $\pm$ 0.03\\
544.01 & 50618703 & 6,32 & 1.54835 $\dagger$ &0.0281 $\pm$ 0.0006 &7.80 $\pm$ 0.14 &0 & 85.103 $\pm$ 0.004 &1.91 $\pm$ 0.05\\
551.01 & 192826603 & 5-6,32 & 2.64730 $\dagger$ &0.3387 $\pm$ 1.1540 &11.60 $\pm$ 0.17 &0 & 84.411 $\pm$ 0.118 &21.32 $\pm$ 72.62\\
552.01 & 44737596 & 4-5,31 & 2.78864 $\dagger$ &0.1552 $\pm$ 0.0014 &11.57 $\pm$ 0.19 &0 & 87.675 $\pm$ 0.003 &10.05 $\pm$ 0.20\\
555.01 & 170849515 & 5,31-32 & 1.94163 $\dagger$ &0.1548 $\pm$ 0.0028 &8.92 $\pm$ 0.20 &0 & 87.380 $\pm$ 0.008 &10.57 $\pm$ 0.32\\
557.01 & 55488511 & 5,31 & 3.34499 $\dagger$ &0.0388 $\pm$ 0.0025 &13.43 $\pm$ 0.21 &0 & 86.264 $\pm$ 0.002 &2.46 $\pm$ 0.16\\
562.01 & 413248763 & 8 & 3.930792 &0.0329 $\pm$ 0.0007 &20.86 $\pm$ 0.41 &0 & 88.691 $\pm$ 0.002 &1.27 $\pm$ 0.03\\
620.01 & 296739893 & 8 & 5.098373 &0.0597 $\pm$ 0.0015 &18.65 $\pm$ 0.35 &0 & 87.394 $\pm$ 0.001 &3.56 $\pm$ 0.11\\
654.01 & 35009898 & 9 & 1.527419 &0.0513 $\pm$ 0.0019 &9.44 $\pm$ 0.16 &0 & 87.873 $\pm$ 0.010 &2.49 $\pm$ 0.10\\
656.01 & 36734222 & 9 & 0.813470 &0.1608 $\pm$ 0.0005 &4.67 $\pm$ 0.03 &0 & 81.396 $\pm$ 0.002 &12.11 $\pm$ 0.19\\
663.01 & 54962195 & 9 & 2.598654 &0.0408 $\pm$ 0.0017 &12.24 $\pm$ 0.17 &0 & 88.685 $\pm$ 0.010 &2.32 $\pm$ 0.11\\
663.02 & 54962195 & 9 & 4.698465 &0.0433 $\pm$ 0.0023 &18.16 $\pm$ 0.25 &0 & 88.488 $\pm$ 0.004 &2.46 $\pm$ 0.14\\
672.01 & 151825527 & 9-10 & 3.633618 &0.0889 $\pm$ 0.0008 &14.44 $\pm$ 0.20 &0 & 87.932 $\pm$ 0.002 &5.45 $\pm$ 0.10\\
674.01 & 158588995 & 9-10 & 1.977238 &0.1174 $\pm$ 0.0009 &11.35 $\pm$ 0.17 &0 & 86.352 $\pm$ 0.002 &5.67 $\pm$ 0.11\\
696.01 & 77156829 & 4-5,31-32 & 0.86024 $\dagger$ &0.0221 $\pm$ 0.0008 &7.96 $\pm$ 0.16 &0 & 84.721 $\pm$ 0.004 &0.79 $\pm$ 0.03\\
698.01 & 141527579 & 1-5,7-13,27-32 & 15.08666 $\dagger$ &0.0436 $\pm$ 0.0010 &42.17 $\pm$ 0.68 &0 & 89.058 $\pm$ 0.001 &2.26 $\pm$ 0.07\\
700.01 & 150428135 & 1,3-11,13,27-28,30-31 & 16.05110 $\dagger$ &0.0573 $\pm$ 0.0010 &47.21 $\pm$ 0.89 &0 & 88.902 $\pm$ 0.000 &2.66 $\pm$ 0.07\\
700.02 & 150428135 & 1,3-11,13,27-28,30-31 & 37.42475 $\dagger$ &0.0272 $\pm$ 0.0008 &82.17 $\pm$ 1.66 &0 & 90.000 $\pm$ 5.307 &1.26 $\pm$ 0.04\\
700.03 & 150428135 & 1,3-11,13,27-28,30-31 & 9.97701 $\dagger$ &0.0181 $\pm$ 0.0009 &34.16 $\pm$ 0.66 &0 & 89.885 $\pm$ 0.014 &0.84 $\pm$ 0.04\\
\hline
\end{tabular}
\begin{minipage}{\linewidth}
\vspace{0.1cm}
\textbf{Notes:} Periods denoted by $\dagger$ are not as reported by ExoFOP, and have been refitted here. These are overwhelmingly systems with TESS extended mission data, thus having longer time baselines with which to constrain orbital periods. Our fitted periods however are generally consistent within uncertainties of their ExoFOP values, and as such we do not report new uncertainties here. Additionally, our least squares fits to 7 of our light curves proved insenstive to non-edge-on inclinations. As such, we report conservative uncertanties of $\pm5$\degree~ for these planets.\\
\end{minipage}
\end{table*}

\begin{table*}
\centering
\contcaption{Final results for TESS candidate exoplanets}
\begin{tabular}{ccccccccc}
\hline
TOI & TIC & Sector/s & Period & $R_p/R_{\star}$ & $a/R_{\star}$ & $e$ flag & $i$ & $R_p$ \\
 &  &  & (days) &  &  &  & ($^{\circ}$) & ($R_{\oplus}$) \\
\hline
702.01 & 237914496 & 1-4,7,11,27,29-31 & 3.56809 $\dagger$ &0.0280 $\pm$ 0.0010 &14.46 $\pm$ 0.24 &0 & 87.421 $\pm$ 0.002 &1.73 $\pm$ 0.07\\
704.01 & 260004324 & 1-5,7-13,27-32 & 3.81431 $\dagger$ &0.0208 $\pm$ 0.0004 &15.38 $\pm$ 0.28 &0 & 87.419 $\pm$ 0.002 &1.21 $\pm$ 0.03\\
711.01 & 38510224 & 1-5,7-8,11-12,28-32 & 18.38384 $\dagger$ &0.0301 $\pm$ 0.0012 &39.90 $\pm$ 0.64 &0 & 89.221 $\pm$ 0.001 &2.09 $\pm$ 0.09\\
713.01 & 167600516 & 1-8,10-13,27-32 & 35.99988 $\dagger$ &0.0315 $\pm$ 0.0006 &62.37 $\pm$ 0.95 &0 & 89.724 $\pm$ 0.001 &2.19 $\pm$ 0.05\\
713.02 & 167600516 & 1-8,10-13,27-32 & 1.87150 $\dagger$ &0.0155 $\pm$ 0.0007 &8.69 $\pm$ 0.13 &0 & 84.872 $\pm$ 0.003 &1.07 $\pm$ 0.05\\
714.01 & 219195044 & 4-8,11-12,28,31-32 & 4.32378 $\dagger$ &0.0260 $\pm$ 0.0010 &18.59 $\pm$ 0.31 &0 & 88.556 $\pm$ 0.003 &1.33 $\pm$ 0.06\\
714.02 & 219195044 & 4-8,11-12,28,31-32 & 10.17742 $\dagger$ &0.0307 $\pm$ 0.0011 &32.90 $\pm$ 0.51 &0 & 89.108 $\pm$ 0.001 &1.57 $\pm$ 0.06\\
727.01 & 149788158 & 8 & 4.726090 &0.0293 $\pm$ 0.0020 &18.76 $\pm$ 0.31 &0 & 89.435 $\pm$ 0.014 &1.60 $\pm$ 0.11\\
731.01 & 34068865 & 9 & 0.321941 &0.0133 $\pm$ 0.0005 &3.29 $\pm$ 0.07 &0 & 85.081 $\pm$ 0.031 &0.67 $\pm$ 0.03\\
732.01 & 36724087 & 9 & 0.768418 &0.0311 $\pm$ 0.0011 &6.59 $\pm$ 0.13 &0 & 90.000 $\pm$ 5.000 &1.30 $\pm$ 0.05\\
732.02 & 36724087 & 9 & 12.254218 &0.0607 $\pm$ 0.0022 &41.84 $\pm$ 0.74 &0 & 88.868 $\pm$ 0.001 &2.53 $\pm$ 0.10\\
741.01 & 359271092 & 9-10 & 7.576262 &0.0160 $\pm$ 0.0009 &25.25 $\pm$ 0.62 &0 & 88.472 $\pm$ 0.002 &0.91 $\pm$ 0.05\\
756.01 & 73649615 & 10-11 & 1.23952 $\dagger$ &0.0548 $\pm$ 0.0019 &7.43 $\pm$ 0.10 &0 & 85.039 $\pm$ 0.005 &3.12 $\pm$ 0.12\\
761.01 & 165317334 & 10 & 10.563348 &0.0417 $\pm$ 0.0016 &27.53 $\pm$ 0.41 &0 & 89.196 $\pm$ 0.003 &2.86 $\pm$ 0.12\\
782.01 & 429358906 & 10 & 16.047203 &0.0659 $\pm$ 0.0040 &47.81 $\pm$ 0.69 &0 & 89.070 $\pm$ 0.001 &2.97 $\pm$ 0.19\\
789.01 & 300710077 & 1-3,5-13,27-32 & 5.44693 $\dagger$ &0.0274 $\pm$ 0.0011 &24.99 $\pm$ 0.44 &0 & 89.127 $\pm$ 0.003 &1.11 $\pm$ 0.05\\
797.01 & 271596225 & 1-13,27-32 & 1.80078 $\dagger$ &0.0256 $\pm$ 0.0006 &10.21 $\pm$ 0.17 &0 & 86.580 $\pm$ 0.003 &1.33 $\pm$ 0.04\\
797.02 & 271596225 & 1-13,27-32 & 4.14002 $\dagger$ &0.0292 $\pm$ 0.0011 &17.78 $\pm$ 0.30 &0 & 87.236 $\pm$ 0.001 &1.52 $\pm$ 0.06\\
806.01 & 33831980 & 1-3,5-6,8-9,12-13,27-29,32 & 21.91625 $\dagger$ &0.0347 $\pm$ 0.0011 &47.14 $\pm$ 0.66 &0 & 89.479 $\pm$ 0.001 &2.24 $\pm$ 0.08\\
824.01 & 193641523 & 11-12 & 1.392930 &0.0436 $\pm$ 0.0007 &6.69 $\pm$ 0.10 &0 & 83.665 $\pm$ 0.004 &3.29 $\pm$ 0.07\\
833.01 & 362249359 & 9-11 & 1.042241 &0.0190 $\pm$ 0.0007 &6.02 $\pm$ 0.10 &0 & 89.977 $\pm$ 1.790 &1.24 $\pm$ 0.05\\
836.01 & 440887364 & 11 & 8.593935 &0.0346 $\pm$ 0.0006 &23.32 $\pm$ 0.30 &0 & 88.727 $\pm$ 0.001 &2.49 $\pm$ 0.06\\
836.02 & 440887364 & 11 & 3.817115 &0.0240 $\pm$ 0.0008 &13.59 $\pm$ 0.23 &0 & 87.727 $\pm$ 0.003 &1.72 $\pm$ 0.06\\
870.01 & 219229644 & 3-5,30-32 & 22.03813 $\dagger$ &0.0330 $\pm$ 0.0012 &45.25 $\pm$ 0.81 &0 & 89.013 $\pm$ 0.001 &2.22 $\pm$ 0.09\\
873.01 & 237920046 & 1-4,11,28-31 & 5.93122 $\dagger$ &0.0279 $\pm$ 0.0012 &21.26 $\pm$ 0.38 &0 & 90.000 $\pm$ 5.000 &1.58 $\pm$ 0.07\\
875.01 & 14165625 & 5-6 & 11.020153 &0.0277 $\pm$ 0.0021 &26.47 $\pm$ 0.36 &0 & 90.000 $\pm$ 5.000 &2.08 $\pm$ 0.16\\
876.01 & 32497972 & 5-6,32 & 38.69629 $\dagger$ &0.0367 $\pm$ 0.0044 &65.85 $\pm$ 1.39 &0 & 89.254 $\pm$ 0.001 &2.46 $\pm$ 0.30\\
900.01 & 210873792 & 12 & 4.844050 &0.0426 $\pm$ 0.0026 &15.81 $\pm$ 0.26 &0 & 90.000 $\pm$ 5.000 &3.01 $\pm$ 0.19\\
904.01 & 261257684 & 12-13 & 18.35654 $\dagger$ &0.0342 $\pm$ 0.0017 &43.55 $\pm$ 0.79 &0 & 90.000 $\pm$ 5.000 &2.00 $\pm$ 0.10\\
910.01 & 369327947 & 12-13,27 & 2.02911 $\dagger$ &0.0311 $\pm$ 0.0007 &15.53 $\pm$ 0.31 &0 & 87.229 $\pm$ 0.002 &0.94 $\pm$ 0.03\\
912.01 & 406941612 & 12-13 & 4.679100 &0.0414 $\pm$ 0.0008 &20.71 $\pm$ 0.32 &0 & 88.819 $\pm$ 0.002 &1.93 $\pm$ 0.05\\
929.01 & 175532955 & 30-31 & 5.83010 $\dagger$ &0.0314 $\pm$ 0.0015 &17.32 $\pm$ 0.28 &0 & 89.396 $\pm$ 0.012 &2.37 $\pm$ 0.12\\
932.01 & 260417932 & 28-29,31-32 & 19.310700 &0.0337 $\pm$ 0.0009 &38.00 $\pm$ 0.62 &0 & 89.463 $\pm$ 0.002 &2.59 $\pm$ 0.08\\
1067.01 & 201642601 & 13,27 & 3.13167 $\dagger$ &0.1071 $\pm$ 0.0012 &11.13 $\pm$ 0.20 &0 & 89.174 $\pm$ 0.012 &8.42 $\pm$ 0.17\\
1073.01 & 158297421 & 13,27 & 3.92282 $\dagger$ &0.1799 $\pm$ 0.0039 &13.88 $\pm$ 0.27 &0 & 86.903 $\pm$ 0.002 &13.04 $\pm$ 0.39\\
1075.01 & 351601843 & 13,27 & 0.60474 $\dagger$ &0.0286 $\pm$ 0.0006 &4.39 $\pm$ 0.08 &0 & 85.023 $\pm$ 0.014 &1.78 $\pm$ 0.05\\
1078.01 & 370133522 & 13,27 & 0.51824 $\dagger$ &0.0274 $\pm$ 0.0004 &5.04 $\pm$ 0.10 &0 & 85.016 $\pm$ 0.010 &1.17 $\pm$ 0.03\\
1082.01 & 261108236 & 12-13,27-28,31 & 16.34646 $\dagger$ &0.0399 $\pm$ 0.0012 &37.72 $\pm$ 0.63 &0 & 89.448 $\pm$ 0.002 &2.65 $\pm$ 0.09\\
1201.01 & 29960110 & 4,31 & 2.49197 $\dagger$ &0.0389 $\pm$ 0.0009 &12.46 $\pm$ 0.24 &0 & 87.967 $\pm$ 0.004 &2.07 $\pm$ 0.06\\
153065527.02 & 153065527 & 3-4,30-31 & 3.30745 $\dagger$ &0.0270 $\pm$ 0.0013 &17.26 $\pm$ 0.33 &0 & 87.940 $\pm$ 0.003 &1.16 $\pm$ 0.06\\
\hline
\end{tabular}
\end{table*}

\section{Discussion}\label{sec:discussion}
\subsection{Radial Velocities}
Just over half our TESS sample have radial velocities in Gaia DR2, with the remaining 42 therefore having an incomplete set of positional and kinematic data. Our RVs are consistent with Gaia DR2 for our overlap sample and accurate to within ${\sim}4.5\,$km$\,$s$^{-1}$ (Section \ref{sec:rv_fitting}), thus providing RVs for the remainder and enabling insight into Galactic population, or kinematic analysis using tools such as \texttt{Chronostar} \citep{crundall_chronostar_2019} to determine ages for those that are found to be members of stellar associations. These results are especially interesting given the planet-hosting nature of these stars. 
\subsection{Standard Star Parameter Recovery}\label{sec:discussion:param_recovery}
Comparing our $T_{\rm eff}$ results to those of \citet{mann_how_2015} reveals excellent agreement for our two parameter fit (Figure \ref{fig:standard_teff_comp}), with the scatter on our residuals being smaller than their mean reported uncertainty of 60$\,$K and only a relatively small systematic of $\sim$$30\,$K observed. Such consistency is encouraging given that this represents our largest uniform set of comparison stars, a set whose temperatures have already been successfully benchmarked against those from interferometry and should be much less sensitive to model limitations than our own. 

When comparing to \citet{rojas-ayala_metallicity_2012}, the results are less consistent, though we observe a similar effect to \citet{mann_how_2015} in that \citet{rojas-ayala_metallicity_2012} overestimates temperatures for the warmest stars. These temperatures, however, come solely from measurement of the H$_2$O-K2 index in the $K$ band in conjunction with BT-Settl model atmospheres - much more limited in wavelength coverage than \citet{mann_how_2015} or our work here.

The interferometric sample shows good agreement, though we observe a $\sim$$70\,$K temperature systematic of the same sign as for the \citet{mann_how_2015} sample. However, due to the bias of interferometry towards close and thus bright targets, these are also the brightest stars we observe and they have correspondingly high photometric uncertainties due to saturation. This is particularly acute in the 2MASS bands, where less than half the sample have the photometric quality flag (Qflg) of `AAA', in contrast to the rest of the standard sample where all but two of 117 stars has Qflg `AAA', and the entirety of the TESS sample. Nonetheless, our derived radii for the interferometric standards (Figure \ref{fig:radius_comp_interferometric}) are consistent when allowing for additional scatter from poor quality photometry on bright stars that will not be present for our science targets. Encouragingly however \citet{mann_how_2015}, which we are in agreement with, integrated their own photometry from low resolution flux calibrated spectra and found a good match between their results and their own interferometric sample. 

Finally, our results are consistent with our sample of mid-K-dwarfs in the temperature range of our warmest science targets. The observed higher scatter (than e.g. the \citealt{mann_how_2015} sample) is to be expected due to inter-study systematics, as these targets were not pulled from a single uniform catalogue. 

While the exact cause of the \citet{mann_how_2015} and interferometric systematic is unclear, its appearance in both samples suggests it is not an artefact. As such, we apply a -30$\,$K correction to the observed temperature systematic. Although our remaining scatter is consistent with the scatter in our external reference catalogues, we add a further $\pm30\,$K $T_{\rm eff}$ uncertainty in quadrature with our statistical uncertainties to account for the unknown origin of the observed systematic. Given these corrections, we are confident our fitting methodology is able to recover both accurate and precise stellar temperatures and radii for stars in the range $3000\,$K$\lesssim T_{\rm eff} \lesssim 4500\,$K - critical for insight into the radii of their transiting planets.

\subsection{Model Limitations}
The inability of cool dwarf atmospheric models to reproduce optical fluxes is significant. Such wavelengths are among the most easily accessible, and understanding them is required to take full advantage of photometry from surveys like Gaia and SkyMapper. Thus anyone relying directly (e.g. spectral fitting) or indirectly (e.g. isochrone fitting with colours) on models for cool stellar atmospheres must do so with caution (for specifics on isochone systematics, see e.g. \citealt{vandenberg_victoria-regina_2006} for the Victoria-Regina models, \citealt{dotter_dartmouth_2008} and \citealt{joyce_not_2018} for DSEP, and \citealt{dotter_mesa_2016} for MIST).

We identify two key areas for improvement with our models and methods as implemented. The first relates to TiO, the dominant opacity source at optical wavelengths. Comparing high-resolution spectra of M-dwarfs to PHOENIX models and TiO templates, \citet{hoeijmakers_search_2015} concluded that `the modelled spectrum of TiO is not representative of the real TiO'. \citet{mckemmish_exomol_2019} confirmed this discrepancy in the process of validating their updated TiO line list, with their comparisons showing significant improvements in both predicted TiO wavelengths and line depths across the optical when using the updated line list. \citet{mckemmish_exomol_2019} was not yet published at the time our MARCS models were generated, and although they note that there remains room for further work, this represents a significant improvement on the previous state of the art. While recomputing our library of synthetic spectra with the new line list would constitute a significant computational effort, we will endeavour to do this in future work.

The second issue concerns proper consideration of the relative abundances of C and O---constituents of the dominant molecular opacity sources in cool dwarf atmospheres---denoted here as [(O-C)/Fe]. As described by \citet{veyette_physical_2016}, it is not just $T_{\rm eff}$ and [Fe/H] that affects the location of the pseudo-continuum, but also [(O-C)/Fe]. The principal reason for this is that [(O-C)/Fe] influences the concentrations of C- and O-based molecules, affecting the flux of the pseudo-continuum and apparent strength of metal lines. They conclude that ultimately the inferred value of [Fe/H] depends on [(O-C)/Fe], and that much better spectral fits are possible when allowing [(O-C)/Fe] to vary. An important note is that empirical calibrations based on FGK-M binaries such as \citet{rojas-ayala_metallicity_2012}, \citet{mann_prospecting_2013}, and our own photometric [Fe/H] relation should remain valid as statistical [Fe/H] indicators due to the tight Solar Neighbourhood [Fe/H]-[C/O] correlation. Per the recommendation of \citet{veyette_physical_2016}, this issue is significant enough to merit new models with [C/Fe] and [O/Fe] as independent parameters.

That said, there is an ever increasing empirical knowledge of M-dwarfs meaning that, even in the absence of accurate models, empirical or data driven approaches should be possible, especially if methods to break the [Fe/H]-[(O-C)/Fe] degeneracy can be found. For instance, see \citet{birky_temperatures_2020} which demonstrates that a data driven approach, at least in the $H$ band, is possible for M-dwarfs. The very small rate of evolution for these low-mass stars means we can rely on mass and chemical composition to derive the fundamental parameters of the star, thus making for a more tractable problem.

\subsection{TESS Input Catalogue Stellar Parameters}\label{sec:tic_params}
\begin{figure}
 \includegraphics[width=\columnwidth]{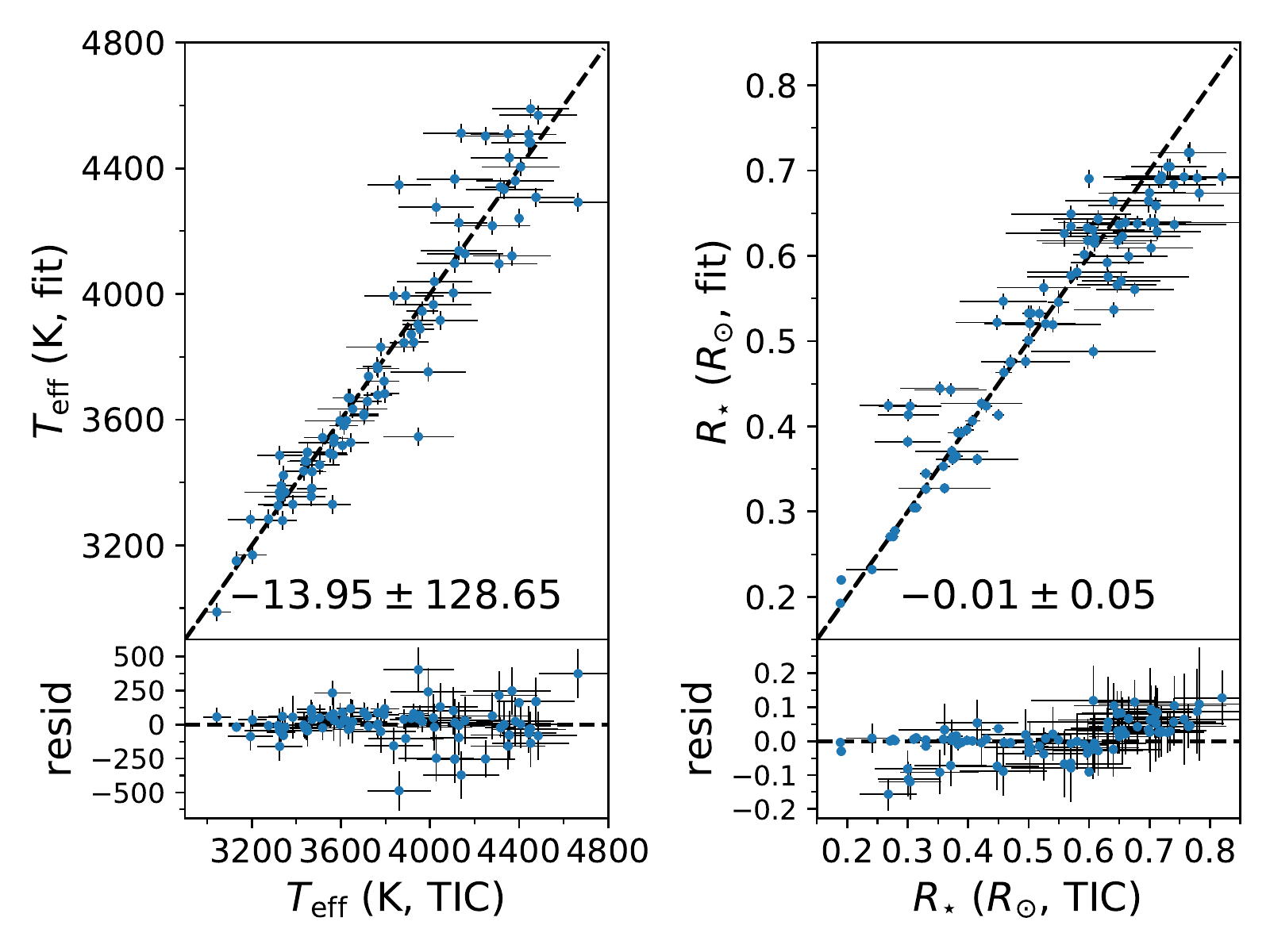}
 \caption{Comparison of $T_{\rm eff}$ and $R_{\star}$ as reported here compared to those from the TESS Input Catalogue. The median and standard deviation of each set of residuals is annotated.}
 \label{fig:tic_comp}
\end{figure}

The TESS Input Catalogue is often the first stellar parameter reference for newly alerted TOIs. As these parameters are mostly derived from empirical relations using literature photometry, we thought it useful to compare these predictions with our fits to inspect for remaining catalogue systematics. Figure \ref{fig:tic_comp} displays this comparison for $T_{\rm eff}$ and $R_{\star}$, and while the TIC temperatures are broadly consistent, TIC radii for the warmest stars in our sample appear systematically large. This stellar radii systematic is noteworthy as it would bias any predicted exoplanet radii around mid-K dwarfs.

\subsection{Emission Features in TESS Candidates}\label{sec:emission}
While model limitations prevented us from taking full advantage of our spectra during fitting, our wide wavelength coverage allows us to look for spectral peculiarities. In the current study, these take the form of emission in the Hydrogen Balmer Series or Ca II H\&K (both signs of stellar activity and youth), as well as absorption in the Li $6,708\,$\SI{}{\angstrom} (another sign of youth). While none of our TESS planet hosts show detectable Lithium absorption, we report H$\alpha$ equivalent widths and $\log R^{\prime}_{\rm HK}$ in Table \ref{tab:final_results_tess}, calculated using the methodology of \citet{zerjal_spectroscopically_2021}. 53 stars in our sample have EWH$\alpha > -0.5\,$\SI{}{\angstrom} (adopted as the limiting bound for activity, noting as well that this is strongly dependent on $T_{\rm eff}$ and thus somewhat approximate), and 35 have $\log R^{\prime}_{\rm HK} > -4.75$ (the lower bound for active stars used in \citealt{gray_contributions_2006}). 

Of particular note are our two most active stars, the first of which is TOI 507 (TIC 348538431). TOI 507 appears substantially overluminous in Figure \ref{fig:hr_diagram}, and presents with strong emission across the Balmer Series and in Ca II H\&K. Visual inspection of its spectrum, along with comparison to the cool dwarf standard HIP 103039 which is very similar in $T_{\rm eff}$, indicates that it is actually a double-lined spectroscopic binary. Transit depths appear similar for both primary and secondary eclipses, which points to the system being composed of roughly equal mass components. Taking a $\sim$0.75$\,$mag offset into account due to binarity, TOI 507 still sits slightly above the main sequence, meaning that it remains a potentially young touchstone system amenable to characterisation as in e.g. \citet{murphy_thor_2020}. The mass, radius, $m_{\rm bol}$, and flux reported in Table \ref{tab:final_results_tess} have been derived for a single component of this binary system, assuming equal mass and brightness.

\begin{figure*}
    \centering
    \includegraphics[width=\textwidth,page=1]{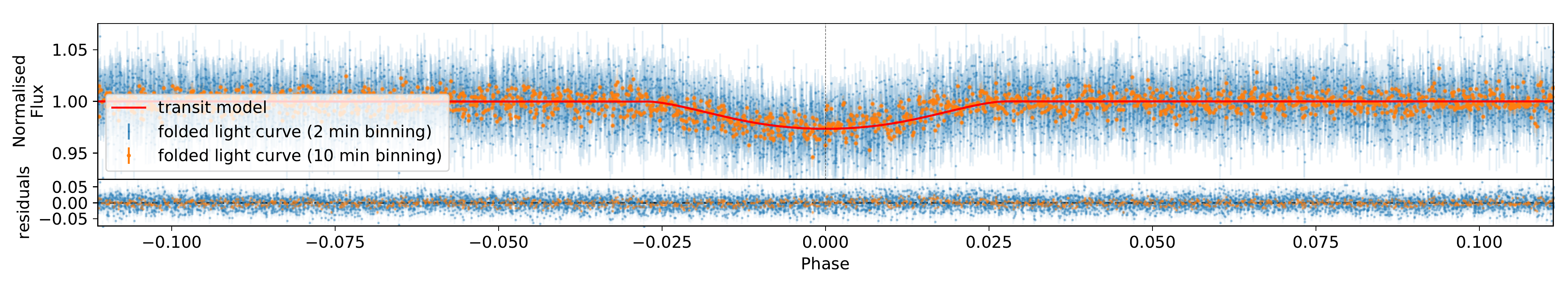}
    \caption{Phase folded light curve with best fitting transit model for TOI 142.01.}
    \label{fig:toi_142}
\end{figure*}

The second star is TOI 142 (425934411) which is also overluminous and shows displays strong emission features. Interestingly, it appears to host a giant planet ($R_P=13.31\pm1.39\,R_\oplus$) on a short period ($T\approx0.85\,$day) - see Figure \ref{fig:toi_142}. While this is unusual for such a cool star, it is not unheard of, such as K2 32b which is a known short period super-Neptune orbiting a pre-main sequence star \citep{david_neptune-sized_2016,mann_zodiacal_2016}. Further characterisation of the system however, whilst scientifically interesting, is likely to be hampered by the faintness of the host star ($G$$\sim$$15.8$).

\subsection{Planet Parameter Recovery}
\begin{figure*}
    \centering
    \includegraphics[width=\textwidth,page=1]{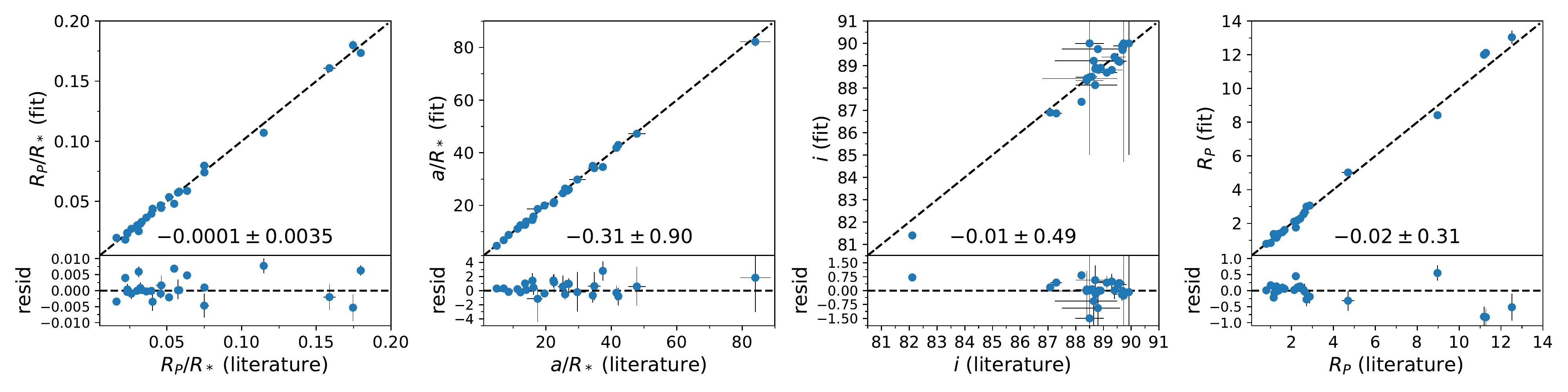}
    \caption{Comparison of $R_P/R_\star$, $a/R_\star$, $i$, and $R_P$ to literature results in Table \ref{tab:planet_lit_params}. Our two largest literature planets, TOIs 129.01 and 551.01, are hot Jupiters in a grazing configuration which leaves their radii poorly constrained. As such, they have been left off for clarity, though our results are consistent within uncertainties. The median and standard deviation of each set of residuals is annotated and excludes these two planets.}
    \label{fig:lit_planet_comp}
\end{figure*}

Table \ref{tab:planet_lit_params} collates literature parameters for previously characterised planets in our sample. These planets have typically had follow-up radial velocity observations which not only allows for planetary mass determination, but helps constrain their orbits when combined with the TESS light curves we use here (or additional time series photometric follow up). Figure \ref{fig:lit_planet_comp} compares these results to our own for $R_P/R_\star$, $a/R_\star$, $i$, and $R_P$. We find our results consistent with the literature, aside from a few exceptions discussed below.

\subsubsection{LHS 3844 b} 

\citet{vanderspek_tess_2019} reports a larger value of $Rp/R_\star$ for LHS 3844 b (TOI 136.01) than we do here, a difference we can attribute to our access to an extra sector of TESS data. While they also have ground based data, the extra TESS sector amounts to some 60 extra transits, which should give us improved precision.

\subsubsection{HATS-48 A b, TOI 178 b/e, LHS 1140 c}
Comparison with HATS-48 A b (TOI 1067.01) from \citet{hartman_hats-47b_2020} shows an inconsistent value of $R_P/R_\star$, indicating a difference in how we have modelled the light curves. While we have access to an additional sector of TESS data, the difference primarily appears to come from a) including RVs in their fit, and b) their use of an additional `dilution factor' when fitting to account for nearby unresolved stars. Such nearby stars have the effect of diluting the transit and making the transit appear shallower than it would were only the flux from the host star observed. Our transit fits, by comparison, rely on the quality of the detrending and correction for crowding already done by the TESS team and provided in their PDCSAP fluxes. 

\citet{leleu_six_2021} reports parameters for six planets orbiting TOI 178, of which only three were alerted on as TOIs. Our parameters are consistent for all but two of these, TOI 178 b (not alerted on) and TOI 178.03, both of which are relatively low SNR detections by TESS. Although our analysis includes an additional TESS sector of data, they employ higher precision data from CHEOPS to which we attribute the difference.

The analysis of LHS 1140 c (TOI 256.02) by \citet{ment_second_2019} results in a value of $R_P/R_\star$ discrepant with our own. While our analysis makes use of an additional sector of TESS data, we consider their results more reliable as they conducted a joint RV and transit photometry analysis, including additional ground based data alongside high precision \textit{Spitzer} data.

\subsubsection{WASP-43 b and HATS-6 b}
We find a consistent $R_P/R_\star$ with \citet{esposito_gaps_2017} for WASP-43 b (TOI 656.01), though our value of $R_P$ is smaller. This difference is attributable to their larger and less precise stellar $T_{\rm eff}$, with which they obtain a smaller stellar radius - resulting in a smaller planetary radii. As discussed, we are confident with our $T_{\rm eff}$ and $R_\star$ recovery, and consider the difference the result of differing approaches to stellar parameter determination.

For HATS-6 b (TOI 468.01) we find our $R_P/R_\star$ and $T_{\rm eff}$ consistent, but a different value for $R_P$ as compared to \citet{hartman_hats-6b_2015}. This difference again arises from a smaller literature value of $R_\star$. We consider our approach to radius determination using stellar fluxes more direct than the modelling based approach used here, especially given our access to precision Gaia parallaxes.

\subsection{Candidate Planet Radii Distribution}
\begin{figure}
 \includegraphics[width=\columnwidth]{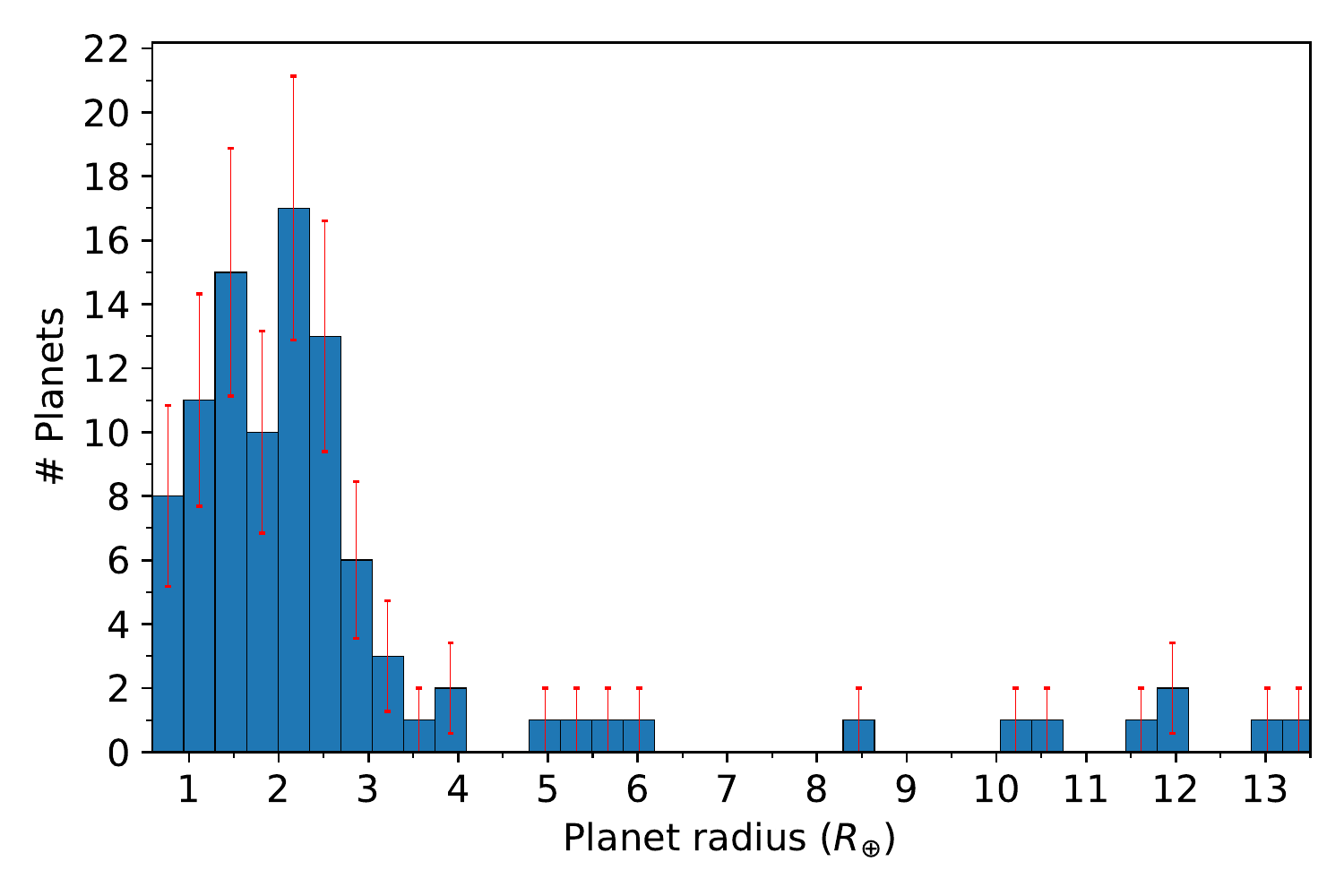}
 \caption{Histogram of candidate planet radii with $R_P<14\,R_{\oplus}$, with $0.35\,R_{\oplus}$ width bins and Poisson uncertainties. Note that we detect the exoplanet radius gap at approximately a $\sim$$1\sigma$ level, though remain limited by our small sample size.}
 \label{fig:planet_radii_hist}
\end{figure}

We plot a histogram of our candidate planet radii in Figure \ref{fig:planet_radii_hist}, which shows the existence of the planet radius gap, first identified by \citet{fulton_california-kepler_2017}, at $\sim$$1.65-2.0\,R_\oplus$ at a $\sim$$1\sigma$ level. As we remain limited by our small sample size, we do not perform any additional analysis and leave such investigations for future studies based on a larger sample of TESS planets. 

Our results however do provide encouraging further evidence for the radius gap being present around planets orbiting low-mass stars. Its detection for the stellar mass range considered here is similar to the work of \citet{cloutier_evolution_2020} who investigated a set of confirmed and candidate planets from Kepler and K2 orbiting stars with $T_{\rm eff}<4,700\,$K, with their sample being roughly a factor of ${\sim}4.5$ larger than our own. Separating their planets into bins of different stellar mass, they demonstrated that the bimodality in the radius distribution vanishes as stellar mass decreases, corresponding to the population of rocky planets beginning to dominate that of their more gas rich counterparts. They note, however, that a much larger sample of planets is required in order to properly distinguish between the various possible formation channels for the radius valley (e.g. photoevaporation, core-powered mass loss), particularly when further subdividing the sample by stellar mass. It is hoped that our results here can contribute to a larger future analysis combining Kepler, K2, \textit{and} TESS planets, perhaps also looking into correlations with stellar activity using activity measures such as we provide.

\section{Conclusions}\label{sec:conclusion}
In the work presented above, we have described our ANU 2.3$\,$m/WiFeS observing program to characterise 92 southern TESS candidate planet hosts with $3,000 \lesssim T_{\rm eff} \lesssim 4,500\,$K in order to precisely determine the radii of 100 transiting planets they host. In the process of doing so we investigated cool dwarf model atmosphere systematics, as well as developed a new photometric [Fe/H] calibration. The main conclusions from our work are as follows:

\begin{itemize}
    \item Cool dwarf MARCS model atmospheres systematically overestimate flux in the optical relative to the well produced spectral regions $5585-6029\,$\SI{}{\angstrom} and 6159-7000$\,$\SI{}{\angstrom}, with agreement being worse the cooler the star or bluer the wavelength. We report a simple linear relation parameterising the offset as a function of the observed Gaia $(B_P-R_P)$ colour, enabling the correction of synthetic Gaia $B_P$, and SkyMapper $g$ and $r$ magnitudes. We recommend that future work consider updated molecular line lists \citep{mckemmish_exomol_2019} and non-solar scaled chemical abundances (see \citealt{veyette_physical_2016}).
    \item Using the same models, a general least squares fitting approach to medium resolution optical spectra and literature photometry is not sufficient to accurately recover [Fe/H] for cool dwarfs. We instead develop an updated photometric [Fe/H] calibration for cool dwarfs, built using a sample of 69 M and K dwarfs with FGK binary companions having reliable [Fe/H] measurements. By relating the position of these isolated main sequence KM stars in $M_{K_S}-(B_P-K_S)$ space to the FGK companion, and thus system, [Fe/H], our relation can determine metallicity to a precision of $\pm0.19\,$dex for stars with $1.51 < (B_P-R_P) < 3.3$. This relation expands on the work of \citet{bonfils_metallicity_2005}, \citet{johnson_metal_2009}, and \citet{schlaufman_physically-motivated_2010}, and takes advantage of precision Gaia parallaxes (for precise distances) and kinematics (for binary identification) for the first time.
    \item We determine $T_{\rm eff}$ and $R_\star$ for our 92 TESS candidate planet hosts with a median precision of 0.8\% and 1.7\% respectively, as well as radial velocities to $\sim$$4.5\,$km$\,$s$^{-1}$. 42 of these targets did not previously have radial velocities from Gaia DR2, thus completing completing the kinematics for these stars.
    \item We report H$\alpha$ equivalent widths and Ca II H\&K $\log R^{\prime}_{\rm HK}$ for our sample, both signs of activity and youth. None of our stars display detectable Lithium $6708\,$\SI{}{\angstrom} absorption.
    \item We use our derived stellar parameters to fit the TESS light curves for our 100 planet candidates in order to determine $R_P$ with a median precision of 3.5\%. Our planet properties are consistent with the 30 already confirmed by other studies. We additionally see evidence of the planet radius gap at a $\sim$$1\sigma$ level for our low-mass stellar sample, with the robustness of the detection only limited by the small sample size.
    \item We report the existence of two likely young systems based on stellar emission and location above the main sequence: TOI 507 (TIC 348538431) and TOI 142 (425934411). The former appears to be a near-equal mass, double-lined eclipsing binary with $T_{\rm eff}$$\approx$3300$\,$K, potentially amenable to characterisation as a pre-main sequence benchmark system. TOI 142 on the other hand has a giant planet ($R_P=13.31\pm1.39\,R_\oplus$) on a short period ($T\approx0.85\,$day) orbit.
\end{itemize}
This is one of the largest uniform analyses of cool TESS candidate planet hosts to date, and the first cool dwarf photometric [Fe/H] calibration based on Gaia data. Given the major difficulties encountered using model atmospheres for [Fe/H] determination, we plan to conduct follow-up work investigating empirical or data driven approaches built upon our now large collection of cool dwarf standard spectra.

\section*{Acknowledgements}\label{sec:acknowledgements}
We acknowledge the traditional custodians of the land on which the ANU$\,$2.3 m Telescope stands, the Gamilaraay people, and pay our respects to elders past and present. We also acknowledge helpful early conversations with George Zhou about target selection and observing strategy, as well as the efforts of Andy Casey in developing a prototype data-driven model which ultimately proved out of scope for this study. We thank the anonymous referee for their helpful comments.

ADR acknowledges support from the Australian Government Research Training Program, and the Research School of Astronomy \& Astrophysics top up scholarship. M{\v Z} and MJI acknowledge funding from the Australian Research Council (grant DP170102233). 
LC is the recipient of the ARC Future Fellowship FT160100402. MJ was supported the Research School of Astronomy and Astrophysics at the Australian National University and funding from Australian Research Council grant No. DP150100250. Parts of this research were conducted by the Australian Research Council Centre of Excellence for All Sky Astrophysics in 3 Dimensions (ASTRO 3D), through project number CE170100013.

This research has made use of the Exoplanet Follow-up Observation Program website, which is operated by the California Institute of Technology, under contract with the National Aeronautics and Space Administration under the Exoplanet Exploration Program. This paper includes data collected by the TESS mission. Funding for the TESS mission is provided by the NASA Explorer Program. This work has made use of data from the European Space Agency (ESA) mission {\it Gaia} (\url{https://www.cosmos.esa.int/gaia}), processed by the {\it Gaia} Data Processing and Analysis Consortium (DPAC, \url{https://www.cosmos.esa.int/web/gaia/dpac/consortium}). Funding for the DPAC has been provided by national institutions, in particular the institutions participating in the {\it Gaia} Multilateral Agreement. This publication makes use of data products from the Two Micron All Sky Survey, which is a joint project of the University of Massachusetts and the Infrared Processing and Analysis Center/California Institute of Technology, funded by the National Aeronautics and Space Administration and the National Science Foundation. The national facility capability for SkyMapper has been funded through ARC LIEF grant LE130100104 from the Australian Research Council, awarded to the University of Sydney, the Australian National University, Swinburne University of Technology, the University of Queensland, the University of Western Australia, the University of Melbourne, Curtin University of Technology, Monash University and the Australian Astronomical Observatory. SkyMapper is owned and operated by The Australian National University's Research School of Astronomy and Astrophysics. The survey data were processed and provided by the SkyMapper Team at ANU. The SkyMapper node of the All-Sky Virtual Observatory (ASVO) is hosted at the National Computational Infrastructure (NCI). Development and support the SkyMapper node of the ASVO has been funded in part by Astronomy Australia Limited (AAL) and the Australian Government through the Commonwealth's Education Investment Fund (EIF) and National Collaborative Research Infrastructure Strategy (NCRIS), particularly the National eResearch Collaboration Tools and Resources (NeCTAR) and the Australian National Data Service Projects (ANDS). This research made use of Lightkurve, a Python package for Kepler and TESS data analysis. 

Software: \texttt{Astropy} \citep{astropy_collaboration_astropy:_2013}, \texttt{batman} \citep{kreidberg_batman_2015}, \texttt{iPython} \citep{perez_ipython:_2007}, \texttt{dustmaps} \citep{green_dustmaps_2018}, \texttt{lightkurve} \citep{lightkurve_collaboration_lightkurve_2018}, \texttt{Matplotlib} \citep{hunter_matplotlib:_2007}, \texttt{NumPy} \citep{harris_array_2020}, \texttt{Pandas} \citep{mckinney_data_2010}, \texttt{SciPy} \citep{jones_scipy:_2016}.

\section*{Data Availability}\label{sec:data_availability}
All fitted stellar and planetary results are available in the article and in its online supplementary material, and stellar spectra will be shared on reasonable request to the corresponding author. All other data used is publicly available.

\section*{Supporting Information}\label{sec:supporting_info}
Supplementary data are available at MNRAS online.




\bibliographystyle{mnras}
\bibliography{references} 



\appendix

\section{Observations}
\begin{table*}
\centering
\caption{Observing log for TESS candidate exoplanet host stars}
\label{tab:observing_log_tess}
\begin{tabular}{ccccccc}
\hline
TIC & UT Date & airmass & exp & RV & 
\multicolumn{2}{c}{SNR} \\
 &  &  & (sec) & (km$\,$s$^{-1}$) & 
(B) & (R) \\
\hline
219338557 & 19-08-25 & 1.1 & 120 & 36.00 $\pm$ 4.50 & 66 & 185 \\
201248411 & 19-08-26 & 1.1 & 200 & 17.92 $\pm$ 4.50 & 52 & 142 \\
262530407 & 19-08-26 & 1.1 & 200 & 8.71 $\pm$ 4.51 & 37 & 114 \\
118327550 & 19-08-26 & 1.0 & 300 & 6.48 $\pm$ 4.53 & 23 & 75 \\
410153553 & 19-08-26 & 1.3 & 600 & -14.96 $\pm$ 4.55 & 14 & 63 \\
62483237 & 19-08-26 & 1.0 & 150 & 6.17 $\pm$ 4.50 & 61 & 164 \\
\hline
\end{tabular}
\end{table*}

\begin{table*}
\centering
\contcaption{Observing log for TESS candidate exoplanet host stars}
\begin{tabular}{ccccccc}
\hline
TIC & UT Date & airmass & exp & RV & 
\multicolumn{2}{c}{SNR} \\
 &  &  & (sec) & (km$\,$s$^{-1}$) & 
(B) & (R) \\
\hline
141527579 & 20-02-03 & 1.4 & 900 & 24.59 $\pm$ 4.52 & 20 & 75 \\
149788158 & 20-02-03 & 1.2 & 360 & 24.03 $\pm$ 4.52 & 19 & 73 \\
34068865 & 20-02-03 & 1.1 & 30 & 50.46 $\pm$ 4.52 & 30 & 99 \\
36724087 & 20-02-03 & 1.1 & 360 & -7.62 $\pm$ 4.53 & 14 & 60 \\
359271092 & 20-02-03 & 1.2 & 20 & 37.39 $\pm$ 4.51 & 24 & 91 \\
429358906 & 20-02-03 & 1.1 & 1200 & -1.21 $\pm$ 4.54 & 6 & 43 \\
374829238 & 20-02-03 & 1.2 & 600 & 15.29 $\pm$ 4.51 & 19 & 73 \\
300710077 & 20-02-03 & 1.3 & 660 & 58.98 $\pm$ 4.53 & 12 & 60 \\
362249359 & 20-02-03 & 1.2 & 180 & 14.78 $\pm$ 4.51 & 32 & 110 \\
210873792 & 20-02-03 & 1.4 & 450 & -75.36 $\pm$ 4.51 & 22 & 78 \\
261257684 & 20-02-03 & 1.6 & 300 & 7.56 $\pm$ 4.51 & 25 & 91 \\
280437559 & 20-02-03 & 1.3 & 180 & -3.64 $\pm$ 4.50 & 25 & 91 \\
370133522 & 20-09-11 & 1.3 & 180 & -0.93 $\pm$ 4.53 & 20 & 70 \\
12421862 & 20-09-11 & 1.0 & 180 & 19.29 $\pm$ 4.51 & 28 & 97 \\
415969908 & 20-09-11 & 1.0 & 450 & 24.75 $\pm$ 4.52 & 20 & 71 \\
92226327 & 20-09-11 & 1.1 & 450 & -15.10 $\pm$ 4.55 & 9 & 44 \\
348538431 & 20-09-11 & 2.0 & 900 & -6.72 $\pm$ 4.56 & 15 & 56 \\
254113311 & 20-09-12 & 1.0 & 180 & -7.33 $\pm$ 4.50 & 39 & 115 \\
29960110 & 20-09-12 & 1.0 & 360 & 26.52 $\pm$ 4.52 & 17 & 63 \\
425934411 & 20-09-12 & 1.2 & 1200 & -47.79 $\pm$ 4.53 & 11 & 52 \\
153065527 & 20-09-12 & 1.0 & 450 & 17.50 $\pm$ 4.54 & 8 & 41 \\
50618703 & 20-09-12 & 1.3 & 120 & 10.43 $\pm$ 4.50 & 36 & 111 \\
77156829 & 20-09-12 & 1.0 & 300 & 9.06 $\pm$ 4.53 & 21 & 77 \\
231728511 & 20-09-12 & 1.1 & 450 & -5.94 $\pm$ 4.53 & 13 & 51 \\
369327947 & 20-09-12 & 1.6 & 300 & -13.78 $\pm$ 4.53 & 23 & 79 \\
175532955 & 20-09-12 & 1.0 & 360 & 76.23 $\pm$ 4.50 & 25 & 79 \\
141527965 & 20-09-13 & 1.4 & 450 & 37.43 $\pm$ 4.50 & 29 & 90 \\
260417932 & 20-09-13 & 1.2 & 300 & 36.85 $\pm$ 4.50 & 42 & 116 \\
\hline
\end{tabular}
\end{table*}

\section{Stellar Standards}\label{sec:standards}
\begin{table*}
\centering
\caption{Observing log for cool dwarf standards}
\label{tab:observing_log_std}
\begin{tabular}{ccccccc}
\hline
Gaia DR2 & UT Date & airmass & exp & RV & 
\multicolumn{2}{c}{SNR} \\
 &  &  & (sec) & (km$\,$s$^{-1}$) & 
(B) & (R) \\
\hline
19316224572460416 & 19-07-22 & 1.5 & 5 & 20.65 $\pm$ 4.50 & 108 & 251 \\
2306965202564506752 & 19-07-22 & 1.0 & 200 & 19.60 $\pm$ 4.52 & 180 & 487 \\
2603090003484152064 & 19-07-22 & 1.1 & 300 & -4.68 $\pm$ 4.53 & 100 & 281 \\
2683023811628007296 & 19-07-22 & 1.3 & 300 & 19.01 $\pm$ 4.51 & 159 & 462 \\
2828928008202069376 & 19-07-22 & 1.5 & 200 & -30.56 $\pm$ 4.51 & 158 & 458 \\
5134635708766250752 & 19-07-22 & 1.1 & 300 & 8.56 $\pm$ 4.51 & 182 & 529 \\
\hline
\end{tabular}
\end{table*}

\begin{table*}
\centering
\contcaption{Observing log for cool dwarf standards}
\begin{tabular}{ccccccc}
\hline
Gaia DR2 & UT Date & airmass & exp & RV & 
\multicolumn{2}{c}{SNR} \\
 &  &  & (sec) & (km$\,$s$^{-1}$) & 
(B) & (R) \\
\hline
3902785109124370432 & 20-02-03 & 1.4 & 300 & 20.95 $\pm$ 4.51 & 106 & 322 \\
4017860992519744384 & 20-02-03 & 1.9 & 120 & 4.28 $\pm$ 4.53 & 33 & 118 \\
4330690742322011520 & 20-02-03 & 1.4 & 120 & -25.48 $\pm$ 4.53 & 62 & 195 \\
4364480521350598144 & 20-02-03 & 1.8 & 60 & 31.88 $\pm$ 4.52 & 39 & 130 \\
5170039502144332800 & 20-02-03 & 1.2 & 15 & -1.55 $\pm$ 4.50 & 58 & 164 \\
5266270443442455040 & 20-02-03 & 1.3 & 90 & 18.99 $\pm$ 4.50 & 70 & 183 \\
\hline
\end{tabular}
\end{table*}

\begin{table*}
\centering
\contcaption{Observing log for cool dwarf standards}
\begin{tabular}{ccccccc}
\hline
Gaia DR2 & UT Date & airmass & exp & RV & 
\multicolumn{2}{c}{SNR} \\
 &  &  & (sec) & (km$\,$s$^{-1}$) & 
(B) & (R) \\
\hline
2848646203058386560 & 20-09-13 & 1.7 & 450 & -48.92 $\pm$ 4.53 & 37 & 116 \\
2868199402451064064 & 20-09-13 & 2.2 & 600 & -15.06 $\pm$ 4.56 & 21 & 85 \\
2911981886751531648 & 20-09-13 & 1.1 & 240 & 4.93 $\pm$ 4.53 & 34 & 106 \\
3011648664439643520 & 20-09-13 & 1.1 & 120 & 21.03 $\pm$ 4.50 & 46 & 127 \\
3131777319158582272 & 20-09-13 & 1.4 & 300 & -6.02 $\pm$ 4.52 & 28 & 97 \\
3266980243936341248 & 20-09-13 & 1.2 & 600 & 82.84 $\pm$ 4.52 & 28 & 93 \\
4706630501049679744 & 20-09-13 & 1.4 & 60 & 39.09 $\pm$ 4.50 & 96 & 243 \\
4871271406553197184 & 20-09-13 & 1.0 & 90 & 26.97 $\pm$ 4.50 & 53 & 140 \\
5142772953804963456 & 20-09-13 & 1.1 & 600 & -2.48 $\pm$ 4.53 & 38 & 119 \\
6391176735363140480 & 20-09-13 & 1.4 & 30 & 4.49 $\pm$ 4.50 & 69 & 192 \\
\hline
\end{tabular}
\end{table*}

\begin{table*}
\centering
\caption{Final results for cool dwarf standards}
\label{tab:final_results_std}
\begin{tabular}{cccccccc}
\hline
Gaia DR2 & $T_{\rm eff}$ & $\log g$ & [Fe/H] & $M$ & $R_\star$ & $m_{\rm bol}$ & $f_{\rm bol}$ \\
 & (K) &  &  & ($M_\odot$) & ($R_\odot$) &  & {{\footnotesize(10$^{-12}\,$ergs s$^{-1}$ cm $^{-2}$)}} \\
\hline
5853498713160606720 & 2885 $\pm$ 30 & 5.14 $\pm$ 0.02 &- &0.124 $\pm$ 0.003 &0.156 $\pm$ 0.003 &7.36 $\pm$ 0.01 &28793.2 $\pm$ 291.5\\
2640434056928150400 & 2927 $\pm$ 30 & 5.20 $\pm$ 0.02 &- &0.120 $\pm$ 0.003 &0.144 $\pm$ 0.003 &11.19 $\pm$ 0.01 &851.2 $\pm$ 8.9\\
2595284016771502080 & 3006 $\pm$ 30 & 5.03 $\pm$ 0.02 &- &0.176 $\pm$ 0.004 &0.212 $\pm$ 0.004 &10.25 $\pm$ 0.01 &2020.5 $\pm$ 21.1\\
2467732906559754496 & 3022 $\pm$ 30 & 5.10 $\pm$ 0.02 &- &0.145 $\pm$ 0.004 &0.178 $\pm$ 0.004 &11.43 $\pm$ 0.01 &680.0 $\pm$ 7.0\\
4393265392167891712 & 3035 $\pm$ 30 & 5.03 $\pm$ 0.02 &- &0.181 $\pm$ 0.004 &0.214 $\pm$ 0.004 &11.71 $\pm$ 0.01 &524.2 $\pm$ 5.4\\
4145870293808914688 & 3061 $\pm$ 30 & 5.08 $\pm$ 0.02 &- &0.153 $\pm$ 0.004 &0.187 $\pm$ 0.004 &10.65 $\pm$ 0.01 &1398.5 $\pm$ 14.2\\
\hline
\end{tabular}
\end{table*}

\begin{table*}
\centering
\contcaption{Final results for cool dwarf standards}
\begin{tabular}{cccccccc}
\hline
Gaia DR2 & $T_{\rm eff}$ & $\log g$ & [Fe/H] & $M$ & $R_\star$ & $m_{\rm bol}$ & $f_{\rm bol}$ \\
 & (K) &  &  & ($M_\odot$) & ($R_\odot$) &  & {{\footnotesize(10$^{-12}\,$ergs s$^{-1}$ cm $^{-2}$)}} \\
\hline
4519789081415296128 & 3541 $\pm$ 30 & 4.94 $\pm$ 0.02 &- &0.319 $\pm$ 0.008 &0.319 $\pm$ 0.005 &9.08 $\pm$ 0.01 &5906.7 $\pm$ 58.9\\
4519789321933481856 & 3543 $\pm$ 30 & 4.94 $\pm$ 0.02 &- &0.318 $\pm$ 0.008 &0.317 $\pm$ 0.005 &9.09 $\pm$ 0.01 &5856.8 $\pm$ 58.4\\
3131777319158582272 & 3545 $\pm$ 30 & 4.65 $\pm$ 0.02 &- &0.528 $\pm$ 0.013 &0.566 $\pm$ 0.010 &10.46 $\pm$ 0.01 &1659.9 $\pm$ 16.5\\
3184351876391975808 & 3555 $\pm$ 30 & 4.75 $\pm$ 0.02 &- &0.470 $\pm$ 0.011 &0.478 $\pm$ 0.008 &8.70 $\pm$ 0.01 &8389.8 $\pm$ 83.5\\
4293318823182081408 & 3555 $\pm$ 30 & 4.76 $\pm$ 0.02 &0.07 &0.474 $\pm$ 0.011 &0.476 $\pm$ 0.008 &7.32 $\pm$ 0.01 &29932.3 $\pm$ 298.2\\
3409711211681795584 & 3560 $\pm$ 30 & 4.73 $\pm$ 0.02 &0.06 &0.489 $\pm$ 0.012 &0.500 $\pm$ 0.009 &8.24 $\pm$ 0.01 &12905.3 $\pm$ 128.4\\
\hline
\end{tabular}
\end{table*}

\begin{table*}
\centering
\contcaption{Final results for cool dwarf standards}
\begin{tabular}{cccccccc}
\hline
Gaia DR2 & $T_{\rm eff}$ & $\log g$ & [Fe/H] & $M$ & $R_\star$ & $m_{\rm bol}$ & $f_{\rm bol}$ \\
 & (K) &  &  & ($M_\odot$) & ($R_\odot$) &  & {{\footnotesize(10$^{-12}\,$ergs s$^{-1}$ cm $^{-2}$)}} \\
\hline
5378886891122066560 & 4611 $\pm$ 30 & 4.66 $\pm$ 0.02 &- &0.664 $\pm$ 0.016 &0.634 $\pm$ 0.008 &7.25 $\pm$ 0.01 &32106.9 $\pm$ 319.6\\
6143175840404555008 & 4612 $\pm$ 30 & 4.57 $\pm$ 0.02 &- &0.706 $\pm$ 0.019 &0.724 $\pm$ 0.010 &7.90 $\pm$ 0.01 &17587.4 $\pm$ 175.1\\
6604147121141267712 & 4636 $\pm$ 30 & 4.61 $\pm$ 0.03 &- &0.690 $\pm$ 0.042 &0.677 $\pm$ 0.009 &5.95 $\pm$ 0.01 &105643.6 $\pm$ 1053.2\\
3520548825260557312 & 4667 $\pm$ 30 & 4.70 $\pm$ 0.02 &- &0.642 $\pm$ 0.016 &0.589 $\pm$ 0.008 &9.00 $\pm$ 0.01 &6412.7 $\pm$ 63.8\\
6412595290592307840 & 4670 $\pm$ 30 & 4.57 $\pm$ 0.03 &- &0.685 $\pm$ 0.041 &0.711 $\pm$ 0.009 &4.22 $\pm$ 0.01 &523617.3 $\pm$ 5227.6\\
19316224572460416 & 4808 $\pm$ 30 & 4.53 $\pm$ 0.03 &- &0.724 $\pm$ 0.043 &0.765 $\pm$ 0.010 &5.42 $\pm$ 0.01 &172178.2 $\pm$ 1715.9\\
3057712223051571200 & 4917 $\pm$ 30 & 4.56 $\pm$ 0.02 &- &0.729 $\pm$ 0.022 &0.739 $\pm$ 0.009 &6.85 $\pm$ 0.01 &46403.6 $\pm$ 463.8\\
2552925644460225152 & 5011 $\pm$ 30 & 4.57 $\pm$ 0.03 &- &0.702 $\pm$ 0.048 &0.717 $\pm$ 0.009 &5.44 $\pm$ 0.01 &169339.3 $\pm$ 1688.1\\
\hline
\end{tabular}
\end{table*}

\section{Photometric [Fe/H] Data}
\begin{table*}
\centering
\caption{Stellar pairs and primary [Fe/H] used for photometric [Fe/H] relation}
\label{tab:cpm_feh}
\begin{tabular}{ccccc}
\hline
Gaia DR2 ID (s) & $B_P-R_P$ & Gaia DR2 ID (p) & [Fe/H] ref & [Fe/H] adopted \\
 &  &  &  &  \\
\hline
853820948481913472 &1.66 & 853819947756949120 &\citealt{valenti_spectroscopic_2005} &$-0.07\pm0.03$\\
3057712188691831936 &1.71 & 3057712223051571200 &\citealt{valenti_spectroscopic_2005} &$0.07\pm0.03$\\
3732696295305334016 &1.73 & 3732696398384550400 &\citealt{mann_spectro-thermometry_2013} &$-0.57\pm0.03$\\
3999917031474342016 &1.94 & 3999916962754865152 &\citealt{ramirez_oxygen_2007} &$-0.72\pm0.05$\\
3936909723803146496 &2.03 & 3936909723803146368 &\citealt{mann_spectro-thermometry_2013} &$-0.12\pm0.03$\\
3266980243936341248 &2.04 & 3266980170921153920 &\citealt{mann_spectro-thermometry_2013} &$-0.92\pm0.03$\\
\hline
\end{tabular}
\end{table*}

\begin{table*}
\centering
\contcaption{Stellar pairs and primary [Fe/H] used for photometric [Fe/H] relation}
\begin{tabular}{ccccc}
\hline
Gaia DR2 ID (s) & $B_P-R_P$ & Gaia DR2 ID (p) & [Fe/H] ref & [Fe/H] adopted \\
 &  &  &  &  \\
\hline
1034999982042706048 &2.53 & 1035000055055287680 &\citealt{mishenina_correlation_2004} &$-0.16\pm0.07$\\
3223332136283271296 &2.53 & 3223331006703460352 &\citealt{bean_accurate_2006} &$0.17\pm0.06$\\
2384348242516852864 &2.66 & 2384342302577019264 &\citealt{mann_spectro-thermometry_2013} &$-0.04\pm0.08$\\
1416123117756120960 &2.82 & 1416124285987232256 &\citealt{mann_spectro-thermometry_2013} &$-0.12\pm0.03$\\
2005804153064109056 &2.85 & 2005798350576886144 &\citealt{fuhrmann_nearby_2008} &$0.04\pm0.05$\\
3641698032930553088 &3.00 & 3641697998570813952 &\citealt{valenti_spectroscopic_2005} &$0.56\pm0.03$\\
1240498400329095168 &3.11 & 1240498331609669120 &\citealt{ramirez_oxygen_2007} &$-0.50\pm0.05$\\
2669364200719508352 &3.30 & 2669361142702790272 &\citealt{valenti_spectroscopic_2005} &$0.16\pm0.03$\\
2411728182287010816 &2.26 & HD 219617 &\citealt{ramirez_oxygen_2007} &$-1.28\pm0.17$\\
\hline
\end{tabular}
\end{table*}

\section{Limb Darkening}
\begin{table}
\centering
\caption{Nonlinear limb darkening coefficients from \citet{claret_limb_2017}}
\label{tab:ld_coeff}
\begin{tabular}{ccccc}
\hline
TIC & $a_1$ & $a_2$ & $a_3$ & $a_4$ \\
 &  &  &  &  \\
\hline
410153553 & 1.479 & 0.174 & -0.928 & 0.472 \\
200322593 & 1.863 & -1.754 & 1.264 & -0.374 \\
92226327 & 1.589 & -1.152 & 0.657 & -0.152 \\
259962054 & 1.558 & -1.100 & 0.609 & -0.135 \\
348538431 & 3.672 & -5.444 & 4.411 & -1.347 \\
369327947 & 3.002 & -4.406 & 3.796 & -1.249 \\
\hline
\end{tabular}
\end{table}

\begin{table}
\centering
\contcaption{Limb darkening coefficients}
\begin{tabular}{ccccc}
\hline
TIC & $a_1$ & $a_2$ & $a_3$ & $a_4$ \\
 &  &  &  &  \\
\hline
261108236 & 2.771 & -4.830 & 4.877 & -1.733 \\
37749396 & 3.184 & -5.778 & 5.803 & -2.064 \\
165317334 & 2.436 & -4.170 & 4.313 & -1.549 \\
179985715 & 3.242 & -6.008 & 6.095 & -2.182 \\
33831980 & 2.889 & -5.229 & 5.357 & -1.921 \\
141527965 & 1.202 & -1.424 & 1.681 & -0.610 \\
229111835 & 1.276 & -1.586 & 1.836 & -0.661 \\
36734222 & 1.383 & -1.809 & 2.038 & -0.725 \\
254113311 & 1.652 & -2.370 & 2.551 & -0.888 \\
219338557 & 1.617 & -2.234 & 2.373 & -0.821 \\
50618703 & 1.580 & -2.170 & 2.328 & -0.808 \\
440887364 & 1.619 & -2.285 & 2.461 & -0.854 \\
101948569 & 1.667 & -2.443 & 2.644 & -0.919 \\
167600516 & 1.535 & -2.143 & 2.353 & -0.820 \\
210873792 & 1.578 & -2.258 & 2.474 & -0.863 \\
19025965 & 1.692 & -2.547 & 2.774 & -0.969 \\
251848941 & 1.576 & -2.286 & 2.519 & -0.882 \\
14165625 & 1.748 & -2.745 & 3.007 & -1.057 \\
38510224 & 1.616 & -2.431 & 2.685 & -0.943 \\
62483237 & 1.707 & -2.645 & 2.897 & -1.016 \\
158297421 & 1.728 & -2.685 & 2.923 & -1.021 \\
175532955 & 1.767 & -2.777 & 3.012 & -1.051 \\
280437559 & 1.877 & -3.038 & 3.280 & -1.147 \\
260417932 & 1.864 & -3.025 & 3.277 & -1.147 \\
201642601 & 1.870 & -3.048 & 3.303 & -1.157 \\
122613513 & 1.804 & -2.898 & 3.163 & -1.111 \\
201248411 & 3.659 & -6.702 & 6.491 & -2.171 \\
193641523 & 4.062 & -7.523 & 7.206 & -2.398 \\
\hline
\end{tabular}
\end{table}

\section{Literature Planets}
\begin{table*}
\centering
\caption{Summary of literature properties for already confirmed planets}
\label{tab:planet_lit_params}
\begin{tabular}{cccccccc}
\hline
TOI & TIC & Name & $R_P/R_\star$ & $a/R_\star$ & i & $R_P$ & Reference \\
 &  &  &  &  & (\degree) & ($R_\oplus$) &  \\
\hline
122.01 & 231702397 & TOI 122 b & $0.07500^{+0.00300}_{-0.00300}$ & $25.20^{+1.50}_{-1.50}$ & $88.40^{+0.60}_{-0.40}$ & $2.720^{+0.180}_{-0.180}$ & \citet{waalkes_toi_2020}\\
129.01 & 201248411 & HIP 65A b & $0.28700^{+0.08800}_{-0.06800}$ & $5.29^{+0.05}_{-0.04}$ & $77.18^{+0.92}_{-1.00}$ & $22.800^{+6.800}_{-5.500}$ & \citet{nielsen_three_2020}\\
136.01 & 410153553 & LHS 3844 b & $0.06350^{+0.00090}_{-0.00090}$ & $7.11^{+0.03}_{-0.03}$ & $88.50^{+0.51}_{-0.51}$ & $1.303^{+0.022}_{-0.022}$ & \citet{vanderspek_tess_2019}\\
175.01 & 307210830 & L 98-59 c & $0.03960^{+0.00110}_{-0.00080}$ & $22.50^{+1.10}_{-1.40}$ & $89.30^{+0.40}_{-0.50}$ & $1.350^{+0.080}_{-0.070}$ & \citet{kostov_l_2019}\\
175.02 & 307210830 & L 98-59 d & $0.04620^{+0.00300}_{-0.00360}$ & $37.40^{+1.10}_{-1.20}$ & $88.50^{+0.20}_{-0.50}$ & $1.570^{+0.140}_{-0.140}$ & \citet{kostov_l_2019}\\
175.03 & 307210830 & L 98-59 b & $0.02340^{+0.00090}_{-0.00080}$ & $16.20^{+0.80}_{-1.00}$ & $88.70^{+0.80}_{-0.70}$ & $0.800^{+0.050}_{-0.050}$ & \citet{kostov_l_2019}\\
\hline
\end{tabular}
\end{table*}



\bsp	
\label{lastpage}
\end{document}